\documentclass[fleqn,usenatbib]{rasti}
\usepackage{newtxtext,newtxmath}
\usepackage[T1]{fontenc}
\usepackage{graphicx}	
\usepackage{amsmath}	
\usepackage{lipsum}
\usepackage[english]{babel}
\usepackage{algorithm,algorithmicx}
\usepackage{algpseudocode}
\usepackage{soul}
\usepackage[dvipsnames]{xcolor}
\usepackage[flushleft]{threeparttable}
\usepackage{pdflscape}

\usepackage{ulem}
\usepackage{orcidlink}

\usepackage[most]{tcolorbox}
\usetikzlibrary{patterns}
\pgfdeclarepatternformonly{mystrikeout}{\pgfqpoint{-1pt}{-1pt}}{\pgfqpoint{11pt}{11pt}}{\pgfqpoint{10pt}{10pt}}%
{
  \pgfsetlinewidth{0.4pt}
  \pgfpathmoveto{\pgfqpoint{0pt}{0pt}}
  \pgfpathlineto{\pgfqpoint{10.1pt}{10.1pt}}
  \pgfusepath{stroke}
}
\newtcolorbox{tcbstrikeout}{breakable,
 enhanced jigsaw,
 opacityback=0,
 parbox=false,
 boxrule=0mm,
 top=0mm,bottom=0pt,left=0pt,right=0pt,
 boxsep=0pt,
 frame hidden,
 finish={\fill[pattern=mystrikeout] (frame.north west) rectangle (frame.south east);}
}
\graphicspath{{./}{figures/}}
\newcommand\arcdeg{\mbox{$^\circ$}}%
\newcommand{\tonalli}{{\tt tonalli}}

\newcommand{\BT}{BT-NextGen~}
\newcommand{\BOSZ}{BOSZ~}
\newcommand{\Phoenix}{PHOENIX~}
\newcommand{\SM}{SpecModels~}

\newcommand{\marcs}{MARCS}
\newcommand{\apo}{APOGEE-2}
\usepackage[dvipsnames]{xcolor}

\title[tonalli I]{\tonalli : an asexual genetic code to {characterise} \apo\ stellar spectra. I. Validation with synthetic and solar spectra.}

\author[L. Adame et al.]{Luc\'ia Adame\,\orcidlink{0000-0002-6328-6099}$^{1}$\thanks{E-mail: ladame@astro.unam.mx},
Carlos Rom\'an-Z\'u\~niga\,\orcidlink{0000-0001-8600-4798}$^{1}$,
Jes\'us Hern\'andez\,\orcidlink{0000-0001-9797-5661}$^{1}$,
Ricardo L\'opez-Valdivia\,\orcidlink{0000-0002-7795-0018}$^{1}$ and
\newauthor{Edilberto Sanchez$^{1}$}
\\
$^{1}$Universidad Nacional Aut\'onoma de M\'exico. Instituto de Astronom\'ia. A.P. 106, 22800. Ensenada, B.C., M\'exico
}

\date{Accepted XXX. Received YYY; in original form ZZZ}

\pubyear{2024}

\begin{document}
\label{firstpage}
\pagerange{\pageref{firstpage}--\pageref{lastpage}}
\maketitle

\begin{abstract}
We present \tonalli, a spectroscopic analysis {\tt python} code that efficiently predicts effective temperature, stellar surface gravity, metallicity, $\alpha$-element abundance, and rotational and radial velocities for stars with effective temperatures between 3200 and 6250~K, observed with the Apache Point Observatory Galactic Evolution Experiment 2 (\apo). \tonalli\ implements an asexual genetic algorithm to optimise the finding of the best comparison between a target spectrum and the continuum-normalised synthetic spectra library from the Model Atmospheres with a Radiative and Convective Scheme (\marcs), which is interpolated in each generation. Using simulated observed spectra and the \apo\ solar spectrum of Vesta, we study the performance, limitations, accuracy and precision of our tool. Finally, a Monte Carlo realisation was implemented to estimate the uncertainties of each derived stellar parameter. The ad hoc continuum-normalised library is publicly available on Zenodo (DOI 10.5281/zenodo.12736546).
\end{abstract}



\section{Introduction}

Multi-object, fibre spectrographs are game changers in our understanding of stellar populations. Their capabilities to acquire, simultaneously, hundreds of spectra in optical and/or near-infrared wavelengths within ample fields of view, engage the characterisation of practically all stellar ecosystems and the classification of stars across the whole Hertzsprung-Russell (H-R) diagram. Examples of these multi-object fibre spectroscopic telescopes in the optical are LAMOST \citep[Large sky Area Multi-Object fibre Spectroscopic Telescope, $5\arcdeg$ field of view,][]{Lamost12}, GALAH \citep[the Galactic Archaeology with HERMES, High Efficiency and Resolution Multi-Element Spectrograph, $2\arcdeg$ field of view,][]{galah}, and in the near infrared (near-IR) the Sloan Digital Sky Survey (SDSS) \apo\, \citep[Apache Point Observatory Galactic Evolution Experiment-2, with $0.95-1.5\arcdeg$ field of view,][]{apo2,DR17}.
Nowadays, the evolution from discrete to synoptic sky coverage \citep[e.g. the SDSS Milky Way Mapper survey,][]{Kollmeier17}, will probably redefine many aspects of our current knowledge of stellar astrophysics.

Large scale spectroscopic surveys will be successful as long as we are able to extract reliable stellar properties from the spectroscopic data. Such task is far from being simple. Various atmosphere models \citep[e.g.][]{MARCS08,Castelli03} have proven to be adequate for successful classification of main sequence and giant stars with spectral types F, G, K and early M \citep{LAMOSTDR8,Jonsson20}. For earlier spectral type (A, B, O) or late type (M,L,T) sources, there are larger discrepancies among available models \citep{birky20,zpayne}. Moreover, the characteristics of stars not located in the main or giant sequences are more difficult to generalise in models.

One particular example, which significantly motivates this work, is the classification of pre-main sequence stars, which poses a challenge for various reasons. Pre-main sequence stars evolve very rapidly (in a few million years) and their spectra are affected by a diversity of processes associated with circumstellar material, magnetic fields and accretion, and those processes are difficult to incorporate in the methods by which we compare models and data.

The algorithm and pipeline we describe in this paper can be currently applied to spectra of the \apo\ program, which was a large scale spectroscopic survey that used two identical multi-object, high-resolution (R $\sim$ 22,500) fibre spectrographs. 
These fibres collect the light while plugged at a plate in the focal plane, from which they are bundled and transmit the signal to the bench spectrographs, where it is collected on three near-IR detectors, which separate each of the spectra into three windows: blue $15145$–$15810$ \AA, green $15860$–$16430$ \AA, and red $16480$–$16950$ \AA\ \citep{Wilson10, Wilson12}.
The first one of those instruments was coupled to the SDSS 2.5m Telescope at Apache Point Observatory in New Mexico to cover fields in the Northern Hemisphere sky, while the second one extended the program to the Southern hemisphere using the Du Pont 2.5m telescope of the Carnegie Institution at Las Campanas Observatory (LCO). 
Both instruments are capable of obtaining spectra for up to 300 objects simultaneously, on a circular field with a radius of 1.5\arcdeg at the North telescope, and 0.95\arcdeg at the austral counterpart.

The \apo\ all-sky survey was designed to provide valuable constrains for the study of the chemical history and evolution of the Milky Way obtaining over 2.6$\times 10^6$ spectra of over 6.5$\times 10^5$ stars, most of them red giants in all components of the Milky Way Galaxy and the Magellanic Clouds \citep{Zasowski17, Beaton21}. Currently, the \apo\ spectrographs continue to provide H-band spectra for the synoptic, all-sky Milky Way Program with a goal to increase the survey by one order of magnitude, with a pan-optic scope that aims to explore all Galactic ecosystems across the H-R diagram \citep{Kollmeier17}.

The main SDSS \apo\ pipeline, ASPCAP \citep[APOGEE Stellar Parameter and Chemical Abundances Pipeline,][]{aspcap} was optimised for classifying spectra of their main target sample, composed of red giant stars, and thus is clearly not suitable for pre-main sequence stars classification.
Despite this difficulty, young stars were observed as ancillary project sources during Phases III and IV of the SDSS, and efforts were made to provide reliable parameters. The IN-SYNC (INfrared Spectra of Young Nebulous Clusters) project and associated pipeline \citep{Cottaar14} used a forward-modelling methodology to determine the best fit model for an observed spectra against a grid of synthetic data and provide a set of spectral parameters. They set the terrain for such work in several studies \citep[e.g.][]{dario16, yao18,Kounkel19} that provided determinations of atmospheric parameters (effective temperature, surface gravity, radial and rotational velocities, continuum veiling) for hundreds of sources in the Orion, Monoceros and Perseus star forming regions.  
More recently, the works of \citet{anet} and \citet{anet2} presented a new, data driven approach that made use of a convolutional neural network, named APOGEE Net, to estimate stellar parameters for \apo\ spectra using a collection of spectral labels based on previous determinations of parameters using both direct fitting and other data driven approaches. 
As described by \citeauthor{anet2}, APOGEE Net is computationally more efficient, and offer results comparable to those of direct fitting, but in the specific case of the surface gravities, the use of photometric labels based on evolutive models allowed them to apply a renormalisation that improved the agreement with isochrone loci for pre-main sequence stars in the $T_{\mathrm{eff}}$-$\mathrm{\log(g)}$ space, and significantly reduced systematic effects.
The APOGEE Net catalogues were used to successfully select reliable temperatures and gravities for a relatively large sample \citep[$\sim$3500 young stars in 16 star forming regions;][]{roman23}. Still, other parameters like average metallicity and derived properties like ages showed some undesired systematic trends that suggested that if we want to work on aspects like the precise determination of atomic abundances or reduce the dispersion in properties like ages and masses, the direct comparison of observed and synthetic spectra is still needed. 

In this paper we describe the code \tonalli, written in {\tt python}, that attempts simultaneous fitting of various parameters on SDSS near-infrared spectra from the \apo\ survey, against a model grid.
In general terms, this is an approach with a global philosophy similar to that of IN-SYNC in the sense that both adopt the $\chi^2$~minimisation to determine the best-fitting model. 
However, the methodologies of IN-SYNC and \tonalli\ differ in the application of distinct optimisation algorithms. 
IN-SYNC used a differential evolution algorithm to find a global minimum in the parameter space, and then applied a Markov-Chain Monte Carlo routine to determine the convergence to the optimal fit.
Our methodology is, instead, based on an unsupervised application of the Asexual Genetic Algorithm (thereafter AGA) of \citet{Canto09}, whose implementation is described in this paper. The AGA is able to converge mathematically to the closest group of models that fit the observed spectra {and also} allows to estimate realistic uncertainties with minimal biasing. Our method has the ultimate goal to deal with pre-main sequence stars but, it actually can provide simultaneous parameter fitting for all types of unevolved stars within the limitations of the models.

The content of the paper is as follows. We present and describe thoroughly the algorithm and pipeline of \tonalli\ in Section~\ref{sec:tonalli}, and discuss its performance and limitations in Section~\ref{Section:Bias}. The accuracy and precision of the method, showing that it is able to provide reliable parameter fitting plus uncertainties within 3000-6500 K and $3.0\leq\mathrm{log(g)}\leq6.0$. We present in Section~\ref{Section:Results} the results of \tonalli\ for the \apo\, solar spectrum reflected by the asteroid Vesta. Finally in Section~\ref{sec:conclusions} we present a final overview of our work.

\section{tonalli}\label{sec:tonalli}
Genetic Algorithms \citep{Fraser57}, in short, randomly select a sample of individuals from a plausible pool. 
Individuals are then compared to the target, measuring the goodness of fit of this comparison through a previously selected fitness function. 
The best individuals or \textit{parents}, are defined as those individuals having the closest resemblance to the target, and they are then selected to create the next generations of individuals by combining their features (or value of their parameters). 
AGA differs in this specific step from the classical Genetic Algorithms: the next generation of individuals is selected within the vicinity of each of the best individuals. 
This vicinity diminishes in size with each generation, and a final best-fitting individual is obtained once a stopping criterion or criteria is achieved. 

\tonalli\ compares a given \apo\ spectrum with monochromatic fluxes $F_{\lambda,\text{o}}$ and observed error $\sigma_{\lambda,\text{o}}$ to a collection of synthetic spectra.
Each synthetic spectrum is also represented by their monochromatic fluxes $F_{\lambda,\text{s}}$, and have an associated set of stellar parameters: overall metallicity ($\mathrm{[M/H]}$), abundance of $\alpha$-elements ($\mathrm{[\alpha/\text{M}]}$), logarithmic surface gravity ($\mathrm{\log(g)}$), and effective temperature ($\mathrm{T_{eff}}$). 
The spectrum can be altered with routines in \tonalli\ to simulate a projected rotational velocity ($\mathrm{v\sin i}$), limb darkening  ($\epsilon$) of the stellar atmosphere model, and radial velocity ($\mathrm{RV}$).

The code implements AGA to optimise a figure of merit (FOM). In astronomy, the usual FOM is the reduced $\chi^2$ \citep[see for example][]{Andrae2010}, closely related to the $\chi^2$ goodness of fit statistic, defined as:
\begin{equation}\label{eq:FOM}
\chi^2 = \sum_\lambda\frac{(F_{\lambda,\text{o}} - F_{\lambda,\text{s}})^2}{(\sigma_{\lambda,\text{o}})^2}.
\end{equation}
Equation~(\ref{eq:FOM}) is the implemented fitness function in our framework.
We minimise the sum of $N$ squared differences between the observed spectrum $F_{\lambda,o}$ and the synthetic spectrum $F_{\lambda,s}$, taking into account the quality of the spectrum in the model fitting by weighting the differences with the observed errors $\sigma_{\lambda,o}$. The best-fitting model is, in our scheme, the model with the minimum $\chi^2$, that is, the model with the smallest deviations from the observed spectrum in a given run.
Other fitness functions, such as the Root Mean Square Error (RMSE), can be implemented. For example, we adopt the RMSE as the figure of merit for some experiments to test the accuracy and precision of the code, as detailed in Section~\ref{Section:MinimumExp}.
With the minimisation of the FOM (eq.~\ref{eq:FOM}) we aim to find the synthetic spectrum most similar to the observed \apo\ spectrum and, therefore, determine the best set of physical parameters associated to the observed star.
Being an heuristic algorithm, \tonalli\ might not reach the optimal solution in a given single trial, and as such, we devise a repeating procedure to obtain statistical parameters to describe the stellar parameters.

The detailed implementation of AGA at \tonalli, is described in the next subsections.

\subsection{Parameters controlling the algorithm}\label{sec:parameters}

The user of \tonalli\ is allowed to modify several input parameters, three of which are used to control how quickly convergence is reached, namely: the number of individuals in the zero-$\mathrm{th}$ generation $N_0$, the number of asexual parents per generation $N_p$, and the $p$ parameter, which controls the rate of decrease, per generation, of the hyper-volume close to each asexual parent.
 
The number of individuals in the zero-$\mathrm{th}$ generation, $N_0$, is the sample of initial individuals with stellar parameters within the region allowed by the selected synthetic spectra library. This sample of $N_0$ individuals is randomly drawn, much like a Monte Carlo experiment. 

We found that with sufficiently large $N_0$ values (e.g. $N_0\gtrsim100$), \tonalli\ is able to close in the solution at earlier generations, avoiding the algorithm to approach and subsequently select sub-optimal solutions, which is possible to occur in any heuristic algorithms such as the genetic algorithms. 

Once the fitness of each individual in the zero-$\mathrm{th}$ generation is computed, we select the $N_p$ individuals with the best fitness of this generation. 
The $N_p$ individuals become the parents of the next generation.
Each subsequent generation will have $N_p\times N_p$ individuals: \tonalli\ will generate randomly, from the vicinity of each parent, $N_p-1$ offspring. 
The parents of the previous generation are also included in the fitness computation.
While the control parameters $N_0$ and $N_p$ are independent from each other, the only requirement is that $N_0 \ge N_p$.

While the zero-$\mathrm{th}$ generation is randomly generated from the entire range of parameters allowed by the synthetic library (or from a limited user-defined range), the parameters of the individuals in the subsequent generations are drawn from an increasingly smaller vicinity, or hyper-cube, centred in each parent. 
The length of the hyper-cube side corresponding to the parameter $x_i$ at generation $n$ is $\Delta x_{i,n}$. If $\Delta x_{i,0}$ is the initial search range of the parameter $x_i$, the decreased length is given by:
\begin{equation}\label{eq:deltax}
    \Delta x_{i,n} = \bigl(\Delta x_{i,0}\bigr)p^n,
\end{equation}
with $p\in(0,1)$. 
The parameter $p$ is called a \textit{convergence factor} \citep{Canto09}: larger $p$ values result in a slow convergence of \tonalli, but they guarantee a better fit to the observed spectrum compared to the results obtained with smaller $p$ values. 

For instance, a selection of $p=0.4$ implies that the sides of the search hyper-volume fall to $\sim10\%$ of their original lengths at merely the second generation, whereas the same occurs at the sixth generation for $p=0.7$.
The latter allows an assortment of physical parameters in the offspring, while the former restricts the offspring to be akin to the parents.

\subsection{Parameters controlling the synthetic spectrum interpolation}\label{ninterpolspectra}

Once the physical parameters of the offspring are known, their associated synthetic spectrum can be interpolated from a set of synthetic spectra with parameters close to those of the offspring.
Each synthetic spectrum is characterised by four parameters, namely,  $\mathrm{[M/H]}$, $\mathrm{[\alpha/\text{M}]}$, $\mathrm{\log(g)}$ and $T_{\mathrm{eff}}$, hence we need to deal with a fourth-dimensional interpolation. 
We adopt the interpolation routine \texttt{griddata} from the  library \texttt{scipy} \citep{2020SciPy-NMeth}, which constructs $N$-{D} dimensional simplexes on which it performs a linear interpolation, {being $D=4$ the dimension of the synthetic grid.}
Rather than being preset within the code, the value of $N_{\mbox{interpol}}$ (the number of synthetic spectra involved in the interpolation) is selected by the user by taking into account the number of grid points available in the synthetic library for each parameter and how coarse or fine the interpolation needs to be: $N_{\mbox{interpol}}=N_{\mathrm{[M/H]}}\times N_{\mathrm{[\alpha/\text{M}]}}\times N_{\mathrm{\log(g)}}\times N_{T}$, where $N_{\mathrm{[M/H]}}$, $N_{\mathrm{[\alpha/\text{M}]}}$, $N_{\mathrm{\log(g)}}$, and $N_{T}$ are the number of the nearest grid points to the offspring parameters in $\mathrm{[M/H]}$, $\mathrm{[\alpha/\text{M}]}$, $\mathrm{\log(g)}$ and $\mathrm{T_{eff}}$, respectively.
This input parameter can noticeably impact in the precision of the results obtained by \tonalli, and it is library dependent. 
For the synthetic library adopted in this paper \citep[\marcs,][]{MARCS08,Jonsson20}, the minimum possible value of $N_{\mbox{interpol}}$ is $N_{\mbox{interpol}}=2\times2\times2\times2=16$. 
We adopt this value in the first step of \tonalli; in the last and subsequent refinement, this value is changed to $N_{\mbox{interpol}}=2\times2\times4\times4=64$ to obtain the final, best-fitting solution, allowing a finer interpolation in the $\mathrm{\log(g)}-\mathrm{T_{eff}}$ space. 
The final value of $N_{\mbox{interpol}}$ has a noticeable impact in both the computation time and, less strongly, the resulting $\mathrm{\log(g)}-\mathrm{T_{eff}}$ values. 
The obtained stellar parameters compares better with the expected values when we increase the number of synthetic spectra to perform the interpolation; nevertheless, the discrepancy between coarse and finer results should disappear when the numerical experiment is repeated enough times.

\subsection{The algorithm}\label{sec:algorithm}

We now explain the core of \tonalli: the code aims to obtain the stellar parameters of an observed \apo\ spectrum by finding the best-fitting interpolated counterpart from an spectral library. 
We also include the pseudo-codes of the AGA implementation in \tonalli\ in the Appendix \ref{appendix:codes}.


\subsubsection{Synthetic stellar spectra libraries}\label{Section:marcsSpectra}

The best-fitting synthetic spectrum is interpolated from a library of synthetic spectra selected by the user. 
To achieve the above, the user can select one library from the five available, namely: \BT \citep{Allard11,Allard12}, \BOSZ \citep{Bohlin17}, \marcs\ \citep{Jonsson20}, \Phoenix \citep{Husser13}, and \SM \citep{Coelho05}.
Thus, the best-fitting model and its associated parameters depend strongly on the selected synthetic library. 
For the purpose of this work, we adopt \marcs\ to present the implementation of \tonalli. 
The discussion of the dependence of the stellar parameters obtained by \tonalli\ from different libraries is deferred to a subsequent work.

\begin{figure*}
\includegraphics[alt={Graphs comparing selected continuum normalised spectra to the unit horizontal line.}, width=2\columnwidth]{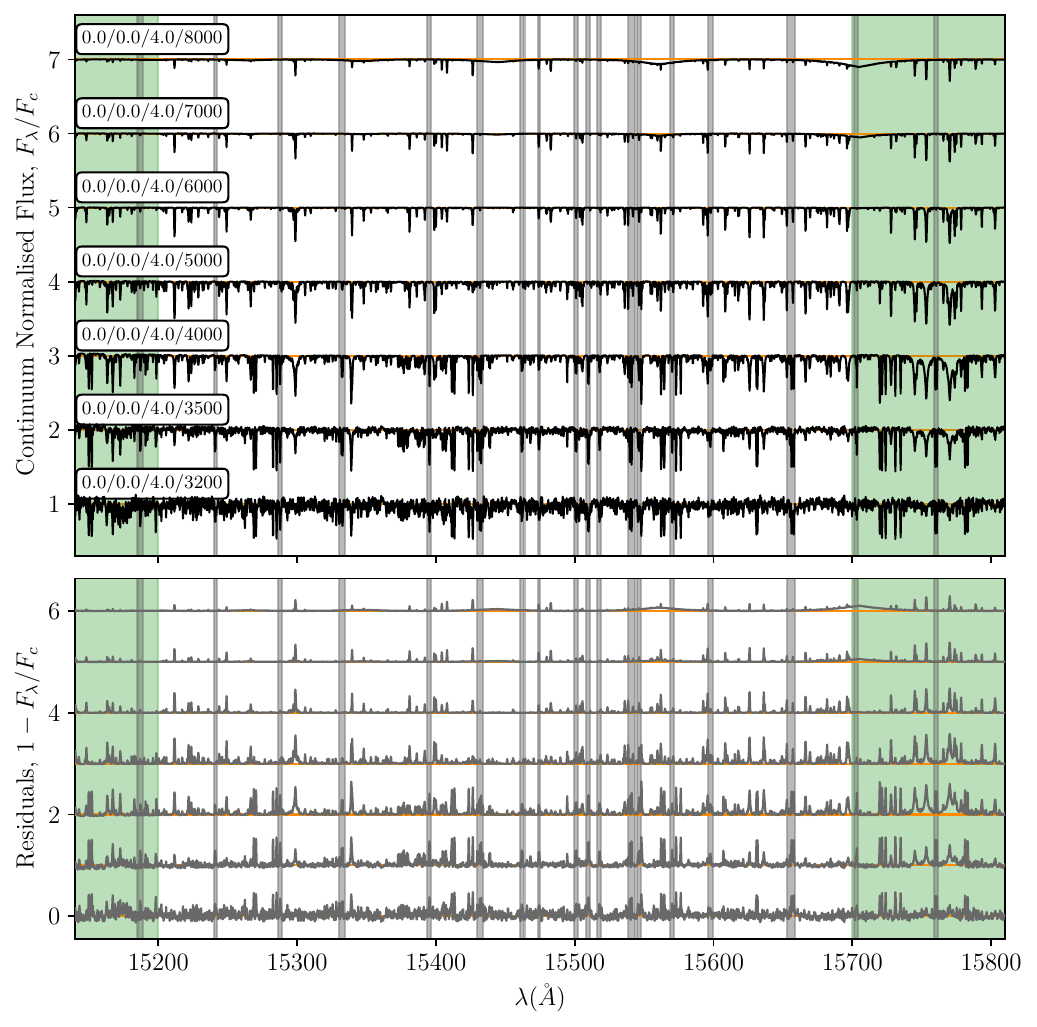}
\caption{Continuum normalised \marcs\ spectra (\textit{top panel}: \textit{black} line: continuum normalised flux) and residuals (\textit{bottom panel}, \textit{gray} line: residuals) for the \apo\ \textit{blue} chip wavelength region. The physical parameters of the star are listed above the spectrum: metallicity/$\alpha$-elements abundance/$\mathrm{\log(g)}$ (dex) /effective temperature (in K). The telluric absorption lines (\textit{grey} vertical regions) and the wavelength regions at the end of each chip (\textit{pale green} vertical regions) are shown; these regions are excluded from the fitness computation (Section~\ref{section:normalisationSpectra}). {\textit{Solid orange} horizontal lines represents the lines $y=1+c$ (\textit{top panel}) and $y=0+c$ (\textit{bottom panel})}.\label{fig:normspectra_blue}}
\end{figure*}

The synthetic spectra have been normalised in advance.
Each spectrum in a synthetic library is convolved with a Gaussian profile to match the \apo\ resolution ($R\sim 22,500$) and then normalised following the same iterative procedure we devise to normalise the observed spectra (see Section \ref{section:normalisationSpectra}, below). 
Figure~\ref{fig:normspectra_blue} shows the resulting \textit{blue} chip continuum normalised spectra for a representative set of atmospheric parameters. 

\subsubsection{Preparing the observed spectrum}\label{section:normalisationSpectra}

The observed \apo\ spectrum is cleaned and then continuum-normalised, if it is not already. 
For this procedure, we start by removing the telluric emission lines and those pixels with bad bits as specified by
{\tt \apo\_PIXMASK} bit mask flags \citep[see][]{Holtzman15} from the spectrum. 
Then, we reject the pixels with a signal-to-noise ratio \footnote{SNR$_\lambda=F_{\lambda,\text{o}}/\sigma_{\lambda,\text{o}}$, where $F_{\lambda,\text{o}}$ is the measured monochromatic flux in the pixel and $\sigma_{\lambda,\text{o}}$ its measured error} smaller than a user-input value; we set this minimum SNR to 50. 
In addition, we reject pixels with measured errors $\sigma_\text{o}(\lambda)<0$.
We then smooth the spectrum by convolving it with a box filter kernel 15 wavelength units wide, using the routines provided by \texttt{astropy.convolution} \citep{astropy:2013,astropy:2018}.

The resultant, cleaned and smoothed spectrum can be normalised as follows: the normalisation is applied in an iterative way and performed separately for each of the three \apo\ detectors. 
First, we find a best-fitting polynomial for each chip of the observed spectrum by means of the Bayesian Information Criterion \citep[BIC, ][]{Schwarz78}. 
{We fit each chip spectrum with a collection of polynomials $P_n(\lambda)$, with degrees ranging from $p=1$ to $p=16$ \citep[using the \texttt{polyfit} function from \texttt{numpy},][]{2020NumPy-Array}. With each fitting polynomial, we construct 16 continuum normalised spectra $S_n(\lambda)$. As our target function is the function $S(\lambda)=1$, we compute the residual sum of squares (RSS) for each $S_n(\lambda)$. Intuitively, we expect that the model with the smallest value of RSS, but we also want to apply the parsimony principle. Enter the BIC, which is one of the criteria in model selection to minimise the RSS and to take into account the parsimony principle, as it penalises the increase in the number of free parameters (in our case, the degree of the fitting polynomials):
\begin{equation*}
\mathrm{BIC}(p)= N_{cl}\log(\mathrm{RSS}/N_{cl})+(p+1)\log(N_{cl})
\end{equation*}
where $N_{cl}$ is the number of wavelengths in the spectrum $S_n(\lambda)$, and $p+1$ is the number of free parameters of the polynomial fit. The best-fitting polynomial is then the one having the smallest value of the Bayesian Information Criterion.}

Secondly, the observed spectrum is normalised by dividing each pixel flux with the value of the polynomial function {(obtained above with the BIC)} at the given pixel. 
Third, the normalised spectrum is then subjected to an asymmetric $\sigma$-clipping procedure in order to remove \textit{noisy} pixels (absorption lines or other outliers) from the real continuum of the spectrum: we remove these pixels from the observed spectrum. 
The $\sigma$-clipping calculates the standard deviation $\sigma[S(\lambda)]$ and the median $\mathrm{Med}[S(\lambda)]$ of the continuum normalised spectrum, and then removes all the pixels with fluxes outside the range $[\mathrm{Med}[S(\lambda)]-\delta_l\times \sigma[S(\lambda)], \mathrm{Med}[S(\lambda)]+\delta_u\times \sigma[S(\lambda)]]$, that is, normalised fluxes below $\delta_l$ and above $\delta_u$ times the standard deviation from the median of the flux distribution. We adopt the function {\tt sigma\_clip} from {\tt Astropy} \citep{astropy:2018}.
The {described procedure} is repeated until the $\sigma$-clipped spectrum is equal to the $\sigma$-clipped spectrum of the previous iteration.
Once the latter condition is reached, we divide (pixel by pixel) the {original }observed spectrum by the {best-fitting polynomial constructed with the }$\sigma$-clipped spectrum, obtaining the normalised observed spectrum.
The pseudo-codes detailing the iterative $\sigma$-clipping are shown in Appendix~\ref{appendix:codes} (algorithms~\ref{alg:normspec} and \ref{alg:zeroclip}).
In addition, we show the continuum normalisation iterative process and the resulting continuum normalised spectra for some iterations in Figures~\ref{fig:polinomioiterations} and~\ref{fig:normfluxiter}, respectively.

The continuum normalisation procedure has some potential pitfalls owing to the presence of the Brackett series lines close to the extremes of the chips.
To minimise their influence, we restrict the wavelength region where the FOM is computed: \textit{blue} chip: $15200\le\lambda\le15700$~\AA; \textit{green} chip: $15950\le\lambda\le16300$~\AA; \textit{red} chip: $16600\le\lambda\le16850$~\AA.
We blindly apply this rule for any input spectrum, unless there is strong evidence that the spectrum corresponds to an early type star (Section \ref{Section:MLClass}); if this is true, we then either further restrict the FOM computing region to the \textit{blue} chip solely or decide to stop the code at this point. After this step, the code can proceed to AGA to obtain the best-fitting spectrum and its associated atmospheric parameters from the cleaned and normalised observed spectrum.

\subsubsection{Spectrum classification: identifying high temperature stars}\label{Section:MLClass}

We implement a supervised machine-learning spectrum classifier, the k-neighbours classifier \texttt{KNeighborsClassifier} from the machine learning library \texttt{scikit-learn} \citep{scikit-learn}, using the equivalent widths of three prominent absorption features in the H band as labels. 
From the absorption features identified by \citet{Covey10} and \cite{Newton15}, we select Mg~{\sc i} (at $\lambda1.57\umu$m), Al~{\sc i} (Al-a at $\lambda1.67\umu$m), and the CO(6,3) band-head at $1.62\umu$m \citep[{which is} strong for giant M stars,][]{Origlia93}. 
The aim of this machine-learning spectrum classifier is to identify high temperature stars ($\mathrm{T_{eff}}\gtrsim 6000$~K), or stars with emission lines in their \apo\ spectrum.
If the classifier detects a possible high temperature star, the user can flag \tonalli\ to restrict the comparison of the observed spectrum to synthetic stars with $\mathrm{T_{eff}}\ge 4000$~K, and to compute the FOM masking the green and red \apo\ chips, as we find the continuum normalisation of the \apo\ chips does break down at the window extremes for some combinations of $\mathrm{T_{eff}}-\mathrm{\log(g)}$.

On the other hand, if the classifier detects a possible emission-line star, \tonalli\ terminates, allowing the user to inspect the spectrum to verify the classification.
For the remaining star classification, the code continues computing the FOM using the 3 chips.

We construct an input set for the spectrum classifier by selecting stars with \apo\ spectra from the Pleiades, a relatively young open cluster, and the W3/4/5 complexes, a massive star forming region; the classifier was developed with young stars in mind.
The Pleiades set comprises 82 stars with known spectral type and 6 stars with recent determination of their temperature; those stars with spectral type F0 or later are assigned to the label 0 (74 stars), whereas the earlier spectral type stars comprise the label 1 stars of our input set (14 stars).
For the stars in the W3/4/5 regions, we inspected the H-band spectra of three \apo\ plates.
Those stars with conspicuous early type features were selected, resulting in 227 label 1 stars and 12 label 2 stars. 
We assign the stars with line emission in the H-band \apo\ spectra the label 2 in our scheme.
A subsequent literature revision show that of these 239 W3/4/5 stars, 216 stars have spectral type A4 or earlier; the rest of the stars (23) are yet to be classified \citep{RomanLopes19}.
We compute the equivalent widths of the Mg~{\sc i}, Al~{\sc i}, and CO features of the input set as described next.

The equivalent width $W_\lambda$ of each absorption feature is obtained following \cite{Hillenbrand95} and \cite{Hernandez04}; the continuum at the centre of the spectral feature is constructed by the interpolation of the fluxes of the two nearest bands to feature (the \textit{blue} and the \textit{red} continuum bands): 
\begin{equation}
    F_\text{c,f}= F_{b}+\frac{\lambda_\text{f}-\lambda_\text{b}}{\lambda_\text{r}-\lambda_\text{b}}(F_\text{r}-F_\text{b}),
\end{equation}
where the subscripts $\text{b}$ and $\text{r}$ refers to the blue and red bands, respectively.
The equivalent width of the feature is then:
\begin{equation}
    W_\text{f} = \biggl(1-\frac{F_{\lambda,\text{f}}}{F_\text{c,f}}\biggr)\Delta \lambda_\text{f},
\end{equation}
where $F_{\lambda,\text{f}}$ and $\Delta \lambda_\text{f}$ are the flux and the width of the feature band, respectively.
Table~\ref{table:specfeatmlc} lists the wavelengths of the feature and adjacent bands. 
The observed equivalent widths of the input set, together with their internal class and spectral types (if available), are listed in Table~\ref{table:ewobsmlc}.

\begin{figure*}
\includegraphics[alt={Graphs showing the relations within the equivalent widths of Magnesium I, Aluminium I, and CO of the training set for the machine learning classifier.}, width=1.5\columnwidth]{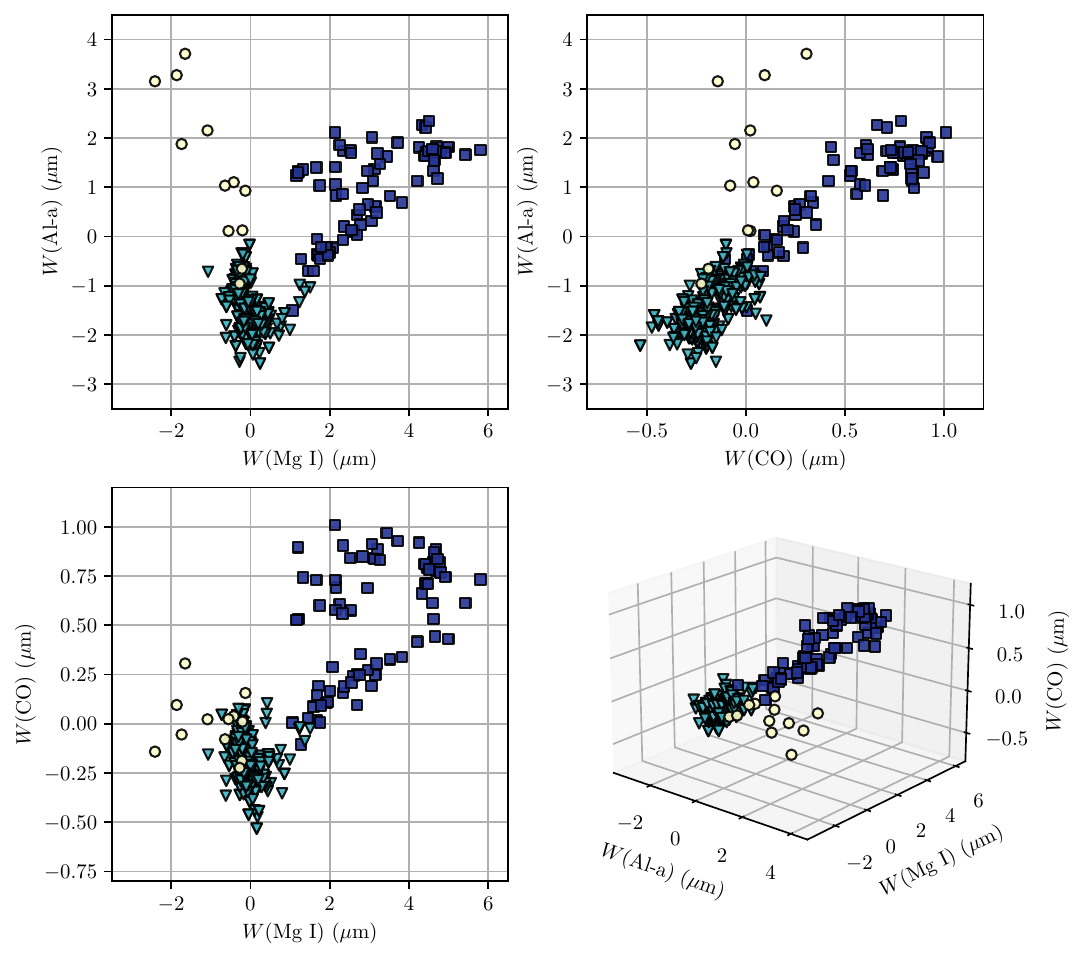}
\caption{Observed equivalent widths $W$(Mg~{\sc i},), $W$(Al~{\sc i},), and $W$(CO) of the stars comprising the machine-learning classifier training set implemented in \tonalli; \textit{blue squares}: label 0 (late type) stars; \textit{green triangles}: label 1 (early type) stars;  \textit{pale yellow circles}: label 3 (emission line) stars. \textit{Top-left panel}: correlation between the Al-a and Mg~{\sc i} equivalent widths. \textit{Top-right panel}: correlation between the Al-a and CO equivalent widths. \textit{Bottom-left panel}: correlation between the CO and Mg I equivalent widths. \textit{Bottom-right panel}: 3D plot of the correlation between the CO, Al-a, and Mg ~{\sc i} equivalent widths.}\label{fig:trainingset}
\end{figure*}

Figure~\ref{fig:trainingset} shows the 2D and 3D correlations between the Mg~{\sc i}, Al-a, and CO equivalent widths of the input set. 
It is evident that most of the late type stars have positive equivalent widths. However, there is no clear division between the three classes in the 2D plots. 
In contrast, the 3D correlation plot (bottom-right panel of Figure~\ref{fig:trainingset}) shows a somewhat clean separation between the classes. 
For the k-neighbours classifier, \texttt{KNeighborsClassifier}, our pre-classified 327 stars represent the input set, which is split into a test set ($20\%$ of the input stars) and a train set ($80\%$ of the input stars).
We then construct the k-neighbours classifier with the inverse of the Euclidean distance between the {observed and the $k$-neighbours coordinates (the triad of equivalent widths) as a weighting (setting the option {\tt weigthts} to \textit{'distance'} in the {\tt KNeighborsClassifier})}.
With this model, the classifier can predict whether an input spectrum corresponds to either a low-temperature star or a high-temperature/emission line star, and therefore \tonalli\ can constrain the search range of temperature of the synthetic spectra accordingly. 
For label~0 (low-temperature) stars, we constrain the search to spectra with $\mathrm{T_{eff}}\le7000$~K, whereas for label~1 (possible high temperature) stars, the search is limited to $\mathrm{T_{eff}}\ge4000$~K

Clearly, for genuine early type temperature candidates, the determination of spectral parameters using the $4000-8000$~K \marcs\ library provides acceptable results for super-solar type stars (A6V and later). 
At this point, \tonalli\ is not able to provide parameters for exemplars with types earlier
than A5 using the limited $\mathrm{T_{eff}}$ range of \marcs. 
As shown by \citet{RomanLopes18} and \citet{RamirezPreciado20}, the correct classification of O, B and possibly A type stars using the \apo\ spectral  range is constrained to a few spectral features, as synthetic model libraries are not sufficiently well calibrated in the near-IR. 
The works of \citet{zpayne} and \citet{anet2} are examples of recent efforts to provide acceptable spectral labels for \apo\ spectra of early type stars using machine learning techniques. 

\begin{landscape}
\begin{table}
\caption{Spectral Features for the Machine Learning Classifier}\label{table:specfeatmlc}
\begin{threeparttable}
\begin{tabular}{c|cc|cc|cc}
Feature &\multicolumn{2}{c}{Feature Window} & \multicolumn{2}{c}{Blue Window} &\multicolumn{2}{c}{Red Window}\\
& $\lambda_0$ & $\lambda_f$ & $\lambda_0$ & $\lambda_f$ & $\lambda_0$ & $\lambda_f$ \\
\hline
Mg~{\sc i} ($1.57\umu$m)\tnote{a}&  1.5737 & 1.5790 & 1.5640 & 1.5680 & 1.5790 &1.5815 \\
Al-a ($1.67\umu$m)\tnote{b}& 1.6714 & 1.6741 & 1.6580 & 1.6630 & 1.6780 & 1.6815 \\
CO ($1.61\umu$m) & 1.6245 & 1.6265 & 1.6120 & 1.6150 & 1.6265 & 1.6295\\\hline
\end{tabular}
  \begin{tablenotes}
    \item[a] Window wavelengths from \citet{Newton15}.
    \item[b] Window wavelengths from \url{https://github.com/ernewton/nirew}.
  \end{tablenotes}
\end{threeparttable}
\end{table}

\begin{table}
\caption{Training Set for the Machine Learning Classifier}\label{table:ewobsmlc}
\begin{threeparttable}
\begin{tabular}{llllcllllrrrc}
ID\tnote{a} & 2MASS & R.A.\tnote{b} & Dec.\tnote{b} & Region & H & Main Type\tnote{c}  &  Sp. Type\tnote{d} & Sp. Type\tnote{e} & $W$(Mg~{\sc i}) & $W$(Al-a) & $W$(CO) & Label\\
& & (deg) & (deg) & & (mag) & & & & ($\umu$m) & ($\umu$m) & ($\umu$m) & \\\hline 
LS I +60 226 & 2MASS J02175321+6111129 & 34.4717 & 61.1869 & W34 & 8.245 & Star & O8.5-O9IV-V & O8.5-O9IV-V (18) & -0.1135 & -0.6577 & -0.0139 & 1 \\
TYC 4046-1396-1 & 2MASS J02175521+6055410 & 34.4801 & 60.9281 & W34 & 9.638 & Star & B9V &  & -0.1111 & -1.5665 & -0.2410 & 1 \\
GSC 04046-00261 & 2MASS J02175787+6046104 & 34.4911 & 60.7696 & W34 & 9.979 & Star & A0V &  & -0.1715 & -2.0416 & -0.3586 & 1 \\
HD 14061 & 2MASS J02185723+6108453 & 34.7385 & 61.1459 & W34 & 8.581 & Star & B9V & B9V (13) & 0.3878 & -1.6646 & -0.2704 & 1 \\
TYC 4046-1130-1 & 2MASS J02191646+6103295 & 34.8186 & 61.0582 & W34 & 10.063 & D/M Star &  &  & -0.0857 & -2.1342 & -0.1963 & 1 \\
TYC 4046-232-1 & 2MASS J02193602+6131517 & 34.9001 & 61.5310 & W34 & 9.859 & Star & B5 &  & -0.0208 & -1.4875 & -0.1936 & 1 \\
TYC 4046-835-1 & 2MASS J02194353+6041543 & 34.9314 & 60.6984 & W34 & 10.395 & Star & B3 &  & -0.3792 & -1.0018 & -0.2825 & 1 \\
BD+59 465 & 2MASS J02194779+6039137 & 34.9491 & 60.6538 & W34 & 8.905 & EL Star & B0 & OB-e (4) & -1.7359 & 1.8785 & -0.0548 & 2 \\
BD+60 464 & 2MASS J02195769+6129139 & 34.9904 & 61.4872 & W34 & 9.505 & Star & A0V & A0V (13) & -0.1101 & -1.7376 & -0.3334 & 1 \\
TYC 4046-1453-1 & 2MASS J02200031+6030271 & 35.0013 & 60.5075 & W34 & 8.992 & Star & B9V &  & -0.2149 & -0.6554 & -0.1885 & 2\\\hline
\end{tabular}
\begin{tablenotes}
\item[a] Table \ref{table:ewobsmlc} is published in its entirety in the machine-readable format. A portion is shown here for guidance regarding its form and content.
\item[b] Right Ascension and Declination coordinates are J2000.0.
\item[c] Main Type and Spectral Type from SIMBAD Astronomical Database \citep{Simbad}
\item[d] Spectral Type references in parenthesis.
(1): \citet{Breger84}
(2):\citet{Cannon93}
(3): \citet{Fehrenbach66}, (4):\citet{Hardorp59}, (5):\citet{Haro64}, (6):\citet{He19}, (7):\citet{Ishida70}, (8):\citet{Kiminki15}, (9):\citet{Koenig11}, (10):\citet{Kounkel19}, (11):\citet{Carrera19}, (12):\citet{Maiz19}, (13):\citet{McCuskey74}, (14):\citet{Mendoza56}, (15):\citet{Nesterov95}, (16):\citet{Prosser91}, (17):\citet{Raddi13}, (18):\citet{RomanLopes19}, (19):\citet{Voroshilov85}.
\end{tablenotes}
\end{threeparttable}
\end{table}
\end{landscape}

\subsubsection{Construction of the zero-$\mathrm{th}$ generation}\label{Section:ZerothG}

The zero-$\mathrm{th}$ generation consists of $N_0$ individuals, with attributes (the parameters $x_1=\mathrm{[M/H]}$, $x_2=\mathrm{[\alpha/\text{M}]}$, $x_3=\mathrm{\log(g)}$, $x_4=\mathrm{T_{eff}}$, $x_5=\mathrm{v\sin(i)}$, $x_6=\epsilon$ and $x_7=\mathrm{RV}$) randomly and independently generated from a uniform distribution with limits either given by the selected synthetic library or by the user. 
Thus each individual $j$ is characterised by the parameters $x_{(j,i)}$, with $i=1$ to $7$.

The synthetic spectrum of the individual is interpolated from the nearest $N_{\mbox{interpol}}$ spectra with star parameters close to the parameters $x_{(j,1)}$, \ldots, $x_{(j,4)}$ of that individual, as explained in Section~\ref{ninterpolspectra} above.

The interpolated synthetic spectrum of the $j-\mathrm{th}$ individual, $S_j(\lambda)$, is then convolved to consider rotational velocity broadening, as set by the $x_{(j,5)}$ attribute, using the routine \texttt{fastRotBroad} from the PyAstronomy collection \citep{pya}.

Next, the synthetic spectrum is {Doppler shifted} using the routine \texttt{ dopplerShift} from \textit{PyAstronomy}, with the Doppler shift attribute $x_{(j,7)}$. Once the synthetic spectrum of the individual $j$ is rotationally broadened and Doppler shifted, the fitness of the individual is measured using the FOM (eq. \ref{eq:FOM}), which gives the value of $\chi_j^2$ of the spectrum.

We set the limb-darkening parameter to a constant value, $\epsilon=0.4$. 
While the limb-darkening depends on both the observation window and the star effective temperature \citep{Magic15}, the adopted value of $0.4$ is appropriate for the infrared \apo\ spectra of M dwarfs \citep{Gilhool2018}. 
During extensive tests performed to probe the accuracy and precision of \tonalli, we chose to keep the parameter fixed.  However, we allow the parameter $\epsilon$ to be modified along with the input parameters to run the code.

The parameters $x_{(j,i)}$, the spectrum $S_j(\lambda)$ and the aptitudes $\chi_j^2$ of all the $N_0$ individuals are collected in a matrix $V_0$ of dimension $N_0\times 8$. 
The matrix $V_0$ is then sorted by ascending $\chi^2$. 
From this matrix, the best $N_p$ individuals, those with the smallest $\chi^2$ values, are then selected to become the parents of the next generation. 
The parameters, spectra and aptitudes of this parent set are stored in the matrix $V_{(0,\text{best})}$ (with dimension $N_p\times 8$).

\subsubsection{Subsequent generations}\label{Section:subgen}

The matrix $V_{(k-1,\text{best})}$ contains the information of the fittest individuals of the previous generation, which are the parents of the current generation $k$.
The volume of the search hyper-cube, which is centred in each $j$ parent, decreases with each generation, and their sides $\Delta x_{(i,k)}$ are reduced following equation~(\ref{eq:deltax}). 
Notice that the values $\Delta x_{(i,k)}$ are the same for all the $N_p$ parents.

Once the lengths $\Delta x_{(i,k)}$ are known, the asexual reproduction of the $N_p$ parents can proceed (see Algorithm~\ref{alg:agtonalli} in Appendix~\ref{appendix:codes}).
The parameters of the $j$-parent are denoted by $x_{(j,i,\text{best})}$.
The offspring of the $j$-parent is also randomly and independently generated from a uniform distribution within the limits $x_{(j,i,\text{min})}=x_{(j,i,\text{best})}-\Delta x_{(i,k)}/2$ and $x_{(j,i,\text{max})}=x_{(j,i,\text{best})}-\Delta x_{(i,k)}/2$. 
Each one of the $N_p-1$ offspring is characterised by the parameters $x_{(j,i)}$.
The interpolation of the spectra of the offspring, the rotational broadening, the Doppler shifting, and the fitness computation proceed as described above for the individuals of the zero-$\text{th}$ generation.

The stellar parameters, the spectra and their fitness, of all the $N_p\times N_p$ individuals in generation $k$, are collected in the matrix $V_k$, of dimensions $(N_p\times N_p)\times 8$. 
The parents are also included in this matrix: they are competing against their progeny to be the parents of the next generation. 
Again, the matrix $V_k$ is sorted with increasing $\chi^2$. The $N_p$ individuals with the smallest $\chi^2$ values are then collected in the matrix $V_{(k,\text{best})}$ (with dimension $N_p\times 8$).

\subsubsection{The best-fitting model}\label{Section:bestfit}
The code \tonalli\ finishes when the length of the temperature side of the hyper-volume $(\Delta T)_k=(\Delta T)_0 p^k$ at generation $k$ is less than a critical value $(\Delta T)_c$, that is, $(\Delta T)_k\le 1$~K. 

For a given stellar spectrum, the convergence criteria (such as the difference between the FOM of the fittest and of the worst individuals in a given generation being smaller than a preset value, or the difference between the values of the FOM of the fittest individual of the current generation and of the previous generation being smaller than another preset value) may not be fulfilled before $(\Delta T)_k$ reaches~1~K. 
The preset convergence values can be relaxed, or, since the convergence factor $p$ controls the rate of the temperature decrease per generation, the criterion could be fulfilled by increasing the convergence factor before the length $(\Delta T)_k$ reaches 1K, which in turn rises the number of computed generations.
Thus the value of the optimal convergence factor $p$ would need to be determined on a case by case basis; for this work, we adopt the minimum length in temperature as a fixed convergence criteria. 
However, the matrix $V_{(k=\text{last,best})}$ typically contains offspring with the same or very close stellar parameters values and spectra.

At any rate, the best-fitting model we select is the model with the smallest $\chi^2$ value from the last generation, which is also the model with the smallest $\chi^2$ of all generations.

\subsubsection{Selection of the input parameters}

Both the accuracy and the precision of the results obtained by \tonalli\ depend on the input parameters $N_p$ (the number of asexual parents per generation), $p$ (the convergence factor that modulates search hyper-volume decrease rate), $N_{\text{interpol}}$ (the number of the nearest spectra needed in the interpolation routine), and on the selected comparison library (in the present work, the library \marcs).

To obtain the appropriate combination of input parameters $N_0$, $N_p$, $p$, and $N_{\text{interpol}}$, we explored the input parameter space and found that the number of spectra employed in the interpolation had the largest impact on the quality of the resultant best-fitting model. {Increasing the \textit{fine} $N_{\text{interpol}}$ implies that synthetic spectra characterised by physical parameters distant to the parameters of the generated individual will have an input in the interpolated spectra, effectively worsening its fitness.}
{We defer to Appendix~\ref{appendix:inputparameters} for the analysis of the influence the input parameters $N_0$, $N_p$, and $p$ have in the recovery of the solar atmospheric parameters. From the experiments carried out and detailed in Appendix~\ref{appendix:inputparameters}, we}
suggest the following values for the input parameters for the \marcs\ synthetic spectra library {as the minimum values to still obtain accurate results}: $N_0=240$, $N_p=10$, $p=0.4$. For the \textit{coarse} interpolation: $N_{\mathrm{[M/H]}}=N_{\mathrm{[\alpha/\text{M}]}}=N_{\mathrm{\log(g)}}=N_{T}=2$, while for the \textit{fine} interpolation: $N_{\mathrm{[M/H]}}=N_{\mathrm{[\alpha/\text{M}]}}=2$, $N_{\mathrm{\log(g)}}=N_{T}=3$ or $4$. 

For the selection of $N_{\mathrm{[M/H]}}=N_{\mathrm{[\alpha/\text{M}]}}=2$, $N_{\mathrm{\log(g)}}=N_{T}=3$, we find the input parameters $N_0=240$, $N_p=10$, and $p=0.4$ to provide quick and accurate results, albeit at the expense of the precision. 
    The experiments with the solar spectrum, detailed in Appendix~\ref{appendix:inputparameters}, show the combination $N_p$-$p$ rules both the quality of the results and the computing time, as they regulate the total number of offspring individuals, $N_{\mathrm{off}}$ computed through a \tonalli\ run. 
    A number of $N_{\mathrm{off}}\gtrsim10^4$ in a single run can decrease the bias (the difference between the expected and the optimised stellar parameter, see Section~\ref{Section:Bias}), but the computational cost increases above $\sim8$ minutes per optimisation (with 25 allocated CPUs of the multicore AMD Ryzen 399X 64-Core Processor), following the trend $t\sim\sqrt{\mathrm{N_0+N_{off}}}$ (see Figure~\ref{fig:timetot}). As we detail Sections~\ref{Section:MinimumExp} and~\ref{Subsection:VestaMinRep}, we should carry out a Monte Carlo scheme to obtain credible intervals for the stellar parameters. The Monte Carlo simulation is fundamentally a repetition of AGA with the \textit{fine} interpolation. At worst, the total time would scale as $t_{\mathrm{total}}\sim N_{\mathrm{rep}}\times\sqrt{N_0+N_\mathrm{{off}}}$ (where $N_{\mathrm{rep}}$ represents the number of repetitions), hence our adopted parameters.

\subsubsection{Working flow of \tonalli}

\begin{figure*}
\includegraphics[alt={Flowchart of the code tonalli, detailing the implemented procedure.}, width=1.9\columnwidth]{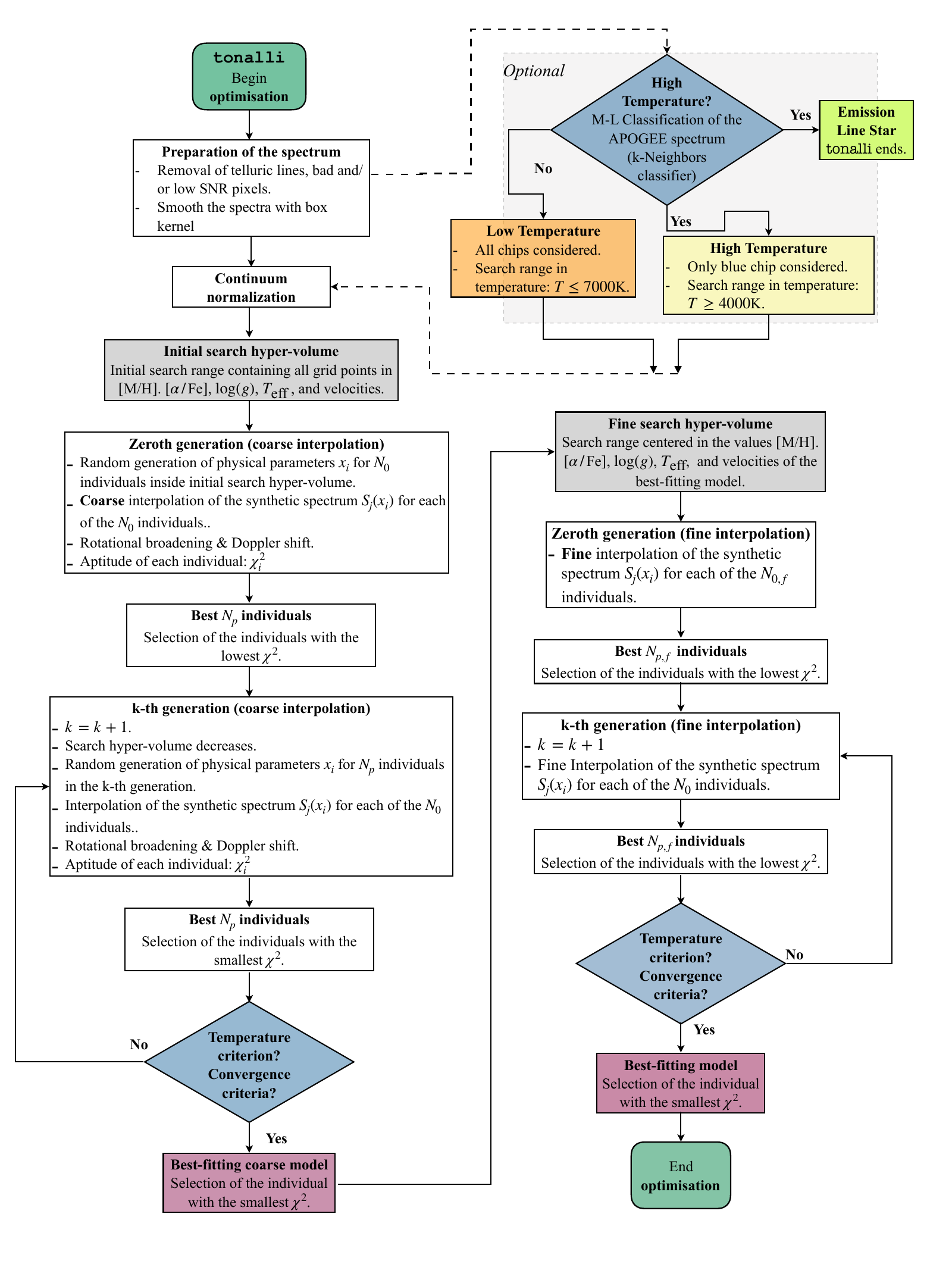}
\caption{Flowchart of the code \tonalli. \textit{Left}: the optimisation begins with the preparation of the observed spectrum, and then a \textit{coarse} best-fitting model is found by the comparison of randomly generated synthetic spectra with known parameters with the observed spectrum. \textit{Bottom right}: the hyper-volume centred at the \textit{coarse} best-fitting model parameters serves as a new search region. The optimisation is repeated with new parameters and the spectrum interpolation requires four times the spectra used in the coarse interpolation. \textit{Top right}: the optional k-neighbours classifier proceeds before the continuum normalisation of the observed spectrum.} \label{fig:flowchart}
\end{figure*}

Figure~\ref{fig:flowchart} shows the flowchart of the code \tonalli\ explained in length above. 
In summary, the optimisation code \tonalli\ follows this scheme: the observed input spectrum is prepared by removing telluric lines, bad and low S/N pixels, then it is continuum-normalised by an iterative sigma-clipping procedure (Section~\ref{section:normalisationSpectra}).
Before the continuum normalisation, the optional supervised machine-learning classification of the spectrum (low or high temperature, emission line star) is available (Section~\ref{Section:MLClass}). 

After the continuum normalisation procedure, the initial search hyper-volume is set to be $V_{0} = \Delta{\mathrm{[M/H]}}\times \Delta\mathrm{[\alpha/\text{M}]}\times \Delta\mathrm{\log(g)}\times \Delta \mathrm{T_{eff}} \times \Delta \mathrm{v\sin(i)}\times\Delta\epsilon\times \Delta\mathrm{RV}$, and $N_0$ individuals are randomly spawned from within this volume.

For each of these random individuals, the synthetic spectrum is interpolated from $N_{\text{interpol}}=N_{\mathrm{[M/H]}}\times N_{\mathrm{[\alpha/\text{M}]}}\times N_{\mathrm{\log(g)}}\times N_{T}$ spectra with parameters close to those of the spawned individual.
The interpolated spectrum is then compared to the observed spectrum; the aptitude or fitness of each individual is computed by the preset FOM (eq.~\ref{eq:FOM}). 
The above $N_0$ individuals constitute the zero-$\mathrm{th}$ generation (Section~\ref{Section:ZerothG}).
We sort the $N_0$ based on their aptitude and select the $N_p$ individuals having the smallest $\chi^2$ values to be the \textit{asexual} parents of the next generation. 
The search hyper-volume diminishes as $V_{k}=V_0\times p^{6k}$ (cf. eq.~\ref{eq:deltax}), where $k$ is the generation counter and $p$ the \textit{convergency} factor; $(N_p-1)\times10$ offspring are spawned within the volume $V_{k}$ centred in each of the $N_p$ parents. 
This constitutes the so-called \textit{asexual} reproduction (Section~\ref{Section:subgen}).

The resemblance of the progeny with the asexual parent increases as the search hyper-volume reduces. The iterative procedure repeats the asexual reproduction (within an ever shrinking search hyper-volume) of the best $N_p$ individuals in the previous $k-1$ iteration.
In absence of spectra model degeneration, the $10\times N_p$ individuals in the last iteration will have equal or close parameters. 
The code \tonalli\ stops when the temperature length of the search hyper-volume is $\lesssim 1$~K (see Section~\ref{Section:bestfit}). 
The best-fitting model will have the smallest $\chi^2$ of all the generations, and we label this best-fitting model as the \textit{coarse} best-fitting model.

We then repeat the steps described in Sections~\ref{Section:ZerothG}, \ref{Section:subgen} and \ref{Section:bestfit} in the search for the \textit{fine} best-fitting model spectrum: the zero-$\mathrm{th}$ generation search volume is now centred in the \textit{coarse} best-fitting model parameters. The sides have lengths: $\Delta\mathrm{[M/H]}=1$~dex, $\Delta\mathrm{[\alpha/\text{M}]}=1$~dex, $\Delta\mathrm{\log(g)}=1.5$~dex, $\Delta \mathrm{T_{eff}}=2000$~K, $\Delta \mathrm{v\sin(i)}= 5$~km s$^{-1}$ and $\Delta \mathrm{RV} = 5$~km s$^{-1}$ if $\mathrm{v\sin(i)}<10$~km s$^{-1}$, $\mathrm{RV}<10$~km s$^{-1}$;  $\Delta\mathrm{v\sin(i)}= 10$~km s$^{-1}$ and $\Delta \mathrm{RV} = 10$~km s$^{-1}$ otherwise. 
The \textit{convergency} factor $p$ is increased by 0.15, and the number of the zero-$\mathrm{th}$ generation individuals is fixed to be either $10\%$ of the search grid or $100$, whichever is bigger. However, the number of parents in the subsequent generations is decreased by 2 (with respect to the $N_p$ value in the \textit{coarse} search). 
We name the best-fitting model of this search as the \textit{fine} best-fitting since we refine the computed offspring spectra by increasing the number of synthetic spectra employed in the interpolation.
After reaching the temperature length limit, the \textit{fine} best-fitting model is found.

The code minimises the differences between the observed spectrum and the synthetic spectrum, and from this best-fitting spectrum we define the physical parameters of the observed star. 
However, it must be emphasised that the values of the stellar parameters obtained by \tonalli~are model dependent, as they may present differences when using distinct synthetic spectrum libraries (Adame et al, in preparation).
Differences can also be expected with respect to other methodologies and samples in distinct wavelength ranges.

\section{Accuracy and precision of \tonalli}\label{Section:Bias}

The next step to ensure the reliability of \tonalli\ is to measure its intrinsic precision and accuracy (bias) when adopting the \marcs\ library.
For that, we need first to obtain the minimum number of repetitions to estimate the performance of \tonalli\ when the input spectrum has known physical parameters. 
To do this, we select a few representative synthetic spectra with $\mathrm{[M/H]}=0$ and $\mathrm{[\alpha/\text{M}]}=0$ and examine the impact of the number of repetitions in the determination of the mean best-fitting parameters (Section~\ref{Section:MinimumExp}). 
Once we estimate the minimum number of repetitions/experiments, we expand the analysis to a complete set of synthetic models with $\mathrm{[M/H]}=0$ and $\mathrm{[\alpha/\text{M}]}=0$ to obtain reliable measures for the intrinsic accuracy and precision of \tonalli\ (Section~\ref{Section:accuracy}).
The reason to restrict our experiments to synthetic models with solar abundances through this section is to detect any potential bias introduced by our continuum normalisation procedure, and ultimately to ensure the correct recovery of parameters from the solar spectrum (Section~\ref{Section:Results}). 
    Also, as we developed \tonalli\ to characterise young main-sequence and pre-main sequence stars in nearby regions ($\lesssim1$~kpc from the Sun), we do not expect the metallicity abundance $[\text{M}/\text{H}]$ (or $[\text{Fe}/\text{H}]$) to fluctuate over $\pm0.3$~dex from the solar value, as indicated by the radial metallicity distribution from young open clusters \citep[e.g.][]{Netopil16,Baratella20,Recio23,Carbajo24} and star forming regions \citep{Santos08,Spina14,Spina17}. 
    Regardless, we applied \tonalli\ to determine the atmospheric parameters of 1600 main sequence stars within 100 parsec from the Sun, and their estimated individual metallicities agree with previous published results \citep{Ricardo24}.


\begin{figure}
\includegraphics[alt={Histograms of the probed synthetic stellar parameters demonstrating the accuracy of tonalli}, width=1\columnwidth]{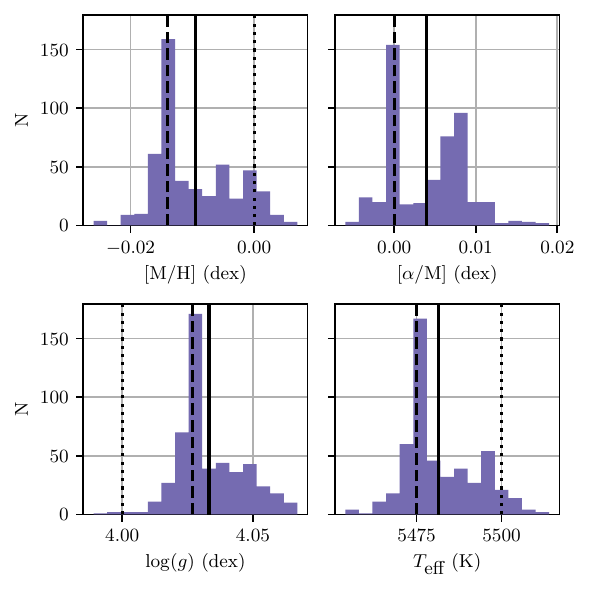}
\caption{Histograms of the 500 best-fitting model parameters for the synthetic star with $\mathrm{[M/H]}=0$, $\mathrm{[\alpha/\text{M}]}=0$, $\mathrm{\log(g)}=4.0$~dex and $\mathrm{T_{eff}}=5500$~K, obtained by \tonalli. Vertical lines depict the \textit{true} value (\textit{dotted} line), the \textit{mean} of the distribution (\textit{solid} line), and the \textit{mode} (\textit{dashed} line), respectively. \textit{Top left panel}: Distribution of $\mathrm{[M/H]}$; \textit{top right panel}: distribution of $\mathrm{[\alpha/\text{M}]}$; \textit{bottom left panel}: distribution of $\mathrm{\log(g)}$; \textit{bottom right panel}: distribution of $\mathrm{T_{eff}}$. \label{fig:histo_rep}}
\end{figure}

\begin{figure}
\includegraphics[alt={Graphs of the probed syntethic stellar parameters, showing the variation and subsequent plateau of the parameter bias as the Monte Carlo experiment number increases.}, width=1\columnwidth]{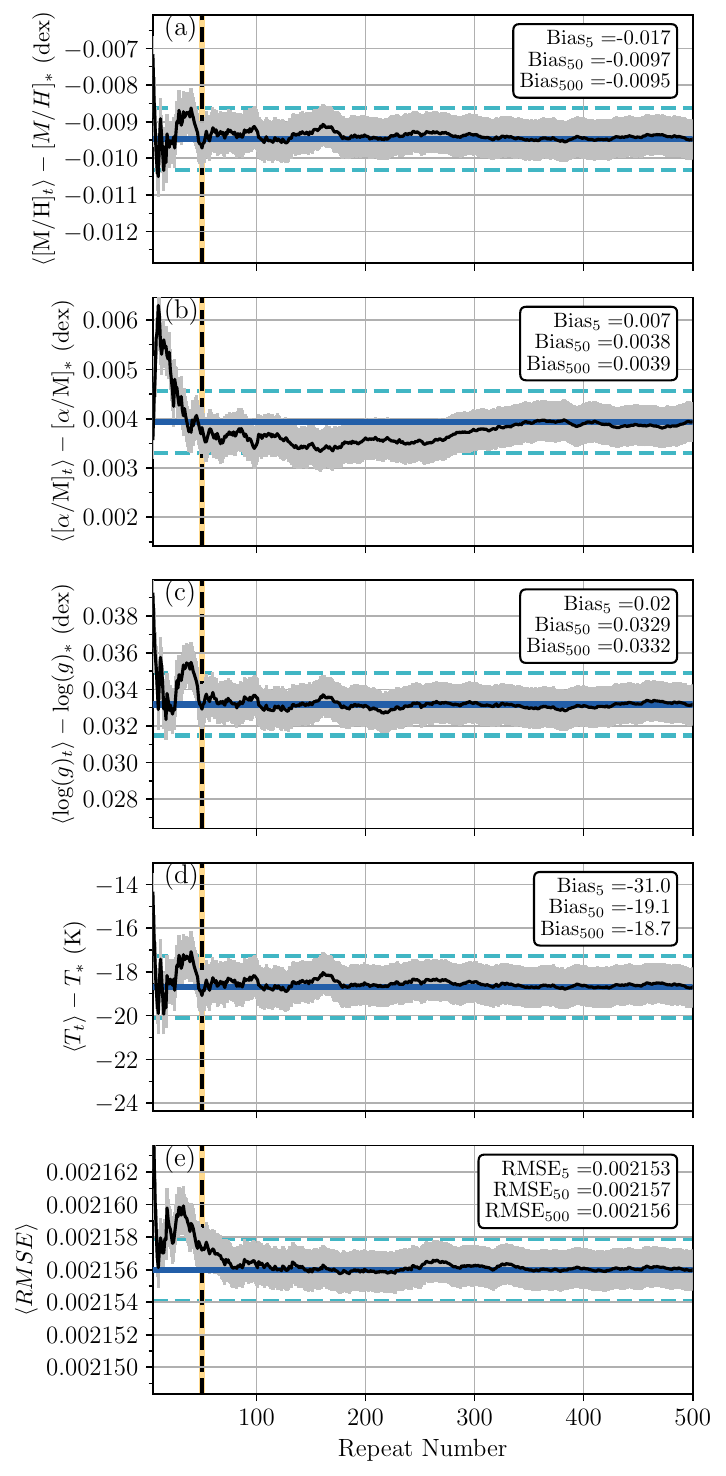}
\caption{\textit{Panels (a)-(d)}: Population mean (\textit{black solid line}) of the difference between  the best-fitting parameter, $\langle X\rangle_t$, and the \textit{true} value, $X_*$, as function of the number of repeats, and \textit{Panel (e):} Mean RMSE (\textit{black solid line}) as function of the number of repeats, for the synthetic star with $\mathrm{[M/H]}=0$, $\mathrm{[\alpha/\text{M}]}=0$, $\mathrm{\log(g)}=4.0$~dex and $\mathrm{\mathrm{T_{eff}}=5500}$~K. The \textit{vertical dashed} line marks the location of 50-th repetition; the population mean (\textit{solid blue line}) and the $3\sigma$ errors (\textit{dashed cyan lines}) at 500 repetitions are shown as a guide. The \textit{grey} area represents the 95\% confidence interval of the population mean.\label{fig:repetitions}}
\end{figure}

\subsection{Minimum number of experiments}\label{Section:MinimumExp}

For this experiment, we select the synthetic spectrum of a star with $\mathrm{[M/H]}=0$~dex, $\mathrm{[\alpha/\text{M}]}=0$~dex, $\mathrm{\log(g)}=4.0$~dex, and $\mathrm{T_{eff}}=5500$~K.
The spectrum is convolved with a Gaussian profile to match the continuum-normalised library resolution. 
For this experiment we do not add noise to the spectrum; we adopt the root-mean-square error (RMSE) as the figure of merit instead of the $\chi^2$ value. 
We run \tonalli\ 500 times in the following fashion: at first, \tonalli\ finds the \textit{coarse} best-fitting model; this step is only performed once. 
The hyper-volume around the \textit{coarse} best-fitting model serves as the initial search region for each of the subsequent 500 repetitions of the \textit{refinement} step of \tonalli.
With each independent \textit{refinement} repetition, we obtain a \textit{fine} best-fitting model. 
Ideally, the 500 \textit{fine} best-fitting models would have the same, or close by, best-fitting parameters (i.e. \textit{Law of large numbers}), but since \tonalli\ is a stochastic optimisation algorithm, we expect the best-fitting parameters to have some degree of dispersion, thus impacting on both the accuracy and the precision of the code.

In Figure~\ref{fig:histo_rep}, we present the histograms of the best-fitting parameters distributions obtained by \tonalli\ for the selected synthetic star.  
The distributions appear to be bimodal, with clear and unique maxima in the distributions of $\mathrm{[M/H]}$, $\mathrm{\log(g)}$ and $\mathrm{T_{eff}}$ (where the maximum peaks are $\sim1.5$ to $\sim3$ times the size of the secondary peaks of the distributions). 
The locations of the \textit{real} parameter value and both the mean and the mode of the distribution are shown in the Figure. 
The mode pinpoints the location of the maximum peak, while the mean falls somewhere close to this peak. 
It is clear that the mean and the mode of the distributions are shifted from the real value of the parameter, but this shift is smaller than half the step of the synthetic library.

We adopt the bias of the mean of a given parameter $X$ as the proxy of the accuracy of \tonalli:
\begin{equation}\label{eq:bias1}
    \mathrm{Bias}(X)=\langle X_t \rangle - X_*,
\end{equation}
where $\langle X_t \rangle$ represents the mean best-fitting parameter obtained by \tonalli, and $X_*$ the expected (\textit{true}) value of the parameter of the synthetic spectrum. 
The top panels of Figure~\ref{fig:repetitions} show the variation of the bias of the parameters $\mathrm{[M/H]}$, $\mathrm{[\alpha/\text{M}]}$, $\mathrm{\log(g)}$, and $\mathrm{T_{eff}}$ with the repetition number for the synthetic star. 
These biases ($\langle\mathrm{[M/H]}_t\rangle - \mathrm{[M/H]}_*$, $\langle \mathrm{\log(g)}_t\rangle - \mathrm{\log(g)}_*$, and $\langle T_t\rangle -T_*$) stabilise after $\sim100$ repetitions, while the bias $\langle\mathrm{[\alpha/\text{M}]}_t\rangle-\mathrm{[\alpha/\text{M}]}_*$ stabilises after $\sim300$ repetitions. 
The latter is a consequence of the previously observed bi-modality of the best-fitting $\mathrm{[\alpha/\text{M}]}$ distribution (Figure~\ref{fig:histo_rep}). {In the plots of Figure~\ref{fig:repetitions} we zoom in the vertical coordinates to emphasise the size of the fluctuations in the differences from repeat to repeat: for the abundances, fluctuation occurs in the ten thousandth place; for the logarithm of the surface gravity, in the thousandth place. The difference of temperature fluctuates in the ones place.}

We can define the bias of \tonalli\ as the mode of the distribution of the best-fitting parameters; in this case, the \textit{mode bias} is independent of the number of repeats (as long as the number is larger than $\sim20$). It is worth to mention that either the mean or the mode bias is way below the grid steps of the \marcs\ library for all the stellar parameters, even for a small number of repetitions.

We quantify the precision of \tonalli\ for the mean of a given parameter as the standard error,
\begin{equation}\label{eq:precisionn}
    \sigma_{\overline{X}}=\frac{\sigma_X}{\sqrt{n}},
\end{equation}
where $\sigma_X$ is the standard deviation of the mean, and $n$ is the number of repetitions. 
The precision is shown as symmetric error bars (the grey area around the bias) in Figure~\ref{fig:repetitions}, and represents the 95\% confidence interval of the bias. 
Thus, the precision of \tonalli, when recovering synthetic spectra, is fairly good, although somewhat biased.
We observe that $\mathrm{\log(g)}$ and the temperature posses the largest bias, when compared to half of the grid step $\Delta_X$ of the synthetic library ($\sim2\%\times\Delta_\mathrm{[M/H]}/2$, $\sim1\%\times\Delta_\mathrm{[\alpha/\text{M}]}/2$, $\sim10\%\times\Delta_\mathrm{\log(g)}/2$, and $\sim15\%\times\Delta_\mathrm{T_{eff}}/2$ for $\mathrm{[M/H]}$, $\mathrm{[\alpha/\text{M}]}$, $\mathrm{\log(g)}$, and $\mathrm{T_{eff}}$, respectively).

Regarding the number of minimum repetitions needed to obtain reliable results, both the bias and the standard error values at the $50-\text{th}$ and the $500-\text{th}$ repetitions are comparable, at least for the synthetic star we discussed above. 
We repeat the above analysis for a set of 30 synthetic stars with abundances $\mathrm{[M/H]}=\mathrm{[\alpha/\text{M}]}=0$, $\mathrm{\log(g)}=3$, $4$ and $5$~dex and effective temperatures $3200$~K, and $3500$ to $7500$~K in steps of $500$~K. Except for some combinations of $\mathrm{T_{eff}}-\mathrm{\log(g)}$ in the mode plots, the bias (or mode) of \tonalli\ remains unchanged regardless the number of repetition adopted. 
This result supports the adoption of minimum repetitions as low as $50$ for the remaining of this work.

\begin{figure*}
\includegraphics[alt={Several heat maps of the bias of tonalli, for the experiments, the probed the stellar parameters, and for the statistical figures detailed in the text.}, width=2\columnwidth]{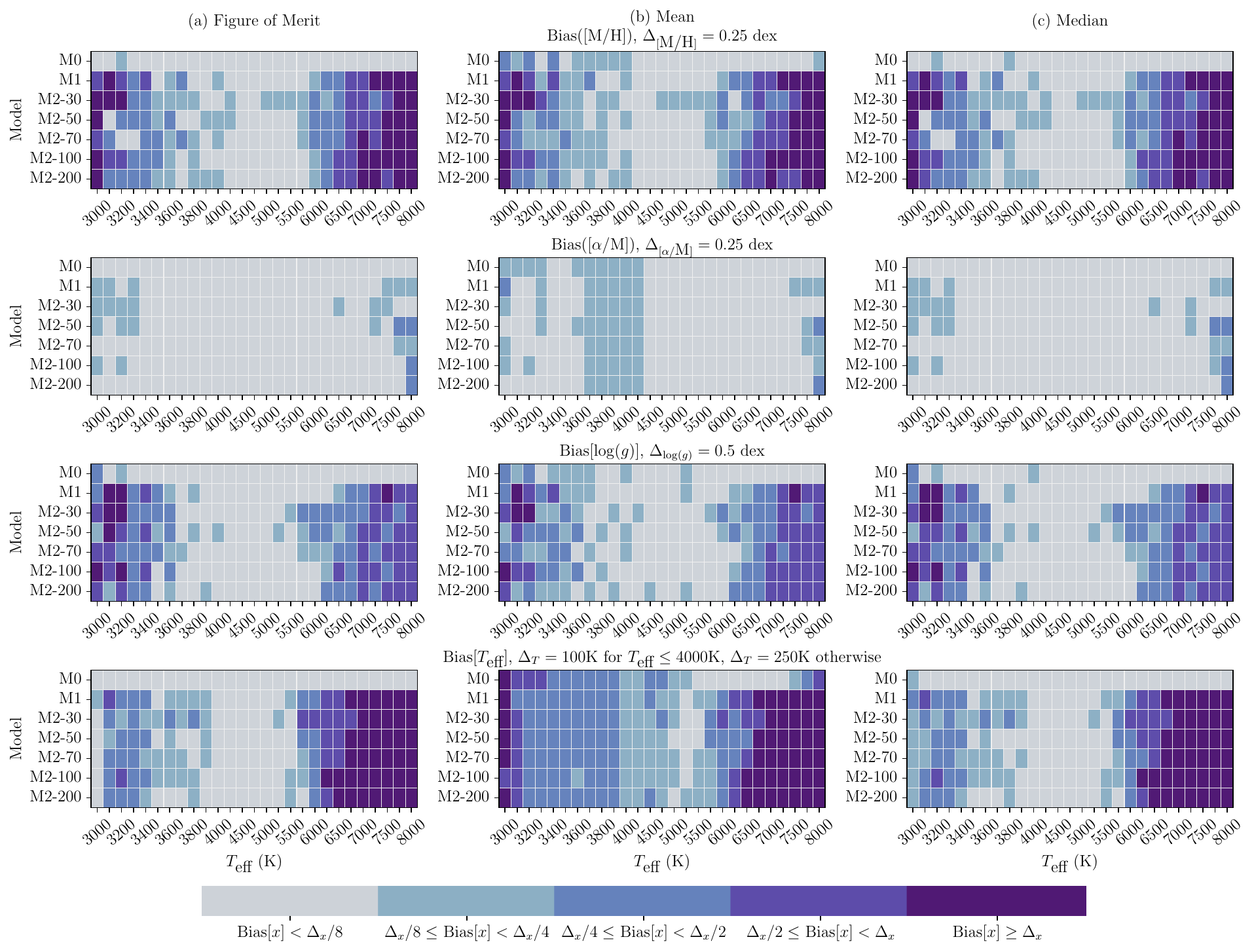}
\caption{Recovery of the synthetic stars parameters: heat maps of the bias $B$ of \tonalli. 
The shades of blue represent the interval where the median of the difference $|X_*-X_{\text{ton}}|$, $B_x$, lies.
The intervals, in fractions of the \marcs\ grid steps $\Delta_x$, are shown in the colour bars. 
Lighter colour shades purport accurate and precise \tonalli\ results. Each column presents the bias of the \tonalli\ results adopting the best-fitting parameters of each repetition as follows: (a) best-fitting parameters from the model with the minimum FOM, (b) best-fitting parameters are the mean of the 1D parameter distribution, and (c) best-fitting parameters are the median of the 1D parameter distribution.\label{Fig:bias_heatmap}}
\end{figure*}

\begin{figure*}
\includegraphics[alt={Several heat maps of the precision of tonalli, for the experiments, the probed the stellar parameters, and for the statistical figures detailed in the text.}, width=2\columnwidth]{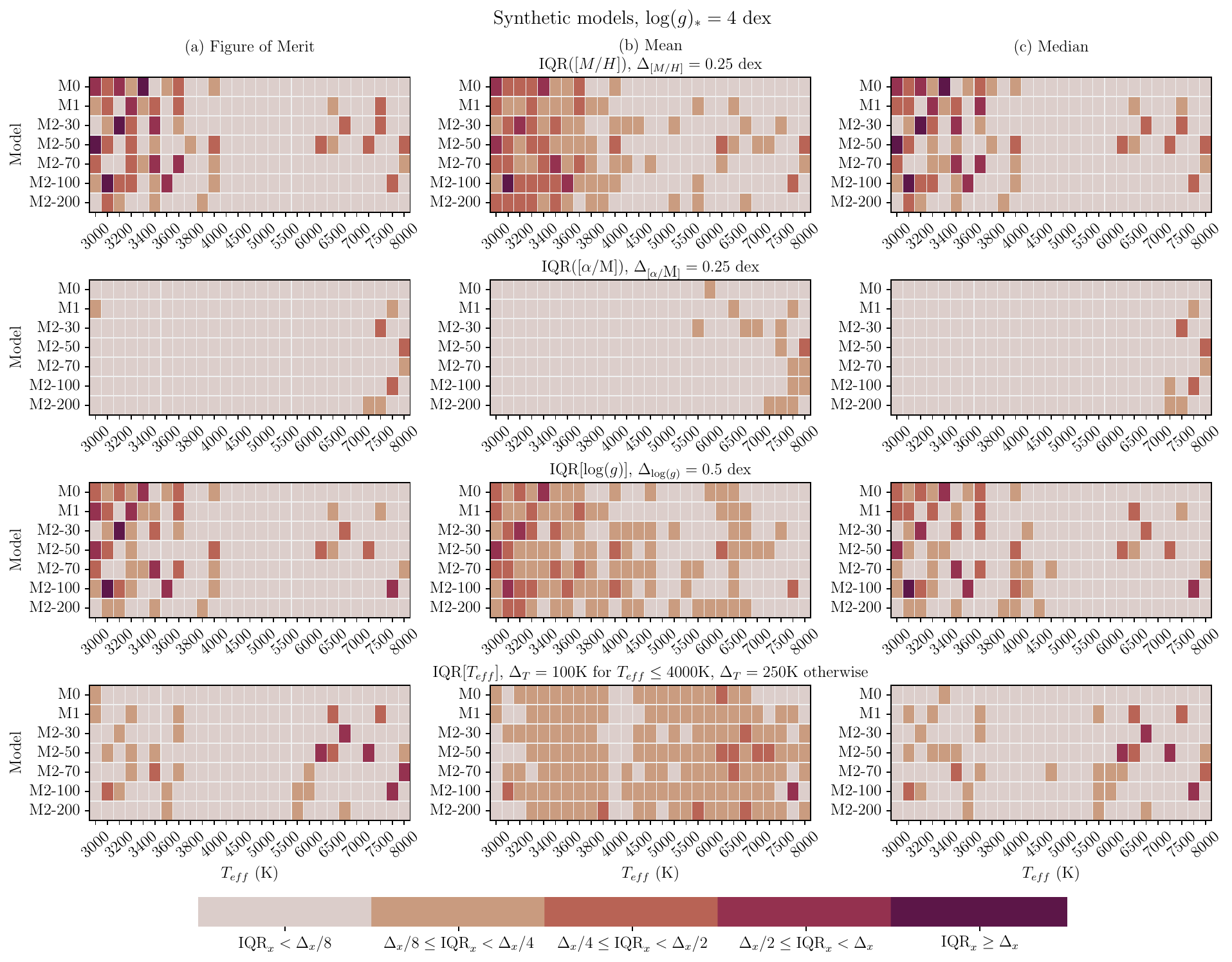}
\caption{Recovery of the synthetic stars parameters: heat maps of the precision of \tonalli, for synthetic models with $\mathrm{\log(g)}_*=4$. 
The shades of red represent the IQR of the distribution of each Monte Carlo realisation lies.
The intervals, in fractions of the \marcs\ grid steps $\Delta_x$, are shown in the colour bars. 
Lighter colour shades purport precise \tonalli\ results. Each column presents the IQR of the \tonalli\ results adopting the best-fitting parameters of each repetition as follows: (a) best-fitting parameters from the model with the minimum FOM, (b) best-fitting parameters are the mean of the 1D parameter distribution, and (c) best-fitting parameters are the median of the 1D parameter distribution.\label{Fig:prec_heatmap}}
\end{figure*}

\subsection{Accuracy and precision}\label{Section:accuracy}
Having established a suitable minimum number of repetitions for our experiments with the synthetic spectra ($N_{\text{rep,min}}=50$), we now explore the effects of our continuum normalisation method and the inclusion of noise in the spectra in the recovery of the spectroscopic parameters. 
For this, we select synthetic spectra with zero metal and $\alpha$-elements abundances, effective temperatures of $3000-4000$~K (in steps of 100~K) and $4250-7000$~K (in steps of 250~K), and $\mathrm{\log(g)}$ of $3, 4, 5$~dex, resulting in a set of 66 synthetic stars.

First, we construct the baseline experiment M0. For this experiment, the set of synthetic spectra is drawn from the \marcs\  library, which is already continuum-normalised.  
Secondly, we construct another experiment to probe the effects of the continuum-normalisation method in the recovery of the parameters, experiment M1. 
In this case, the set of synthetic spectra is not continuum normalised, therefore enabling the normalisation subroutine in \tonalli. The spectra are previously convolved with a Gaussian profile to match the \apo\ resolution.
Finally, we add noise to the spectra of experiment M1 to establish a minimum working signal-to-noise ratio (SNR) of the spectra for \tonalli; these are dubbed M2 models. Table~\ref{table:tonalliaccuracy} and Figure~\ref{Fig:bias_heatmap} show the bias and the precision of \tonalli\ for models M0, M1 and M2 described above.

For a given effective temperature and experiment, we have three synthetic models, each differing in $\mathrm{\log(g)}$. For each one, we compute the $50$ repetitions, and collect the $3\times50$ absolute differences $|X_*-X_{(\text{ton,rep})}|$ in a distribution $D_X$, where $X_*$ is the \textit{true} parameter value of the synthetic spectrum, and $X_{(\text{ton,rep})}$ is the parameter value obtained by \tonalli\ in a given repetition. Moreover, $X_{(\text{ton,rep})}$ can be the value from the model with the best FOM, or the mean/median parameter derived from the distribution of all the offspring computed in the repetition.
We re-define the bias (cf. Eq~\ref{eq:bias1}) as the median of this distribution $D_X$, 
\begin{equation}\label{eq:bias2}
    \mbox{Bias}(X)=\mbox{Median}(D_X).
\end{equation}

The 12 plots in Figure \ref{Fig:bias_heatmap} show the bias (eq.~\ref{eq:bias2}) of the parameters as follows. From top to bottom, each plot in the row presents the bias heat map in $\mathrm{[M/H]}$, $\mathrm{[\alpha/\text{M}]}$, $\mathrm{\log(g)}$ and $\mathrm{T_{eff}}$, respectively. From left to right, the columns present the bias heat map arranged according to the adopted \tonalli\ parameter value $X_{(\text{ton,rep})}$ to compute the bias: the best FOM (left column), the mean of the offspring distribution in each repetition (middle panel), and the median of the offspring distribution in each repetition (right panel).

Each of the 12 heat maps consists of 7 rows and 27 columns. 
A given column presents the results of a probed temperature, while the row indicates the experiment (M0, M1, M2). 
We also vary the SNR for the M2 models, including SNR of 30, 50, 70, 100 and 200.
Thus in a single heat map plot, we have $7\times27$ rectangles, each one presenting the bias, that is, the median of a distribution of $3\times50$ absolute differences $|X_*-X_{\text{ton}}|$ gathered from the models for three synthetic stars sharing a specific effective temperature (specified in the x-axis of the heat map) but differing in $\mathrm{\log(g)}$, computed following the prescription of the given experiment (y-axis of the heat map).

\begin{table*}
\caption{Set of models probed to obtain the accuracy and precision of \tonalli, and suitable temperature ranges for \tonalli\ operation.}\label{table:tonalliaccuracy}
\begin{threeparttable}
\begin{tabular}{lcr|rr|rr|rr}
\multicolumn{3}{c}{} &\multicolumn{6}{c}{Bias$(X)<\Delta_X/2$}\\\cline{4-9}
\multicolumn{3}{c}{} & \multicolumn{2}{c}{FOM} & 
\multicolumn{2}{c}{Mean} & \multicolumn{2}{c}{Median} \\
\cline{4-5}\cline{6-7}\cline{8-9}
Model name & FOM & SNR & 
$T_{\text{min}}$ & $T_{\text{max}}$ & $T_{\text{min}}$ & $T_{\text{max}}$ & $T_{\text{min}}$ & $T_{\text{max}}$\\ 
 &  &  &(K) &(K) & (K)& (K) & (K) & (K)\\

 (1) & (2) & (3) &(4) &(5) & (6)& (7) & (8) & (9)\\\hline
\multicolumn{7}{l}{Continuum normalised \marcs\  spectra}\\
M0 &  RMS & \ldots & 3000 & 8000 & 3400 & 7750 & 3000 & 8000 \\\hline
\multicolumn{7}{l}{\marcs\  spectra, no continuum normalisation}\\
M1 &  " & \ldots  & 3300 & 6000 & 3300 & 6000 & 3500 & 6000\\\hline
\multicolumn{7}{l}{\marcs\  spectra + noise, no continuum normalisation}\\
M2-30 & $\chi^2$ & 30  & 3300 & 6500 & 3300 & 6500  & 3300 & 6000\\
M2-50 &  " & 50  & 3300 & 6000 & 3200 & 6000 & 3500 & 6250 \\
M2-70 &  "& 70   & 3200 & 6000 & 3200 & 6000 & 3200 & 6250 \\
M2-100 &  "& 100   & 3300 & 6000 & 3300 & 6000 & 3500 & 6000 \\
M2-200 &  " & 200 & 3100 & 6250 & 3200 & 6250 & 3300 & 6250\\\hline
\end{tabular}
\begin{tablenotes}\footnotesize
        \item \textbf{Note}: All models computed with an initial population of $N_0=240$, number of parents per generation $N_p=10$, and number of repetitions $N_{\text{rep}}=50$ to obtain statistic figures. Column (2): adopted FOM: RMS or $\chi^2$. Column (3): Signal-to-noise ratio of \marcs\  input spectra with added Gaussian noise. Columns (4) and (5): Minimum and maximum temperature limits of the temperature range where both the bias Bias$(x)$ and IQR of the mean best-fitting parameters are smaller than one-half of the \marcs\ grid step in each parameter; the best-fitting model of each repetition is the model with the smallest FOM. Columns (6) and (7): Same as columns (4) and (5), adopting the mean of the parameter distribution in each repetition as the best-fitting model parameters. Columns (8) and (9): Same as columns (6) and (7), adopting the median value of the distributions.
\end{tablenotes}
\end{threeparttable}
\end{table*}

\subsubsection{Baseline experiment: M0 models}\label{Section:accuracyM0}
The spectra is taken directly from our continuum normalised \marcs\ library, in order to avoid any impact the normalisation procedure may have in the results, therefore allowing to study directly the performance of the asexual algorithm. 
The first row of each heat map of Figure~\ref{Fig:bias_heatmap} represents the bias of the four parameters of the M0 models. 
For the bulk of the models, \tonalli\ results are unbiased, regardless the adopted statistic, ensuring that AGA is working properly to recover the synthetic stellar parameters, with the exception being the effective temperature of cool stars when adopting the mean temperature of the offspring distribution at each repetition. 
Even in this case, the temperature bias ranges between 25 and 50~K, which is still small enough to discriminate between the M spectral sub-types \citep[adopting the temperature scale of][]{PecautMamajek}. 
In the following two sub-sections, we discuss how the mean and the median of the distributions differ from each other due the multi-modal nature of the parameter distributions.

\subsubsection{Effects of the continuum normalisation in \tonalli: M1 models}\label{Section:accuracyM1}

These spectra differ from the spectra employed in the previous experiment (Section~\ref{Section:accuracy}) since for this experiment we employ the \marcs\ spectra (convolved with a Gaussian profile to match the \apo\ resolution) prior to the continuum normalisation process (Section~\ref{Section:marcsSpectra}). 
Therefore, we are probing the accuracy and precision of \tonalli, since the internal continuum-normalisation procedure described in Section~\ref{section:normalisationSpectra} is now active. 
This experiment also allows us to measure the impact our continuum-normalisation procedure might have in the recovery of the stellar parameters.

The second row of each heat map of Figure~\ref{Fig:bias_heatmap} shows the bias (eq.~\ref{eq:bias2}) of the four parameters of the M1 models.
We observe at the high end of the temperature spectrum ($\mathrm{T_{eff}}\gtrsim 6750$~K) that \tonalli\ recovers $\mathrm{T_{eff}}$ with a bias larger than the \marcs\ grid step (in this case, $\Delta_T=250$~K), most likely due to a breakdown in the normalisation procedure owing to lines in the Bracket series, Br11 $\lambda16811.111$, Br13 $\lambda16113.714$, as discussed in Section~\ref{section:normalisationSpectra}. 
We can avoid the issue if the green and red chips, where the Br11 and Br13 lines are prominent, are masked altogether. 
In any case, the bias in temperature have a strong impact in the determination of $\mathrm{[M/H]}$, whereas the bias of $\mathrm{\log(g)}$ is noticeable only for stars with $\mathrm{T_{eff}}>7000$~K.

\subsubsection{Effects of adding noise to the observed spectrum in the accuracy and precision of \tonalli: M2 models}\label{Section:accuracyM2}
As a final test, we add noise to the synthetic spectra of the M1 models. 
The noisy data $F_n(\lambda)$ is obtained by adding $g(\lambda)F(\lambda)$ to the original spectrum $F(\lambda)$, and a synthetic error of $\sigma(\lambda) = F_n(\lambda)/\text{SNR}$.
The value $g(\lambda)$ is fixed through the \textit{common} SNR value per pixel, $\text{SNR}=F_n(\lambda)/g(\lambda) F(\lambda)$, or $g(\lambda) = \pm 1/(\text{SNR}-1)$. 
The sign of $g(\lambda)$ is randomly and independently assigned for each $F(\lambda)$. 

The last five rows in each plot of Figure~\ref{Fig:bias_heatmap} display the results of this experiment.
Each row is identified by the SNR of the spectra: M2-30, M2-50, M2-70, M2-100, and M2-200. 
It is not surprising that the rate of parameter recovery of \tonalli\ is  chiefly determined by the continuum-normalisation procedure, as the large bias trend seeps through the M2 high temperature models regardless the magnitude of the noise added to the spectra. 
However, it is worth to note that \tonalli\ can recover with success the four stellar parameters in this experiment for stars with effective temperatures between $\sim3200$ and $\sim 6250$~K. 
For high temperature stars, the temperature recovery is not ideal due to the normalisation procedure, which we will refine in a posterior work. 
For low temperature stars, while both $\mathrm{T_{eff}}$ and $\mathrm{[M/H]}$ are recovered within an acceptable bias margin, it is $\mathrm{\log(g)}$ the parameter which challenges the spectrum fitting procedure. In the following series paper, we implement a framework to minimise the bias in  $\mathrm{\log(g)}$.

\subsubsection{The precision in the recovery of the parameters}\label{sec:iqr}
We equate the precision of our code with the interquartile range $\text{IQR}=Q_3-Q_1$ of the distribution of the $50$ recovered parameters $X_{\text{ton}}$ per effective temperature, this time distinguishing the distributions in terms of their $\mathrm{\log(g)}_*$ input value. 
Such definition of precision reflects the spread of the recovered parameters of the Monte Carlo realisations (50 per effective temperature and $\mathrm{\log(g)}_*$), and it is robust against any outliers of the distribution.

The heat maps of Figure~\ref{Fig:prec_heatmap} show this precision for the synthetic models with $\mathrm{\log(g)}_*=4$~dex. 
Two general features arise from the heat maps: the IQR of both $\mathrm{[M/H]}$ and $\mathrm{\log(g)}$ is somewhat larger for stars with $\mathrm{T_{eff}}\lesssim 4000$~K when compared to the IQR values of hotter stars, regardless the adopted method to compute the so-called best-fitting value. 
Moreover, if the mean of the distribution is adopted as the best-fitting parameter (middle columns of Figure~\ref{Fig:prec_heatmap}), the IQR values are moderately larger than the IQR of the minimum figure of merit or those of the median of the parameter distribution. 
Both outcomes are not surprising in light of the bias results discussed in the preceding sections. 

Lastly, the precision of the baseline model set M0 shows a slightly larger dispersion for $\mathrm{T_{eff}}\lesssim 4000$~K.
This may be indicative of a $\mathrm{[M/H]}-\mathrm{\log(g)}$ coupling effect in the M-dwarf synthetic spectra of the \marcs\ library that we need to explore further. At any rate, this stresses the need to provide an independent first order estimation of either parameter to limit the search range within the synthetic library in order to obtain unbiased and more precise results.  


\begin{figure*}
\includegraphics[alt={Ridge plots showing the probability density function of selected tonalli Monte Carlo realisations for the temperature, gravity, metal and alpha elements abundances, obtained for the Solar spectrum reflected by Vesta.}, width=1.55\columnwidth]{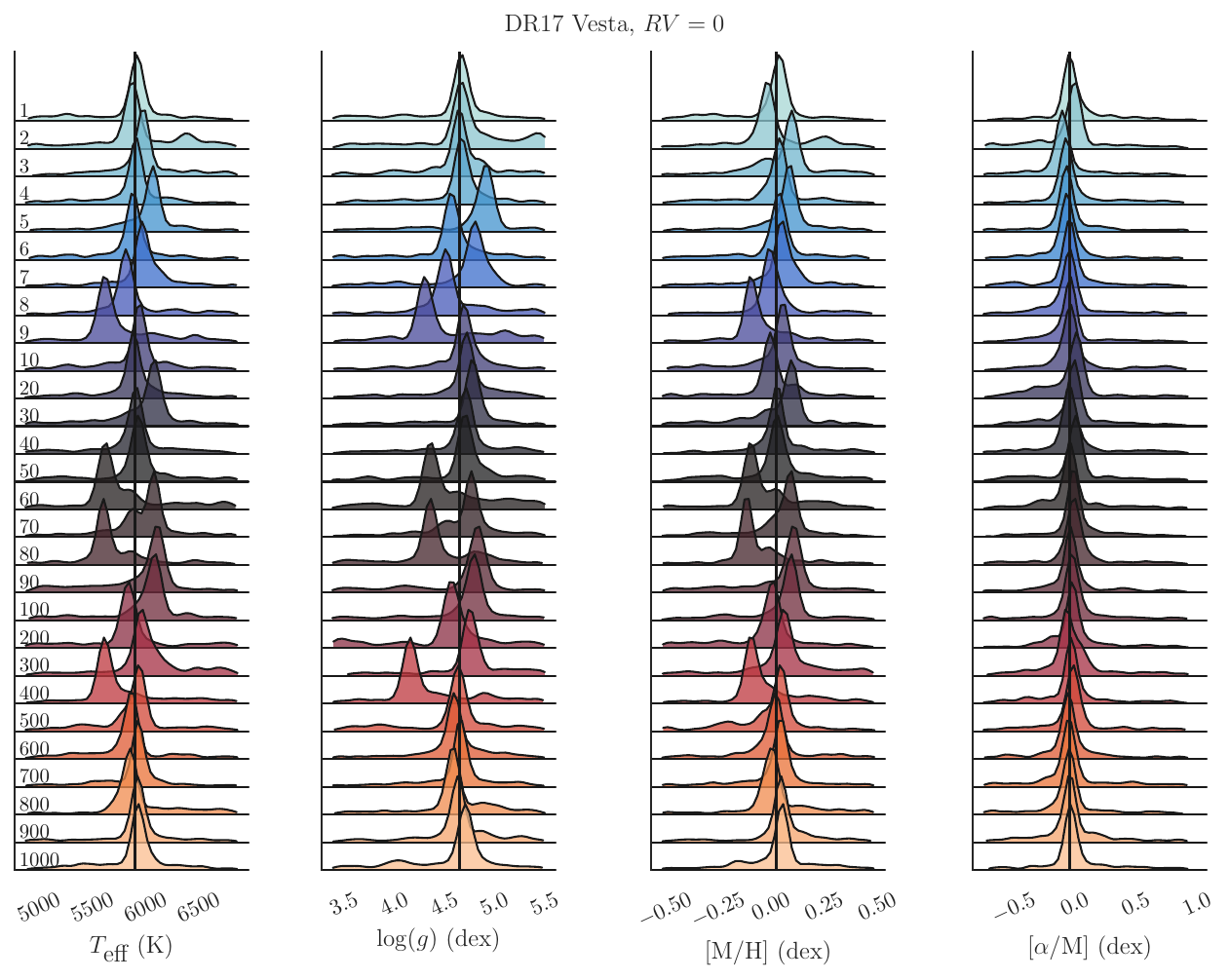}
\caption{Ridge plots showing the PDF of the parameters for a set of selected \tonalli~Monte Carlo realisations (\textit{left label} in each PDF of $\mathrm{T_{eff}}$), for the APOGEE-2 Solar spectrum reflected by Vesta, with the radial velocity of the spectrum restricted to be 0. The \textit{solid line} represents the \textit{mean} value of the median distribution, composed of the median of each Monte Carlo realisation. \label{fig:vesta_ridge_RV0}}
\end{figure*}

\begin{figure*}
\includegraphics[alt={Ridge plots showing the probability density function of selected tonalli Monte Carlo realisations for the temperature, gravity, metal and alpha elements abundances, and radial velocity, obtained for the Solar spectrum reflected by Vesta.}, width=1.8\columnwidth]{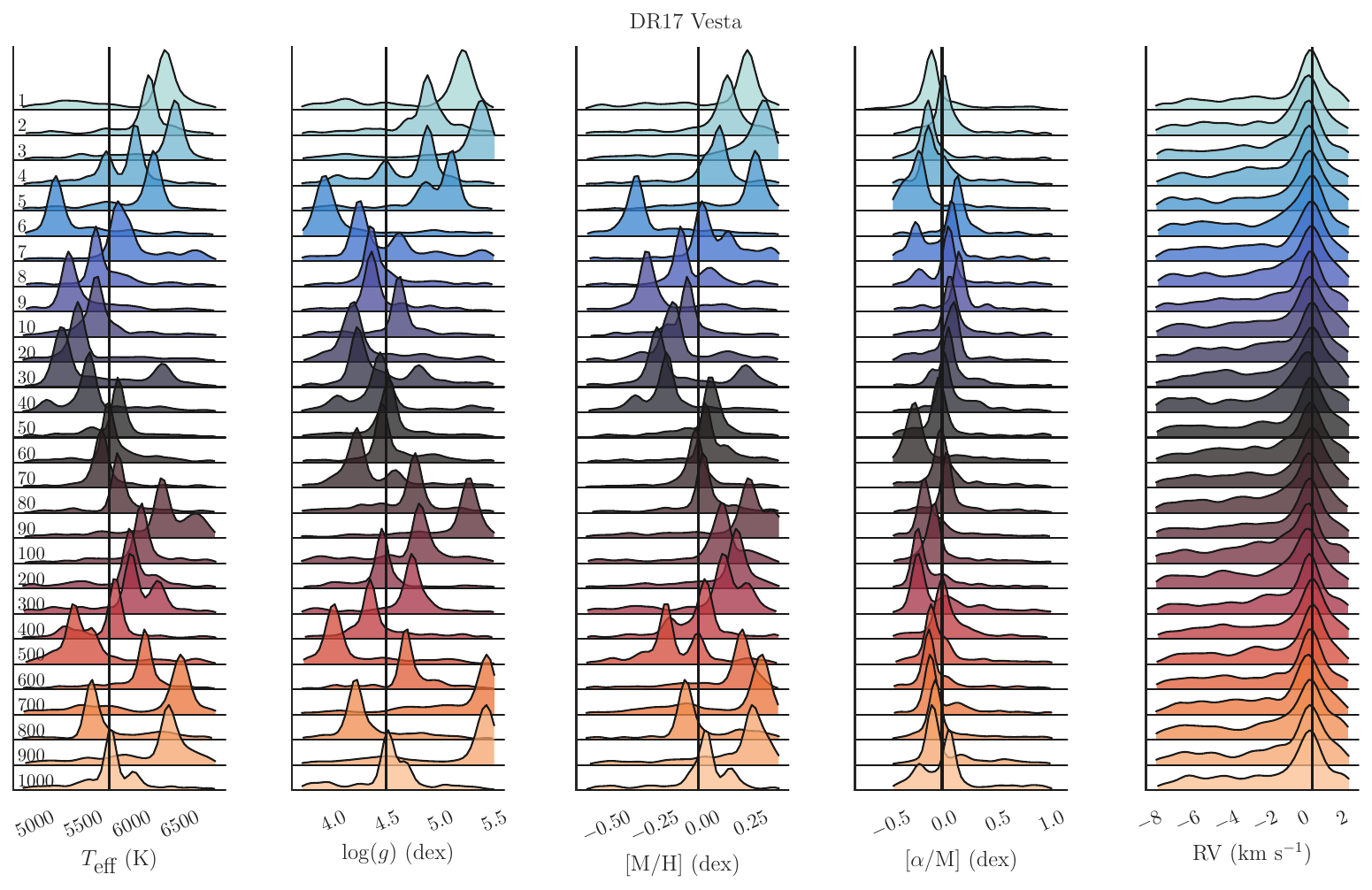}
\caption{Ridge plots showing the PDF of the parameters for a set of selected \tonalli~Monte Carlo realisations (\textit{left label} in each PDF of $\mathrm{T_{eff}}$), for the APOGEE-2 Solar spectrum reflected by Vesta, with $\mathrm{RV}$ optimised. See Figure~\ref{fig:vesta_ridge_RV0}.  \label{fig:vesta_ridge_RV}}
\end{figure*}

\section{The solar spectrum}\label{Section:Results}
We now apply our code \tonalli\ to the solar spectrum reflected by the asteroid Vesta provided in the \textit{apStar} file of \apo\ Data Release 17 \citep[hereafter DR17,][]{DR17}.

\subsection{Monte Carlo realisations}\label{subs:montecarlo1}
We run \tonalli\ optimising either five or six parameters: for the former we assume the observed spectrum is Doppler-shift corrected ($\mathrm{RV}=0$), while for the latter we relax the assumption the radial velocity is fixed, so the radial velocity is also optimised. 
The radial velocity from the header of the \textit{apStar} file is $RV=0$~km s$^{-1}$, meaning the spectrum is already Doppler-shift corrected. Thus, when we optimise the radial velocity, we are effectively applying a small correction, if any, to the observed spectrum. As we shall see in Section~\ref{sec:modeltonallivesta}, this is indeed the case.

For the four runs (two for each \textit{apStar} spectrum), we adopt a limb darkening value of $\epsilon=0.25$ \citep[the value adopted for the solar spectrum by][]{Jonsson20}. 
We set the maximum number of realisations to $1000$. 

In dealing with the results, we choose to ignore the variation of the projected rotational velocity as the derived values are close to the velocity limit imposed by the uniform resolution of our convoluted library \citep[which for $\Delta R=22,500$ is close to 10~km~s$^{-1}$, see][]{Cottaar14}.

{To draw the surface solar parameters from the Monte Carlo realisations, we compute several estimators from the 1D and multidimensional distributions of the optimised parameters. 
The first type its a measure of the accuracy and precision of the algorithm (Section~\ref{Subsection:VestaMinRep}). For a given parameter $\mathrm{X}$, we compute a measurement of the central tendency of the univariate distribution $D_i(\mathrm{X})$ in each $i-\mathrm{th}$ realisation, namely the \textit{mean}, $\langle \mathrm{X}\rangle_i$. This way, we have a univariate distribution $D^*(\langle \mathrm{X}\rangle)$, whose population is comprised by $N_{\text{rep}}$ estimators for the parameter $X$. This mean and their associated standard deviation (of the distribution  $D^*(\langle \mathrm{X}\rangle)$) are tabulated in columns (1) and (2) of Table~\ref{tabla:Vesta} (for 50 and 1000 Monte Carlo experiments, respectively). We also calculate the \textit{median}, $\text{M}(\mathrm{X})_i$ and the \textit{mode}, $\text{Mod}(\mathrm{X})_i$. For the later, the bandwidth of the 1D histogram of the parameter $\mathrm{X}$ is given by the Silverman's rule \citep{Silverman86}.
The ridge plots (Figures~\ref{fig:vesta_ridge_RV0} and~\ref{fig:vesta_ridge_RV}) show some of the 1D distributions $D_i(\mathrm{X})$ of the Monte Carlo realisations.}

{
A second type of estimators are drawn from the distribution $D_s(X)$, which is constructed by combining all the individuals spawned in all the Monte Carlo realisations (Section~\ref{Section:univariate}). We compute three measurements of the central tendency of the univariate distribution $D_s(\mathrm{X})$, namely the \textit{mean}, $\langle \mathrm{X}\rangle$, the \textit{median}, $\text{M}(\mathrm{X})$, and the \textit{mode} $\text{Mod}(\mathrm{X})$ (columns (4), (5), and (6) of Table~\ref{tabla:Vesta}, respectively). It is from these estimators that we draw the Sun physical parameters. 
}

{
The last estimators are computed from the multivariate distribution $D_s$ (Section~\ref{sec:multivariate}). We fit a multidimensional normal distribution to the distribution $D_s$ to obtain the \textit{Gaussian mean} (column (7) of Table~\ref{tabla:Vesta}). We also compute the \textit{halfspace median}, one of the several possible multivariate analogue to the univariate median (column (8) of Table~\ref{tabla:Vesta}).
}

{
The rationale to compute several estimators is to assess the performance of \tonalli, to probe the model degeneracy, and to explore the assumed normality of the 1D distributions. 
The median of the 1D parameter distribution provides an adequate measure of the \textit{true} stellar parameter, as we discuss in Section~\ref{Section:multimodality}.}

Figures~\ref{fig:vesta_ridge_RV0} (fixing $\mathrm{RV}$ to be $0$) and \ref{fig:vesta_ridge_RV} (optimising $\mathrm{RV}$) show the ridge plots of the probability distribution functions (hereafter PDF) of the parameters $\mathrm{T_{eff}}$, $\mathrm{\log(g)}$, $\mathrm{[M/H]}$ and $\mathrm{[\alpha/\text{M}]}$, for selected \tonalli\ Monte Carlo realisations for \apo\  solar spectrum.
The ridge plots are helpful to show the qualitative behaviour of the 1D PDF: both Figures shows how the PDF  changes from repeat to repeat, demonstrating the independence of the Monte Carlo realisations and the exploration of the restricted search hyper-volume of the fine interpolation. 
The Figures also show an striking difference on the temperature, logarithm of the surface gravity, and the metallicity: while the maximum of each of the PDFs are relatively close to the mean value of the median distribution, once we allow the optimisation of the radial velocity, the maxima of the PDFs are scattered over the restricted search hyper-volume.
The results remain accurate, although the precision decreases, at least for the chosen hyper-parameters. This will increase any measure we adopt  for the parameter credible interval, as we see quantitatively in the errors (standard deviations or IQRs) of Table~\ref{tabla:Vesta}.
However, we must emphasise that the credible intervals can be reduced by almost half of the univariate credible intervals if we adopt a Gaussian multivariate distribution to estimate the photopheric stellar parameters (column (7) of Table~\ref{tabla:Vesta}).

\subsection{Minimum number of Monte Carlo realisations}\label{Subsection:VestaMinRep}
We re-examine the minimum efficient number of Monte Carlo realisations (Section \ref{Section:MinimumExp}) in light of the nuances of a real spectrum not captured by the synthetic models.
In addition, we explore the parameter distributions to decide a representative statistic value; for this, we should examine the variation of the mean values of the distributions $D^*$ discussed above in Section~\ref{subs:montecarlo1}, as new means, medians, or modes, are added to them. 

In theory, we should perform $\sim0.6 N_{\lambda}$ Monte Carlo realisations, where $N_{\lambda}$ represents the pixels of the involved spectrum.
Therefore, the Monte Carlo realisations for the \apo\  spectra can exceed a few thousand experiments, which can become computationally costly. 
As we shall see, the minimum number of Monte Carlo realisations are greatly reduced if we impose constrains on the variation of the mean distribution.

\begin{figure*}
\includegraphics[alt={Graphs showing the variation of the statistical figures obtained by tonalli as the number of Monte Carlo simulations increases for the Solar spectrum reflected by Vesta}, width=1.4\columnwidth]{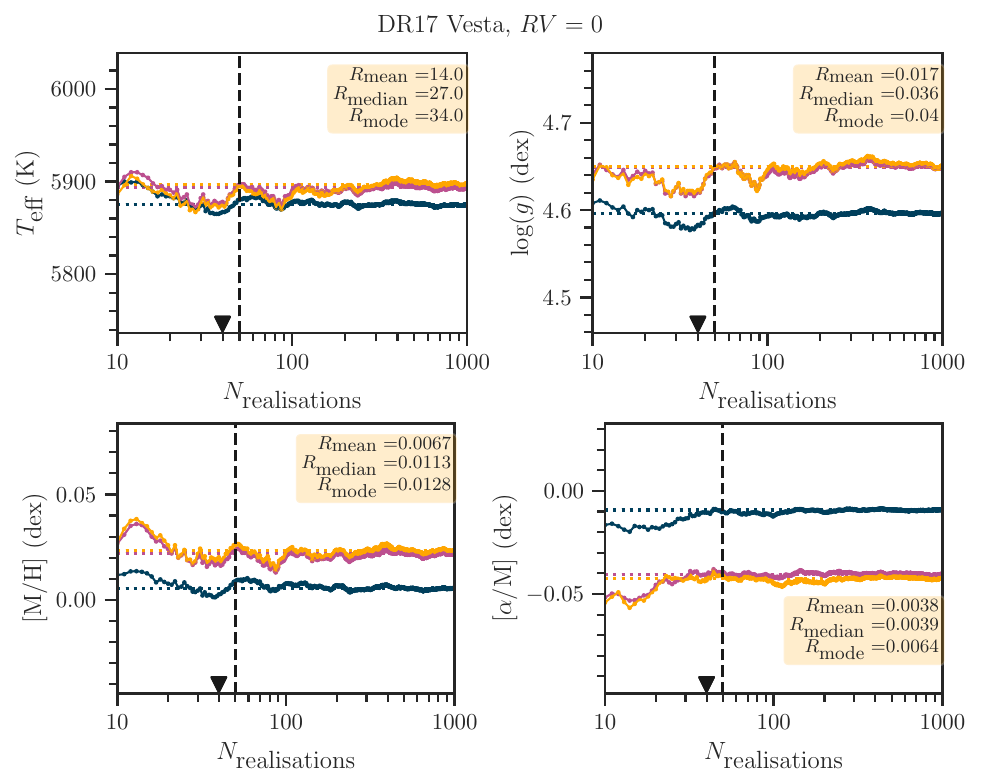}
\caption{Variation of the \textit{mean} value of the mean (\textit{blue solid line}), median (\textit{magenta solid line}), and mode (\textit{yellow solid line}) distributions of the parameters $\mathrm{T_{eff}}$ (\textit{upper left} panel), $\mathrm{\log(g)}$ (\textit{upper right} panel, $\mathrm{[M/H]}$ (\textit{bottom left} panel), and $\mathrm{[\alpha/\text{M}]}$ (\textit{bottom right} panel), as functions of the \tonalli\ Monte Carlo realisations, for the APOGEE-2 Solar spectrum reflected by Vesta, with the radial velocity of the spectrum restricted to be 0. The \textit{dotted lines} represent the \textit{mean} values of the above distributions, computed with the total population of 1000 realisations. The ranges of the average mean $R_{\text{mean}}$, median $R_{\text{median}}$ and mode $R_{\text{mode}}$, between 50 and 1000 realisations, are also shown. The \textit{filled} triangle in the $y=0$ axis points to the minimum Monte Carlo repetition where the accuracy criteria equat.~(\ref{eq:criterion}) is reached: $|\overline{X}_n-\mu|=[2.5~\text{K},0.005~\text{dex}, 0.0025~\text{dex}, 0.0025~\text{dex}]$ for $\mathrm{T_{eff}}$, $\mathrm{\log(g)}$, $\mathrm{[M/H]}$, and $\mathrm{[\alpha/\text{M}]}$, respectively. The \textit{unfilled} triangle points to the minimum Monte Carlo repetition where $|\overline{X}_n-\mu|$ is twice the above values. \label{fig:vesta_varpar_RV0}}
\end{figure*}

\begin{figure*}
\includegraphics[alt={Graphs showing the effect of the optimisation of the radial velocity in the variation of the statistical figures obtained by tonalli as the number of Monte Carlo simulations increases for the Solar spectrum reflected by Vesta}, width=1.4\columnwidth]{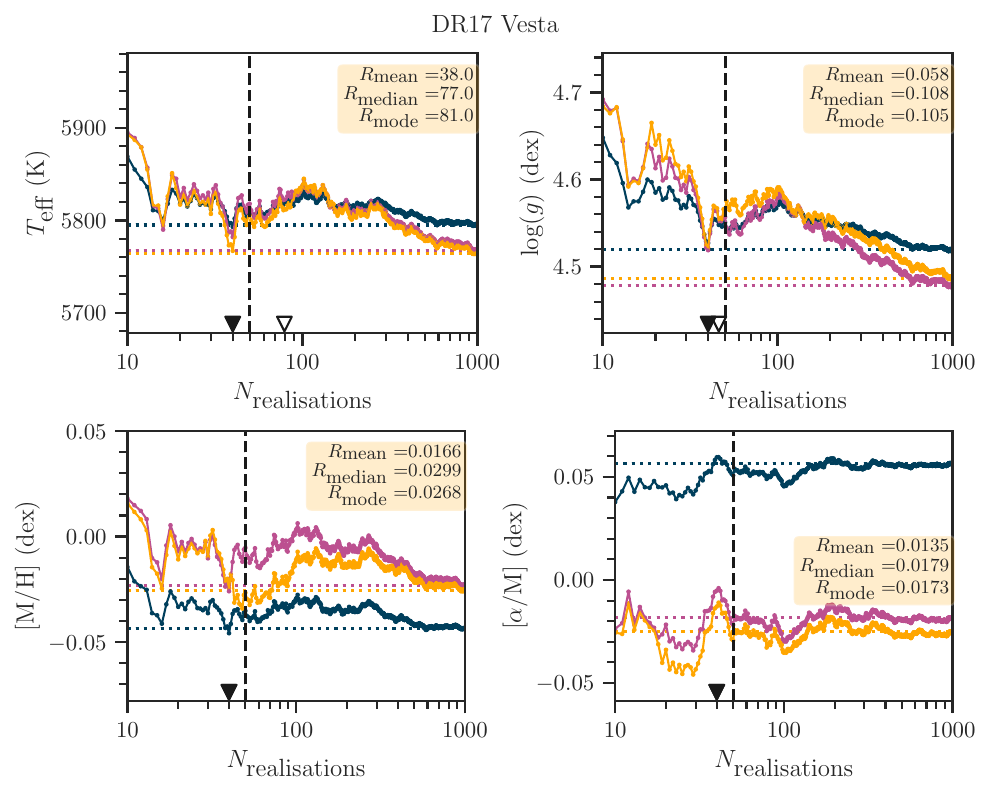}
\caption{Variation of the \textit{mean} value of the mean (\textit{blue solid line}), median (\textit{magenta solid line}), and mode (\textit{yellow solid line}) distributions of the parameters $\mathrm{T_{eff}}$ (\textit{upper left} panel), $\mathrm{\log(g)}$ (\textit{upper right} panel, $\mathrm{[M/H]}$ (\textit{bottom left} panel), and $\mathrm{[\alpha/\text{M}]}$ (\textit{bottom right} panel), as functions of the \tonalli\ Monte Carlo realisations, for the APOGEE-2 Solar spectrum reflected by Vesta, with $\mathrm{RV}\ne0$. See Figure~\ref{fig:vesta_varpar_RV0} for details. \label{fig:vesta_varpar_RV}}
\end{figure*}

Figures~\ref{fig:vesta_varpar_RV0} (assuming $\mathrm{RV}=0$) and \ref{fig:vesta_varpar_RV} ($\mathrm{RV}$ optimised) show the trends of the estimators in four parameters, $\mathrm{T_{eff}}$, $\mathrm{\log(g)}$, $\mathrm{[M/H]}$, and $\mathrm{[\alpha/\text{M}]}$. 
For the models with $\mathrm{RV}=0$, we notice how the parameters of interest stabilise after roughly $\sim30-50$ Monte Carlo repetitions, whereas for the models with optimised radial velocity $\mathrm{RV}$, the parameters of interest stabilise after a few hundreds of repetitions for some stellar parameters.
Nevertheless, Figure~\ref{fig:vesta_varpar_RV0} hints towards a small set of Monte Carlo realisations to draw accurate stellar parameters from the distributions $D^*$ of the parameters of interest, provided $\mathrm{RV}=0$.

We can quantify the minimum number of Monte Carlo realisations assuming the distribution $D^*$ of the means ($\overline{X}_1,\ldots,\overline{X}_i$), for each stellar parameter, tends to the normal distribution (the Central Limit Theorem). 
We require the mean $\overline{X}_n$ of the distribution $D^*$ to be $Z_{\alpha/2}\sigma/\sqrt{n}$ units from the true mean $\umu$, where $\sigma$ is the true standard deviation, $1-\alpha$ the required confidence level, and $Z_{\alpha/2}$ the value of the z-statistic.
An unbiased estimator for the variance $\sigma^2$ is the sample variance 
\begin{equation}
    s_n^2=\frac{1}{n-1}\sum_{i=2}^{i=n}(X_i-\overline{X}_n)^2;
\end{equation}
hence the confidence interval for the mean, 
\begin{equation}\label{eq:criterion}
    Z_{\alpha/2}\frac{s_n}{\sqrt{n}}<|\overline{X}_n-\mu|,
\end{equation}
becomes our convergence criterion, where $n$ is the number of Monte Carlo realisations. 
We impose the following absolute error limits on the stellar parameters: $|\overline{X}_n-\mu|=[2.5~\text{K},0.005~\text{dex}, 0.0025~\text{dex}, 0.0025~\text{dex}]$ for $\mathrm{T_{eff}}$, $\mathrm{\log(g)}$, $\mathrm{[M/H]}$, and $\mathrm{[\alpha/\text{M}]}$, respectively, and $Z_{\alpha/2}=1.96$ for a 95\% interval confidence. 
The above limits correspond to $|\overline{X}_n-\mu|=1\%\times[\Delta_T,\Delta_{\mathrm{\log(g)}},\Delta_{\mathrm{[M/H]}}, \Delta_{\mathrm{[\alpha/\text{M}]}}]$, in terms of the \marcs\ grid resolution.

We compute the convergence criterion, equation~(\ref{eq:criterion}), starting from the Monte Carlo realisation $n=40$ (as to ensure the validity of the Central Limit theorem), and plot the minimum realisation number where the criterion is fulfilled (\textit{filled} triangles in Figures~\ref{fig:vesta_varpar_RV0} and ~\ref{fig:vesta_varpar_RV}).
Except for the model with $\mathrm{RV}\ne0$, convergence is achieved around $\sim40-60$ Monte Carlo realisations.
Effective temperature convergence, however, is reached at $\sim100-200$ realisations.
If we increase the absolute error limits to be $5\%$ the \marcs\  grid resolution, then the minimum number of realisations is $n=40$ for all the stellar parameters and for the two models of Vesta (we recall that $n=40$ is the number from which we start to measure the absolute error $|\overline{X}_n-\mu|$).

We now compare the mean values of the mean distributions $D^*$ at $n=50$ and at $n=1000$ Monte Carlo realisations (columns (2) and (3) of Table~\ref{tabla:Vesta}) to each other, in terms of the \marcs\  grid steps $\Delta_X$. Absolute differences between the models with $RV=0$ are $1.2\%\times\Delta_{\mathrm{[M/H]}}$, $0.4\%\times\Delta_{\mathrm{[\alpha/\text{M}]}}$, $0.02\%\times\Delta_{\mathrm{\log(g)}}$, and $1.8\%\times\Delta_T$, for $\mathrm{[M/H]}$, $\mathrm{[\alpha/\text{M}]}$, $\mathrm{\log(g)}$, and $\mathrm{T_{eff}}$, respectively. For models with $RV\ne0$, absolute differences are $3\%\times\Delta_{\mathrm{[M/H]}}$, $2\%\times\Delta_{\mathrm{[\alpha/\text{M}]}}$, $6.5\%\times\Delta_{\mathrm{\log(g)}}$, and $6.6\%\times\Delta_T$. As expected, the largest differences between the means derive from the model with the radial velocity optimised ($\mathrm{RV}\ne0$).
Thus the differences between the means of the distributions $D^*$ at  $n=50$ and at $n=1000$ are small enough to continue choosing the former value. 

In all, we confirm that with $n=50$ Monte Carlo experiments to obtain 1D estimates of the stellar parameters is a suitable choice, at least for the Vesta \apo\ spectra, when comparing them to the \marcs\ library of synthetic spectra.
However, we should note that fixing the Monte Carlo realisations to compute a large catalogue of stellar parameters is not ideal due to the diversity of the \apo\ spectra (owing to the SNR), and to the several available synthetic libraries to which we can compare the spectra, due to possible model degeneracy (Section~\ref{sec:modeltonallivesta}). 
We defer to future work the implementation of a bootstrap technique to automate the convergence criterion.

\begin{table*}
\caption{Univariate and Multivariate statistics from the Monte Carlo simulation ($n=50$ experiments) for the solar spectrum reflected by Vesta}\label{tabla:Vesta}
\begin{threeparttable}
\resizebox{\textwidth}{!}{
\begin{tabular}{c|rr|rrrc|rr}
 &\multicolumn{2}{c}{Univariate $D^*$ {Mean}} & \multicolumn{4}{c}{Univariate $D_s$}& \multicolumn{2}{c}{Multivariate $D_s$}\\
Input $\mathrm{RV}$ & {$N_\text{rep}=50$} &  {$N_\text{rep}=1000$} & Mean &Median & Binned Mode &$H_0$ & Gaussian Mean & Halfspace Median\\
(1) & (2) & (3) & (4) &(5) & (6) & (7) & (8) & (9)\\\hline
\multicolumn{1}{c}{} & \multicolumn{7}{c}{$\mathrm{T_{eff}}$ (K)}\\\hline
0 & $5880\pm${75}  & {$5875\pm84$} & $5880\pm361$ & $5904_{-159}^{+134}$ & $5942$ &A & $5870\pm197$ & $5899$\\
var & $5812\pm${207} & {$5795\pm 195$} &$5812\pm446$ & $5779_{-304}^{+372}$ & $5652$ &A & $5804\pm234$ & $5781$\\\hline
\multicolumn{1}{c}{} & \multicolumn{7}{c}{$\mathrm{\log(g)}$ (dex)}\\\hline
 0 & $4.60\pm${0.09} & {$4.60\pm0.10$} & $4.60\pm0.41$ & $4.67_{-0.24}^{+0.15}$ & $4.67$ &R & $4.59\pm0.22$ & $4.67$\\
var & $4.55\pm ${0.24}  & {$4.52\pm0.22$} & $4.55\pm0.46$ & $4.51_{-0.33}^{+0.40}$ & $4.25$ &A & $4.56\pm0.24$ & $4.52$\\\hline
\multicolumn{1}{c}{} & \multicolumn{7}{c}{$\mathrm{[M/H]}$ (dex)}\\\hline
0 & $8.48\times10^{-3}\pm${0.03} & {$5.56\times10^{-3}\pm0.04$} & $8.71\times10^{-3}\pm0.18$ & $2.70\times10^{-2}\,_{-0.08}^{+0.06}$ & $4.75\times10^{-2}$ &A & $-0.21\times10^{-3}\pm0.11$ & $2.50\times10^{-2}$\\
var & $-3.61\times10^{-2}\pm${0.10}  & {$-4.36\times10^{-2}\pm0.09$} & $-3.52\times10^{-2}\pm0.23$ & $-3.10\times10^{-2}\pm{-0.16}$ & $3.46\times10^{-2}$ &A & $-4.48\times10^{-2}\pm0.13$ &$-3.34\times10^{-2}$\\\hline
\multicolumn{1}{c}{} & \multicolumn{7}{c}{$\mathrm{[\alpha/\text{M}]}$ (dex)}\\\hline
0 & $-1.03\times10^{-2}\pm${0.03}  & {$-9.28\times10^{-3}\pm0.02$} & $-9.58\times10^{-3}\pm0.28$ & $-4.00\times10^{-2}\,_{-0.06}^{+0.07}$ & $-5.80\times10^{-2}$ &A & $-8.50\times10^{-3}\pm0.14$ & $-4.00\times10^{-2}$\\
var & $5.15\times10^{-2}\pm${0.07} & {$5.65\times10^{-2}\pm0.07$} & $5.08\times10^{-2}\pm0.30$ & $0.00_{-0.14}^{+0.16}$ & $1.10\times10^{-2}$ &R & $6.07\times10^{-2}\pm0.15$ & $ 3.59\times10^{-3}$ \\\hline
\multicolumn{1}{c}{} & \multicolumn{7}{c}{$\mathrm{v\sin(i)}$ (km s$^{-1}$)}\\\hline
0 & $11.06\pm${0.23} & {$11.11\pm0.29$} &$11.06\pm1.66$ & $10.85_{-0.51}^{+0.95}$ & $11$ & A& $11.99\pm5.34$ & $10.86$\\
var & $16.03\pm${0.45} & {$16.00\pm0.45$} &$16.02\pm2.41$ & $13.93_{-0.94}^{+3.20}$ & $14$ & A& $17.34\pm6.30$ & $13.99$\\\hline
\multicolumn{1}{c}{} & \multicolumn{7}{c}{$\mathrm{RV}$ (km s$^{-1}$)}\\\hline
var & $-1.40\pm${0.10}& {$-1.39\pm0.11$} &$-1.40\pm1.32$ & $-0.31_{-2.93}^{+0.82}$ & $0.3$ & A& $-1.4\pm6.38$ & $-0.35$\\\hline
\end{tabular}}
\begin{minipage}{\linewidth}\small
Column (1) indicates if the radial velocity is fixed ($\mathrm{RV}=0$) or was optimised in \tonalli\ (var). 
Columns (2)-(6) report one-dimensional statistics: 
the mean of the distribution of means $D^*$ (columns 2 {and 3, for $N_{\text{rep}}=50$ and $N_{\text{rep}}=1000$, respectively}), the mean, median and the mode (columns 4, 5, and 6, respectively) of all the individuals computed in the realisations, the distribution $D_s$. 
Error values for the mean $D^*$ {are the standard deviation of the distribution of means $D^*$}, while for the mean and median of the distribution $D_s$, the error is the standard deviation of $D_s$ and the differences between the median and the 25 and 75 quartiles of $D_s$, respectively. 
In column (7), we present the result of the Silverman test for multi-modality, where the null hypothesis $H_0$ is the distribution has 1 mode (uni-modal distribution).
Column (8) shows the parameter values when a multivariate normal distribution is fitted to the $D_s$ distribution; error values are the half the standard deviation obtained from the diagonal of the covariance matrix.
Column (9) reports the halfspace median (a multivariate median) of the $D_s$ distribution.
\end{minipage}
\end{threeparttable}
\end{table*}

\subsection{Sun derived parameters from \tonalli}\label{sec:modeltonallivesta}
\begin{figure*}
\includegraphics[alt={Graphs showing unidimensional and bidimensional histograms of the parameters obtained by tonalli for the Solar spectrum reflected by Vesta, with detailed statistical values.}, width=1.5\columnwidth]{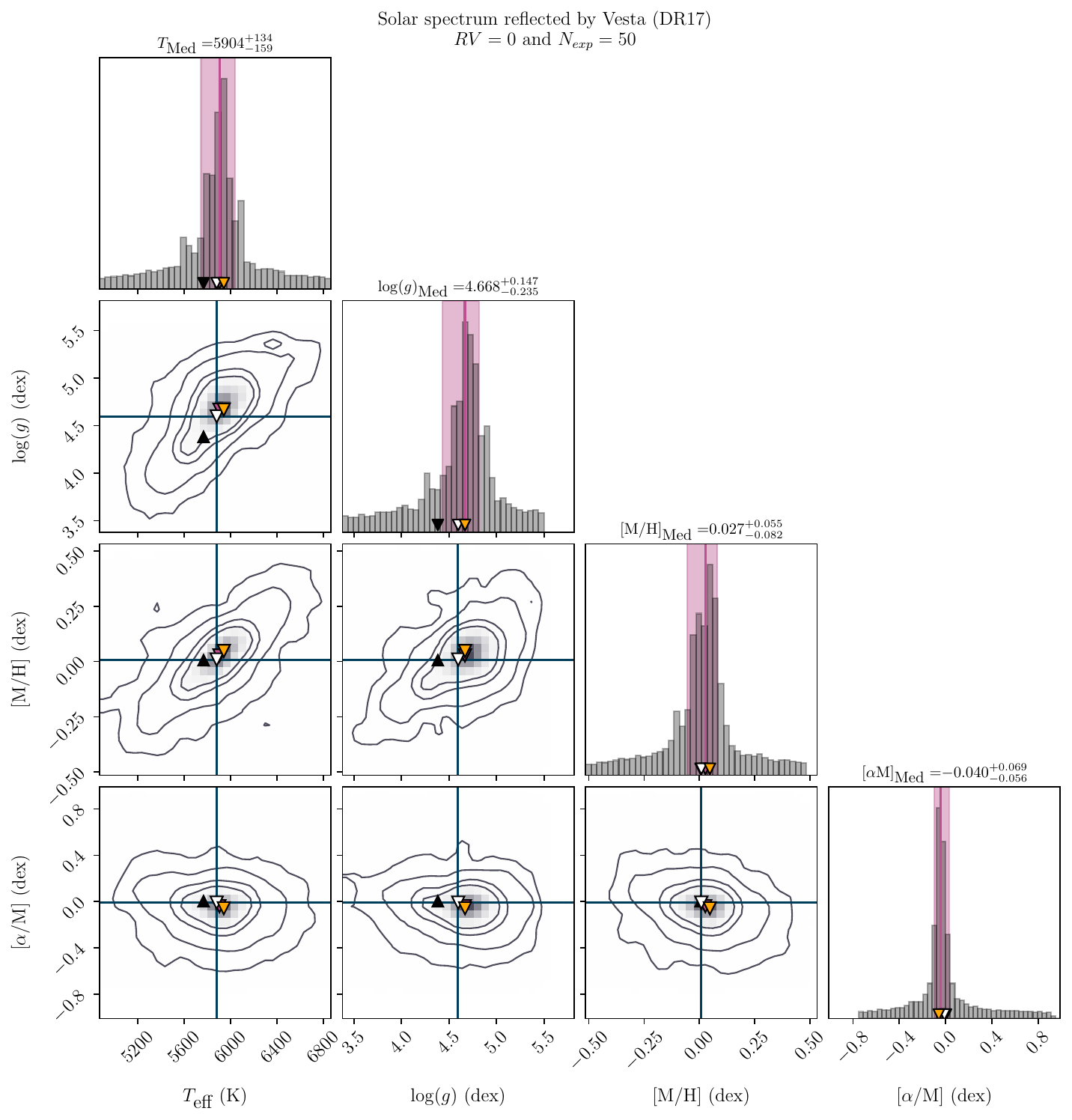}
\caption{Corner plot of the resultant total population of the Monte Carlo realisations in \tonalli\ for the APOGEE-2 Solar spectrum reflected by Vesta, assuming the radial velocity of the spectrum is $\mathrm{RV}=0$, and for $N=50$ realisations.  The \textit{contour lines} encloses the population within the 2D $0.25, 0.5, 0.75$ and $1$ standard deviations of the \textit{population} mean. The \textit{vertical} and \textit{horizontal} lines in the 2D histograms correspond to the mean value of the parameter of the distribution $D^*$. The diagonal displays the 1D histograms of each parameter, along with the median and the corresponding differences between the median and the first and third quartiles. \textit{Triangles} denote the following figures: mean (\textit{white triangle}), median (\textit{magenta triangle}), and mode (\textit{yellow triangle}) of the Monte Carlo distribution $D_*$, along with the spectroscopic results of the solar spectrum reported by DR17 \citep[\textit{black triangle,}][]{DR17}. See text for details. \label{fig:vesta_chi2_RV0}}
\end{figure*}

\begin{figure*}
\includegraphics[alt={Graphs showing unidimensional and bidimensional histograms of the parameters obtained by tonalli for the Solar spectrum reflected by Vesta, with detailed statistical values, when tonalli optimises the radial velocity.}, width=1.5\columnwidth]{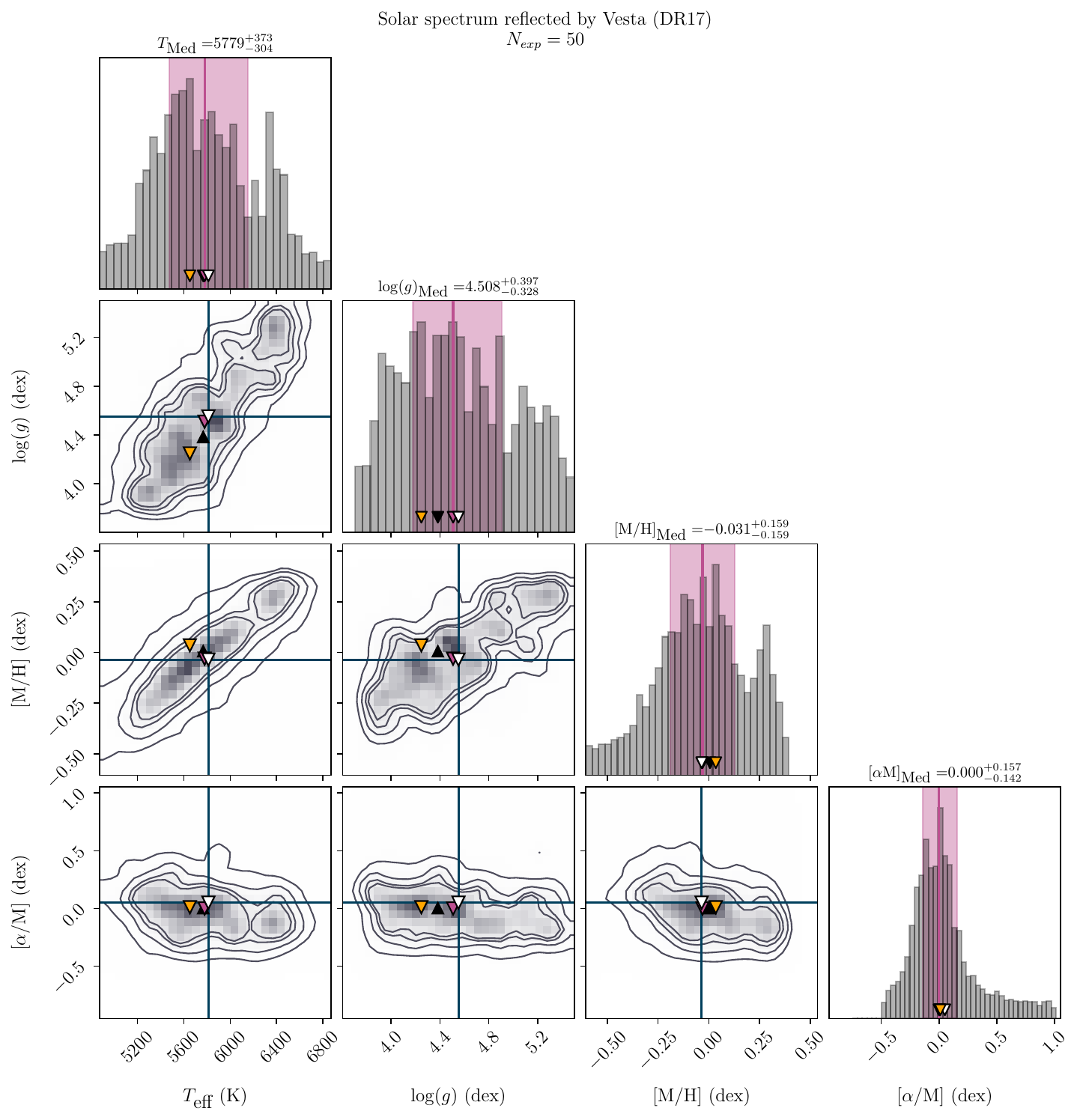}
\caption{Corner plot of the resultant total population of the Monte Carlo realisations in \tonalli\ for the APOGEE-2 solar spectrum reflected by Vesta, assuming the radial velocity of the spectrum is optimised, for $N=50$ realisations. See Figure~\ref{fig:vesta_chi2_RV0} for details. \label{fig:vesta_chi2_RV}}
\end{figure*}

The repetition procedure, which in its core is a Monte Carlo method, aims to obtain both the best value and its associated credible interval.
We can think of all the offspring generated in all the repetitions as a one ensemble of data, $D_s$, regardless of which repetition was each datum generated. 
The corner plots in Figures~\ref{fig:vesta_chi2_RV0} and \ref{fig:vesta_chi2_RV} show the correlation between pairs of parameters and the 1D distribution of each parameter, plotting all the offspring computed in the Monte~Carlo repetition of \tonalli. 

We first discuss the one-dimensional (1D) distribution of each parameter, ignoring any correlation between the parameters.
The median and the interquartile range of the 1D parameter $X$ distribution $D_*$ are shown in the title and as the magenta triangle, the line, and the shaded region of the diagonal plots of Figures~\ref{fig:vesta_chi2_RV0} and \ref{fig:vesta_chi2_RV}, while the mean and the mode of the distribution $D^*$ are shown in blue and yellow triangles, respectively. 
The values are listed in Table~\ref{tabla:Vesta} (mean, median, and binned mode: {columns (4), (5), and (6)}, respectively).

\subsubsection{Multi-modality and parameter degeneracy}\label{Section:multimodality}
We observe that, for a given model and stellar parameter, the means, the medians and the modes differ from each other.   
The absolute differences can probe whether the one-dimensional distribution is multi-modal, which can pinpoint towards a model degeneracy. 
However, we adopt a more rigorous approach, the Silverman test of multi-modality \citep{Silverman81}. 
In our case, the null hypothesis $H_0$ is that the distribution has one mode, versus the alternative $H_1$ that it has two modes. 
In table \ref{tabla:Vesta}, {column (7)} reports if the null hypothesis is accepted or rejected.

For $\mathrm{T_{eff}}$ and $\mathrm{[M/H]}$ our tests show no evidence of bi-modality/multi-modality. 
However, for $\mathrm{[\alpha/\text{M}]}$ we found no evidence of multi-modality when RV is not optimised but the $\mathrm{[\alpha/\text{M}]}$ degenerates when we leave the RV as another optimising parameter. We found the contrary for $\mathrm{\log(g)}$. Having a multi-modal distribution in certain stellar parameter is a direct consequence of the theoretical grid employed and needs to be considered in the error budget as the standard deviation or the interquartile range of the data ensemble. Larger errors are expected -and obtained- for tests where the null hypothesis is rejected.

\begin{figure*}
\includegraphics[alt={Graphs comparing the results of several statistical figures obtained by tonalli with previous published works.}, width=2\columnwidth]{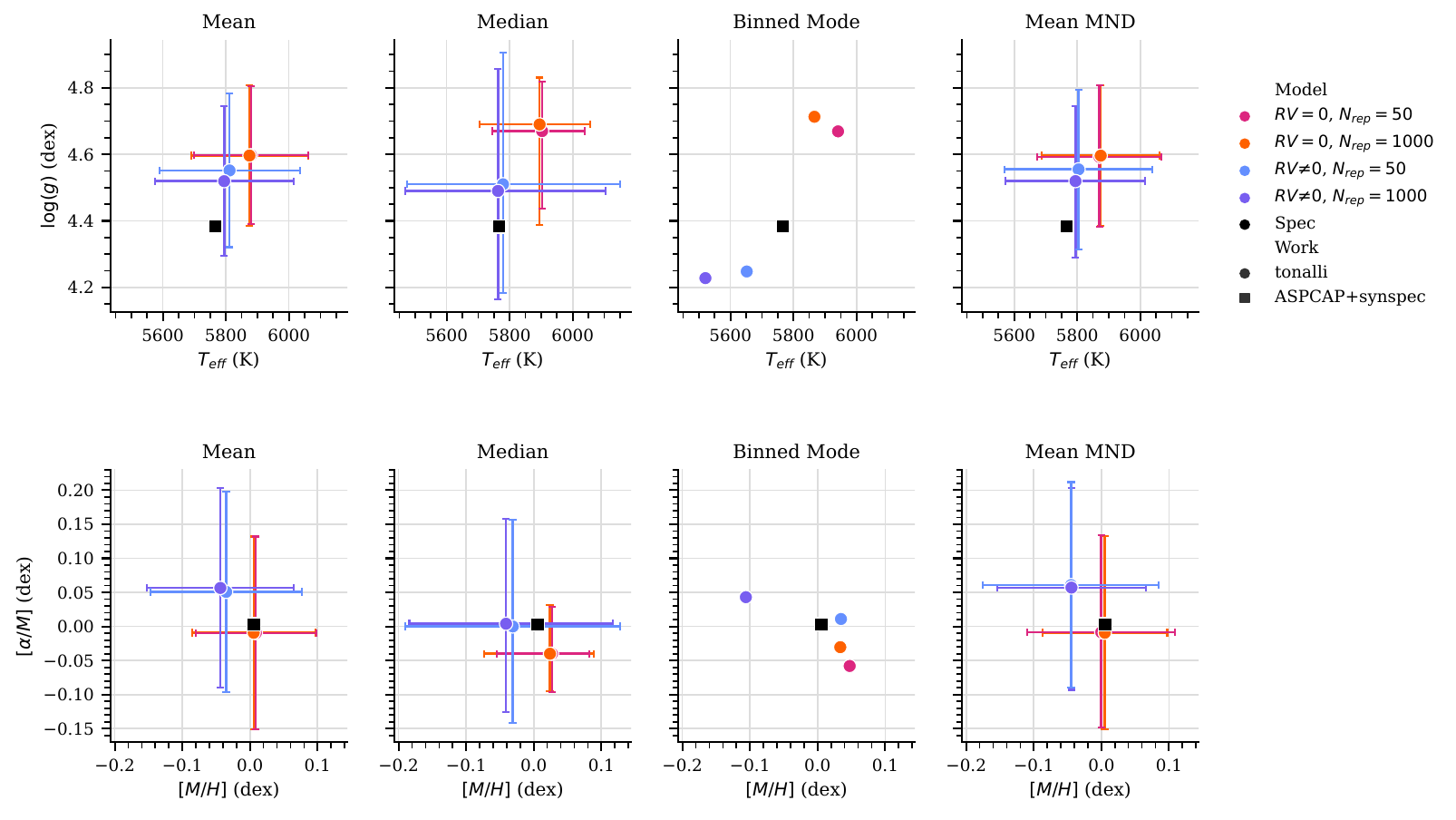}
\caption{Comparison of the diverse statistics figures (1D: mean, median and binned mode; 4D: mean of the multivariate normal distribution, MND, indicated in the titles of each column) obtained by \tonalli\  with the ASPCAP results (Synspec, spectroscopic parameters). \textit{Top panels}: Temperature vs logarithm of the surface temperature plots. \textit{Bottom panels}: Metal vs $\alpha$-elements abundances plots.}\label{fig:vestaparameterscomp}
\end{figure*}

Our results for Vesta differ from those published in the DR17 catalogues, as we show in Figure~\ref{fig:vestaparameterscomp}.  Ideally, we want the three estimators (mode, median, and mean) to match, but in this case, the mismatch between the mean values and the other two figures is prompted by \textit{local minima} in the $\chi^2$ parameter.
The existence of a local minima warrants further inspection of the parameters for a given star spectrum.   
The local minima might not correspond to the \textit{reality} or the \textit{expected values}, thus invalidating their importance. 
However, except for the Sun, we do not presume to know the \textit{reality} -the \textit{actual} values of parameters- a priori.
In this work, we adopt the \textit{median} of the one-dimensional distribution of the parameters, as it is a robust statistic that will lean towards the \textit{most probable} value, or at least, to the centre of the distribution.
Adopting the median and the interquartile range somewhat alleviates the discrepancy between our and the published results, as the quartile 75 of the distributions of each parameter lies close to the \textit{expected} published values. 

When we allow \tonalli\ to optimise the radial velocity, $\mathrm{\log(g)}$ tends towards a smaller value. In turn, both $\mathrm{T_{eff}}$ and $\mathrm{[M/H]}$ shift towards cooler and sub-solar metallicity, respectively, similarly to the results found previously \citep[e.g.][]{Sarmento2020}.
Dealing with this model degeneracies have two possible solutions: provide an initial guide (a prior knowledge) or a later calibration. 
The latter was the adopted approach by \cite{Jonsson20} and by \cite{DR17}.
For this work, we have chosen not to calibrate the \tonalli\ derived parameters for the sake of transparency, thus rendering the Sun-derived parameters to be the result of a direct comparison with the \marcs\ library, or in other words, the \textit{mathematical-correct} result.
We defer to the following paper of the series the adoption of priors.

\begin{table*}
\caption{{Atmospheric parameters of the Sun}}\label{tabla:VestaLiteratura}
\begin{threeparttable}
\resizebox{\textwidth}{!}{
\begin{tabular}{ll|ll|rr|rr}
 \multicolumn{2}{c}{} & \multicolumn{1}{c}{} & \multicolumn{1}{c}{} & \multicolumn{2}{c}{Metallicity} & \multicolumn{2}{c}{$\alpha$-elements abundance}\\
 Work & References and Notes & $\mathrm{T_{eff}}$  & $\mathrm{\log(g)}$ & $\mathrm{[M/H]}$ & $\mathrm{[Fe/H]}$ &$\mathrm{[\alpha/M]}$  & $\mathrm{[\alpha/Fe]}$  \\
&  & (K) & (dex) &(dex) & (dex) & (dex) & (dex) \\
(1) & (2) & (3) & (4) &(5) & (6) & (7) & (8) \\\hline
\tonalli & (this work) Spectroscopic, 1D median & $5779^{+372}_{-304}$& $4.51^{+0.40}_{-0.33}$ & $-0.03\pm0.15$&  & $0.00^{+0.16}_{-0.14}$& \\
& (this work) Spectroscopic, 1D median, $N_0$-$p$\tnote{d} & $5780^{+55}_{-51}$ & $4.44^{+0.06}_{-0.03}$ & $-0.028^{+0.029}_{-0.033}$ & & $-0.024^{+0.020}_{-0.025}$ & \\
J\"onsson et al. SDSS-DR16\tnote{e}& (1) Spectroscopic & $5770$ & $4.49\tnote{b}$ & $-0.001$ &  & $0.000$& \\
& (1) Calibrated & $5712\pm115$\tnote{a} & $4.40\pm0.08$\tnote{b} & $0.002\pm0.009$\tnote{c} &  & $-0.011\pm0.007$\tnote{c} &\\
Abdurro'uf et al. SDSS-DR17\tnote{f} & (2) Spectroscopic & $5767\pm 223\tnote{a}$ & $4.38\pm0.00$\tnote{b} & $0.006\pm0.000$\tnote{c} & & $0.003\pm0.000$\tnote{c} & \\
& (2) Calibrated& $5795\pm 25\tnote{a}$ & $4.43\pm0.02$\tnote{b} & $0.006\pm0.005$\tnote{c} & & $0.004\pm0.007$\tnote{c} & \\
Pr{\v{s}}a et al. & (3) IAU definition & $5772$ & $4.44$ & & $0$ & & \\
Heiter et al. &(4) Measured $\mathrm{L}_\odot$, $\mathrm{G M}_\odot$, and $\mathrm{R}_\odot$ & $5771\pm1$ & $4.44\pm0.00$\tnote{b} & & $0.03\pm0.05$ & & $0.0$ \\
Porto de Mello et al. & (5) Adopted solar values. & $5777$ & $4.44$ & & $0$ & & \\
& (5) Photometric/spectroscopic, Callisto  & $5770$ & $4.52$ & & $-0.04$ & & \\
 & (5) Photometric/spectroscopic, Europa  & $5780$ & $4.48$ & & $-0.02$ & & \\
Takeda et al. &(6)  Fe I and Fe II equivalent widths & $5718\pm25$ & $4.35\pm0.08$ & & & &\\
\hline
\end{tabular}}
\begin{tablenotes}\small
\item[a] Rounded to the nearest integer.
\item[b] Rounded to the nearest hundredth.
\item[c] Rounded to the nearest thousandth.
\end{tablenotes}
\begin{minipage}{\linewidth}\small
$^\mathrm{d}$ Model with the minimum $\chi^2$ (eq.~\ref{eq:chi2nueva}) from the models of Appendix~\ref{app:montecarlo}, adopting the IAU definition for the Sun physical parameters and $[\alpha/M]_\odot=0$. The input parameters of this model are: $N_0=500$, $N_\mathrm{p}=10$, $p=0.8$, $N_{\mathrm{rep}}=50$.\\
$^\mathrm{e}$ Raw spectroscopic results from the comparison with \marcs\ (turbospectrum) library. Calibrated results: temperatures to a photometric scale, surface gravity calibrated with an scale from isochrones. Abundances are calibrated with a zero point offset: $\mathrm{[\alpha/M]}_{\mathrm{off}}=-0.011$, $\mathrm{[M/H]}_{\mathrm{off}}=+0.003$ such as the median in the solar neighborhood is $\mathrm{[X/M]}=0$.\\
$^\mathrm{f}$ Raw spectroscopic results from the comparison with \marcs\ (synspec) library. Calibrated results: effective temperature calibrated with photometric temperatures; surface gravities calibrated with a neural network, using asteroseismology and isochrones data. Abundances are calibrated with a zero point offset, such as the median in the solar neighborhood is $\mathrm{[X/M]}=0$.\\
References: (1): \citet{Jonsson20}, (2): \citet{DR17}, (3): \citet{Prsa16}, (4): \citet{Heiter15}, (5): \citet{Porto14} , (6):\citet{Takeda02}
\end{minipage}
\end{threeparttable}
\end{table*}


\subsubsection{Univariate parameter values}\label{Section:univariate}
Model degeneracy effects in $\mathrm{\log(g)}$ are washed off if we allow the radial velocity to be optimised by \tonalli, as reported by column (7) of Table~\ref{tabla:Vesta}.
As the null hypothesis for uni-modality for $\mathrm{[\alpha/\text{M}]}$ is rejected, we continue to adopt the one-dimensional median as the \textit{true} stellar parameter. 
Thus, the solar parameters, as derived from the comparison of the \apo\ Vesta spectrum with the \marcs\ synthetic stellar library using the methodology presented in this work, are presented in the first row of Table~\ref{tabla:VestaLiteratura}, along with previous published results from the \apo\ solar spectrum reflected by Vesta \citep[comparison of the spectrum to a synthetic spectra library,][]{Jonsson20,DR17}, from the IAU definition \citep{Prsa16}, from measured solar data \citep{Heiter15}, from photometry \citep{Porto14}, and from equivalent widths of Fe I and Fe II \citep{Takeda02}.

The halves of the interquartile ranges {of our adopted \textit{true} solar parameters} are within the \marcs\ grid steps, except for the temperature, which differs by less than $100$~K from the grid step.
We remind the reader the results stated above were obtained with input parameters $N_p=240$, $p=0.4$, and $N_\text{interpol}=2\times2\times3\times3$. 
With this set of input parameters, \tonalli\ calculates the Sun atmospheric parameters with accuracy \citep[compared to those obtained by][]{DR17}, albeit with a relatively large credible intervals. 
If the radial velocity is fixed to zero, the IQRs of the parameter distributions decrease at the expense of the accuracy. 

In addition to the latter effect, the experiments in Appendix~\ref{appendix:inputparameters} demonstrated that the choice of $N_p$ and $p$ impacts the total number of individuals computed in a full \tonalli\ (plus repetitions) run; a larger $N_\text{total}$ procures not only unbiased results but also shorter credible intervals for the median stellar parameters (see Figures~\ref{fig:paramnoffMC} and~\ref{fig:IQRMC}). 
In this context, the above results we present for the APOGEE-2 Vesta spectrum represent the minimum permissible results, due the large credible intervals. 
The precision of \tonalli\ increases once the input parameters $N_p$, $p$, and to some extent the number of Monte Carlo repetitions, are tuned to provide a sufficiently high total number of individuals.
The second row of Table~\ref{tabla:VestaLiteratura} presents the model with the minimum $\chi^2$ of all the models computed in Appendix~\ref{app:montecarlo} with $\mathrm{IQR}\le\Delta_X/2$ (half the grid step) in the four interpolated grid parameters. We define the $\chi^2$ for this step as:
\begin{multline}\label{eq:chi2nueva}    
\chi^2=\frac{(\mathrm{T_{\odot}}-\mathrm{T}_{\mathrm{m}})^2}{[\mathrm{IQR(T)_{\mathrm{m}}}]^2}+\frac{(\mathrm{\log(g)_{\odot}}-\log(g)_{\mathrm{m}})^2}{[\mathrm{IQR(\log(g))_{\mathrm{m}}}]^2}\\
+\frac{(0-\mathrm{[M/H]}_{\mathrm{m}})^2}{[\mathrm{IQR(\mathrm{[M/H]})_{\mathrm{m}}}]^2}+\frac{(0-\mathrm{[\alpha/M]}_{\mathrm{m}})^2}{[\mathrm{IQR(\mathrm{[\alpha/M]})_{\mathrm{m}}}]^2},
\end{multline}
where we adopt the solar values of the IAU definition \citep{Prsa16}, and the subscript $m$ refers to the \tonalli\ model. 
Thus the model with the minimum $\chi^2$ was constructed with input parameters $N_0=500$, $N_\mathrm{p}=10$, $p=0.8$, $N_{\mathrm{rep}}=50$. 
The credible interval of the median of the atmospheric parameters of this model decreased by almost one order of magnitude respect to the model with input parameters $N_0=240$, $N_\mathrm{p}=10$, $p=0.4$, $N_{\mathrm{rep}}=50$. 
The raw abundances values can be calibrated, but, as explained in Section~\ref{Section:multimodality}, the results are the direct comparison with our continuum-normalised \marcs\ synthetic spectrum library. 

At any rate, the typical errors expected for high resolution spectra are $\sim1.1\%\times \mathrm{T_{eff}}$ for efective temperature (which translates to $\sim60$~K for the Sun), $0.10$~dex for $\mathrm{\log(g)}$, and $0.06$~dex for the determination of metallicity abundances \citep{Soubiran16}. These typical errors are defined as the median of the uncertainties of a large sample of stars (the PASTEL catalogue); the uncertainties were computed, as \citeauthor{Soubiran16} remarks, assuming disparately definitions.

The rotational projected velocity, $\mathrm{v\sin(i)}=13.93$~km~s$^{-1}$,  is near the resolution limit for the \apo\ spectra \citep{Cottaar14}, meaning that our estimation is, at best, an upper limit to the real rotational velocity. We ignore this estimation, since the rotational broadening of the synthetic spectra have no impact on our derived parameters.

Figure~\ref{fig:vestaparameterscomp} ($\mathrm{RV}\ne 0$ results in blue symbols) demonstrate the good agreement between our results and the spectroscopic results (black symbols) of \citet{DR17}, regardless their synthesised library. 
Last, we also added the results of the 1000 Monte Carlo experiments to Figure~\ref{fig:vestaparameterscomp} (green symbols); the parameter values overlap those derived from only 50 experiments, which further supports the latter choice in the interest of computational speed.

\subsubsection{Multivariate parameter values}\label{sec:multivariate}
Figures~\ref{fig:vesta_chi2_RV0} and \ref{fig:vesta_chi2_RV} show, along with the one-dimensional histograms, the two-dimensional histograms for each pair combination of $\mathrm{T_{eff}}$, $\mathrm{\log(g)}$, and $\mathrm{[M/H]}$, $\mathrm{[\alpha/\text{M}]}$. 
We ignore both the radial and the rotational projected velocities, as they do not correlate with any of the grid-parameters above. 
The same applies for $\mathrm{[\alpha/\text{M}]}$: the 2D distributions of our models show a weak or no correlation between the parameter $\mathrm{[\alpha/\text{M}]}$ and the remaining parameters.
The two-dimensional histograms of the pairs $\langle \mathrm{T_{eff}},\mathrm{\log(g)}\rangle$, $\langle \mathrm{T_{eff}},\mathrm{[M/H]}\rangle$, and $\langle \mathrm{\log(g)},\mathrm{[M/H]}\rangle$ in Figures~\ref{fig:vesta_chi2_RV0} and \ref{fig:vesta_chi2_RV} show strong positive correlations, warranting a multivariate analysis of the parameter distribution.
Hence we construct either a five or six-dimensional distribution, containing the data $\bigl\langle \mathrm{T_{eff}},\mathrm{\log(g)},\mathrm{[M/H]},\mathrm{[\alpha/\text{M}]},\mathrm{v\sin(i)}\bigr\rangle$ or $\bigl\langle \mathrm{T_{eff}},\mathrm{\log(g)},\mathrm{[M/H]},\mathrm{[\alpha/\text{M}]},\mathrm{v\sin(i)}, \mathrm{RV}\bigr\rangle$, respectively, depending on the radial velocity assumption of the given model.

First, we obtain the Sun parameters by fitting a multivariate Gaussian \citep[\texttt{ R} package \texttt{ mclust},][]{mclust} to the now multivariate distribution $D_s$. 
The mean multivariate Gaussian parameters and their associated errors (half of the standard deviation in the parameter) are tabulated in column (7) of Table~\ref{tabla:Vesta}. 
Except for $\mathrm{[M/H]}$ for the model with $\mathrm{RV}=0$, the multivariate Gaussian parameter values match with their univariate counterparts.

Interestingly, the one-dimensional histograms of each parameter, discussed in the previous subsection, exhibited the need of a robust statistic figure to obtain the spectroscopic solar parameters. 
The figure of a median in a multivariate setting is not as straightforward as is for the one-dimensional case; in fact, the multivariate median has multiple definitions, with no generalisation from the univariate median \citep{Oja83,Small90}. 
Still, we provide the halfspace median, or Tukey median. This median is based on the concept of the multivariate depth statistics \citep[see][for a review in the depth concept]{Small90,Mosler2013}, which, in short, measures a \textit{centrality} of a given multidimensional point in a data distribution. 
Each point of the distribution is ranked by its depth and then ordered, as we would do in the univariate problem, allowing to obtain the central point or median.
In column (8) of Table~\ref{tabla:Vesta}, we report the halfspace median computed using the \texttt{R} package \texttt{ mrfDepth} \citep{mrfDepth}. 
We reiterate the multivariate medians have multiple definitions and therefore multiple solutions. 
However, tests with different depths definitions \citep[using the \texttt{R} package \texttt{DepthProc} of][along with \texttt{mrfDepth}]{DepthProc} with our data show the medians do not differ notably from each other. 
It is clear the halfdepth medians agree with the univariate medians, which supports our previous univariate results from Section~\ref{Section:univariate}.

\section{Conclusions and future work}\label{sec:conclusions}
We put forward a framework to efficiently compare \apo\ spectra against a user-selected synthetic library (continuum normalised and convolved with a Gaussian profile to match the \apo\ resolution of $R\sim 22,500$) by means of an
Asexual Genetic Algorithm \citep{Canto09}. 

We compute the accuracy and precision of our method by running \tonalli\ for a set of
simulated observed spectra derived from the theoretical \marcs\ spectral library \citep{MARCS08, Jonsson20} recovering their parameters, namely $\mathrm{T_{eff}}$, $\mathrm{\log(g)}$, $\mathrm{[M/H]}$, and $\mathrm{[\alpha/\text{M}]}$, without any appreciable bias for low temperature 
($3200\lesssim\mathrm{T_{eff}}\lesssim 6500$~K) stars.
The results demonstrate the reliability of the \tonalli\ framework, which we apply successfully to the solar spectrum reflected by Vesta. 

With a Monte Carlo approach, we obtain the fitted parameters and their associated errors for the solar \apo\ spectrum. 
We find that convergence in the parameter distributions (for both the synthetic set and the solar spectrum) is reached with $\sim50$ Monte Carlo experiments; we defer to automate the convergence criterion to future work.
We compare several statistics figures (mean, median, mode) for the univariate and multivariate distributions of the explored parameters, since the parameter histograms for the model with optimised $RV$ deviate from an unimodal distribution, owing to the synthetic stellar library degeneracy. 
We find that the median univariate values are comparable to the literature data.

This work is the stepping stone to provide reliable spectroscopic stellar parameters for pre-main sequence stars.
The present article is intended as the first of a series, where the second will focus on discussing the performance of \tonalli\ on spectra of pre-main sequence stars and how they improve with a priori delimitation of the parameter space. A third paper of the series will focus on a detailed comparison of results from \tonalli\ using different grids of stellar atmosphere models.

\section*{Acknowledgements}
We thank the reviewers for their careful reading of the manuscript and their thorough and constructive feedback.

L.A. and R. L-V acknowledge support from CONAHCyT through postdoctoral fellowships within the program Estancias Posdoctorales por M\'exico. The authors acknowledge support from CONAHCyT projects CB 2018 A1-S-9754 and CF 86372, as well as projects UNAM DGAPA PAPIIT IN112620 and IG101723. 

This work made use of the high-throughput computing (HTC) infrastructure and the optimisation of computing resources of the project Grid UNAM (\url{https://grid.unam.mx/}) at the Universidad Nacional Aut\'onoma de M\'exico. Participating Institutions of the project Grid UNAM includes Direcci\'on General de C\'omputo y de Tecnolog\'ias de Informaci\'on y Comunicaci\'on (DGTIC), Laboratorio de Modelos y Datos Cient\'ificos (LAMOD), Instituto de Ciencias de la Atm\'osfera y Cambio Clim\'atico (ICAyCC), Instituto de Astronom\'ia (IA) and Instituto de Ciencias Nucleares (ICN). We also thank the members of technical and development staff of the project Grid UNAM for their continuous support. 

Funding for the Sloan Digital Sky Survey IV has been provided by the Alfred P. Sloan Foundation, the U.S. Department of Energy Office of Science, and the Participating Institutions. SDSS acknowledges support and resources from the Center for High-Performance Computing at the University of Utah. The SDSS web site is \url{www.sdss4.org}.

SDSS is managed by the Astrophysical Research Consortium for the Participating Institutions of the SDSS Collaboration including the Brazilian Participation Group, the Carnegie Institution for Science, Carnegie Mellon University, Center for Astrophysics | Harvard \& Smithsonian (CfA), the Chilean Participation Group, the French Participation Group, Instituto de Astrof\'isica de Canarias, The Johns Hopkins University, Kavli Institute for the Physics and Mathematics of the Universe (IPMU) / University of Tokyo, the Korean Participation Group, Lawrence Berkeley National Laboratory, Leibniz Institut für Astrophysik Potsdam (AIP), Max-Planck-Institut f\"{u}r Astronomie (MPIA Heidelberg), Max-Planck-Institut f\"{u}r Astrophysik (MPA Garching), Max-Planck-Institut f\"{u}r Extraterrestrische Physik (MPE), National Astronomical Observatories of China, New Mexico State University, New York University, University of Notre Dame, Observat\'orio Nacional / MCTI, The Ohio State University, Pennsylvania State University, Shanghai Astronomical Observatory, United Kingdom Participation Group, Universidad Nacional Aut\'onoma de M\'exico, University of Arizona, University of Colorado Boulder, University of Oxford, University of Portsmouth, University of Utah, University of Virginia, University of Washington, University of Wisconsin, Vanderbilt University, and Yale University.

This research has made use of the SIMBAD database, operated at CDS, Strasbourg, France. 
This research has made use of the VizieR catalogue access tool, CDS, Strasbourg, France (DOI: 10.26093/cds/vizier). The original description of the VizieR service was published in 2000, A\&AS 143, 23.

\textbf{\textit{Software.}}
The code \tonalli\ and this work made extensive use of the following \texttt{python} \citep{python} libraries/modules:
Astropy \citep{astropy:2013, astropy:2018},
corner.py \citep{corner},
Matplotlib \citep{Matplotlib},
NumPy \citep{2020NumPy-Array},
Pandas \citep{Pandas10},
parmap \citep{parmap},
patchworklib \citep{Patchwork},
PyAstronomy \citep{pya},
SciPy \citep{2020SciPy-NMeth},
seaborn \citep{seaborn}, the software \texttt{ TOPCAT} \citep{topcat}, 
and the following packages in \texttt{ R} \citep{rsoft}: 
mclust \citep{mclust}, mixR \citep{mixr}, and multimode \citep{Ameijeiras2021}.

\section*{Data availability statement}
The astronomical data (\apo\, DR17 solar spectrum reflected by Vesta) used in this work was obtained from the publicly available SDSS Science Archive Server (\url{https://dr17.sdss.org/infrared/plate/search}). The synthetic stellar \marcs\, library is available at \url{https://data.sdss.org/sas/dr16/apogee/spectro/speclib/synth/turbospec/marcs/solarisotopes/}. 

{The \marcs\, continuum normalised library we constructed for this work can be found at \url{https://doi.org/10.5281/zenodo.12736546}.}


\bibliographystyle{rasti}
\bibliography{tonalli}

\begin{thebibliography}{101}
\expandafter\ifx\csname natexlab\endcsname\relax\def\natexlab#1{#1}\fi

\bibitem[{Abdurro'uf} et~al.(2022){Abdurro'uf}, {Accetta}, {Aerts}, {Silva Aguirre}, {Ahumada}, {Ajgaonkar}, {Filiz Ak}, {Alam}, {Allende Prieto}, {Almeida}, {Anders}, {Anderson}, {Andrews}, {Anguiano}, {Aquino-Ort{\'\i}z}, {Arag{\'o}n-Salamanca}, {Argudo-Fern{\'a}ndez}, {Ata}, {Aubert}, {Avila-Reese}, {Badenes}, {Barb{\'a}}, {Barger}, {Barrera-Ballesteros}, {Beaton}, {Beers}, {Belfiore}, {Bender}, {Bernardi}, {Bershady}, {Beutler}, {Bidin}, {Bird}, {Bizyaev}, {Blanc}, {Blanton}, {Boardman}, {Bolton}, {Boquien}, {Borissova}, {Bovy}, {Brandt}, {Brown}, {Brownstein}, {Brusa}, {Buchner}, {Bundy}, {Burchett}, {Bureau}, {Burgasser}, {Cabang}, {Campbell}, {Cappellari}, {Carlberg}, {Wanderley}, {Carrera}, {Cash}, {Chen}, {Chen}, {Cherinka}, {Chiappini}, {Choi}, {Chojnowski}, {Chung}, {Clerc}, {Cohen}, {Comerford}, {Comparat}, {da Costa}, {Covey}, {Crane}, {Cruz-Gonzalez}, {Culhane}, {Cunha}, {Dai}, {Damke}, {Darling}, {Davidson}, {Davies}, {Dawson}, {De Lee}, {Diamond-Stanic}, {Cano-D{\'\i}az}, {S{\'a}nchez},
  {Donor}, {Duckworth}, {Dwelly}, {Eisenstein}, {Elsworth}, {Emsellem}, {Eracleous}, {Escoffier}, {Fan}, {Farr}, {Feng}, {Fern{\'a}ndez-Trincado}, {Feuillet}, {Filipp}, {Fillingham}, {Frinchaboy}, {Fromenteau}, {Galbany}, {Garc{\'\i}a}, {Garc{\'\i}a-Hern{\'a}ndez}, {Ge}, {Geisler}, {Gelfand}, {G{\'e}ron}, {Gibson}, {Goddy}, {Godoy-Rivera}, {Grabowski}, {Green}, {Greener}, {Grier}, {Griffith}, {Guo}, {Guy}, {Hadjara}, {Harding}, {Hasselquist}, {Hayes}, {Hearty}, {Hern{\'a}ndez}, {Hill}, {Hogg}, {Holtzman}, {Horta}, {Hsieh}, {Hsu}, {Hsu}, {Huber}, {Huertas-Company}, {Hutchinson}, {Hwang}, {Ibarra-Medel}, {Chitham}, {Ilha}, {Imig}, {Jaekle}, {Jayasinghe}, {Ji}, {Johnson}, {Jones}, {J{\"o}nsson}, {Katkov}, {Khalatyan}, {Kinemuchi}, {Kisku}, {Knapen}, {Kneib}, {Kollmeier}, {Kong}, {Kounkel}, {Kreckel}, {Krishnarao}, {Lacerna}, {Lane}, {Langgin}, {Lavender}, {Law}, {Lazarz}, {Leung}, {Leung}, {Lewis}, {Li}, {Li}, {Lian}, {Liang}, {Lin}, {Lin}, {Lin}, {Lintott}, {Long}, {Longa-Pe{\~n}a}, {L{\'o}pez-Cob{\'a}}, {Lu},
  {Lundgren}, {Luo}, {Mackereth}, {de la Macorra}, {Mahadevan}, {Majewski}, {Manchado}, {Mandeville}, {Maraston}, {Margalef-Bentabol}, {Masseron}, {Masters}, {Mathur}, {McDermid}, {Mckay}, {Merloni}, {Merrifield}, {Meszaros}, {Miglio}, {Di Mille}, {Minniti}, {Minsley}, {Monachesi}, {Moon}, {Mosser}, {Mulchaey}, {Muna}, {Mu{\~n}oz}, {Myers}, {Myers}, {Nadathur}, {Nair}, {Nandra}, {Neumann}, {Newman}, {Nidever}, {Nikakhtar}, {Nitschelm}, {O'Connell}, {Garma-Oehmichen}, {Luan Souza de Oliveira}, {Olney}, {Oravetz}, {Ortigoza-Urdaneta}, {Osorio}, {Otter}, {Pace}, {Padilla}, {Pan}, {Pan}, {Parikh}, {Parker}, {Peirani}, {Pe{\~n}a Ram{\'\i}rez}, {Penny}, {Percival}, {Perez-Fournon}, {Pinsonneault}, {Poidevin}, {Poovelil}, {Price-Whelan}, {B{\'a}rbara de Andrade Queiroz}, {Raddick}, {Ray}, {Rembold}, {Riddle}, {Riffel}, {Riffel}, {Rix}, {Robin}, {Rodr{\'\i}guez-Puebla}, {Roman-Lopes}, {Rom{\'a}n-Z{\'u}{\~n}iga}, {Rose}, {Ross}, {Rossi}, {Rubin}, {Salvato}, {S{\'a}nchez}, {S{\'a}nchez-Gallego}, {Sanderson}, {Santana
  Rojas}, {Sarceno}, {Sarmiento}, {Sayres}, {Sazonova}, {Schaefer}, {Schiavon}, {Schlegel}, {Schneider}, {Schultheis}, {Schwope}, {Serenelli}, {Serna}, {Shao}, {Shapiro}, {Sharma}, {Shen}, {Shetrone}, {Shu}, {Simon}, {Skrutskie}, {Smethurst}, {Smith}, {Sobeck}, {Spoo}, {Sprague}, {Stark}, {Stassun}, {Steinmetz}, {Stello}, {Stone-Martinez}, {Storchi-Bergmann}, {Stringfellow}, {Stutz}, {Su}, {Taghizadeh-Popp}, {Talbot}, {Tayar}, {Telles}, {Teske}, {Thakar}, {Theissen}, {Tkachenko}, {Thomas}, {Tojeiro}, {Hernandez Toledo}, {Troup}, {Trump}, {Trussler}, {Turner}, {Tuttle}, {Unda-Sanzana}, {V{\'a}zquez-Mata}, {Valentini}, {Valenzuela}, {Vargas-Gonz{\'a}lez}, {Vargas-Maga{\~n}a}, {Alfaro}, {Villanova}, {Vincenzo}, {Wake}, {Warfield}, {Washington}, {Weaver}, {Weijmans}, {Weinberg}, {Weiss}, {Westfall}, {Wild}, {Wilde}, {Wilson}, {Wilson}, {Wilson}, {Wolf}, {Wood-Vasey}, {Yan}, {Zamora}, {Zasowski}, {Zhang}, {Zhao}, {Zheng}, {Zheng}, \& {Zhu}]{DR17}
{Abdurro'uf}, {Accetta}, K., {Aerts}, C., {Silva Aguirre}, V., {Ahumada}, R., {Ajgaonkar}, N., {Filiz Ak}, N., {Alam}, S., {Allende Prieto}, C., {Almeida}, A., {Anders}, F., {Anderson}, S.~F., {Andrews}, B.~H., {Anguiano}, B., {Aquino-Ort{\'\i}z}, E., {Arag{\'o}n-Salamanca}, A., {Argudo-Fern{\'a}ndez}, M., {Ata}, M., {Aubert}, M., {Avila-Reese}, V., {Badenes}, C., {Barb{\'a}}, R.~H., {Barger}, K., {Barrera-Ballesteros}, J.~K., {Beaton}, R.~L., {Beers}, T.~C., {Belfiore}, F., {Bender}, C.~F., {Bernardi}, M., {Bershady}, M.~A., {Beutler}, F., {Bidin}, C.~M., {Bird}, J.~C., {Bizyaev}, D., {Blanc}, G.~A., {Blanton}, M.~R., {Boardman}, N.~F., {Bolton}, A.~S., {Boquien}, M., {Borissova}, J., {Bovy}, J., {Brandt}, W.~N., {Brown}, J., {Brownstein}, J.~R., {Brusa}, M., {Buchner}, J., {Bundy}, K., {Burchett}, J.~N., {Bureau}, M., {Burgasser}, A., {Cabang}, T.~K., {Campbell}, S., {Cappellari}, M., {Carlberg}, J.~K., {Wanderley}, F.~C., {Carrera}, R., {Cash}, J., {Chen}, Y.-P., {Chen}, W.-H., {Cherinka}, B., {Chiappini},
  C., {Choi}, P.~D., {Chojnowski}, S.~D., {Chung}, H., {Clerc}, N., {Cohen}, R.~E., {Comerford}, J.~M., {Comparat}, J., {da Costa}, L., {Covey}, K., {Crane}, J.~D., {Cruz-Gonzalez}, I., {Culhane}, C., {Cunha}, K., {Dai}, Y.~S., {Damke}, G., {Darling}, J., {Davidson}, James~W., J., {Davies}, R., {Dawson}, K., {De Lee}, N., {Diamond-Stanic}, A.~M., {Cano-D{\'\i}az}, M., {S{\'a}nchez}, H.~D., {Donor}, J., {Duckworth}, C., {Dwelly}, T., {Eisenstein}, D.~J., {Elsworth}, Y.~P., {Emsellem}, E., {Eracleous}, M., {Escoffier}, S., {Fan}, X., {Farr}, E., {Feng}, S., {Fern{\'a}ndez-Trincado}, J.~G., {Feuillet}, D., {Filipp}, A., {Fillingham}, S.~P., {Frinchaboy}, P.~M., {Fromenteau}, S., {Galbany}, L., {Garc{\'\i}a}, R.~A., {Garc{\'\i}a-Hern{\'a}ndez}, D.~A., {Ge}, J., {Geisler}, D., {Gelfand}, J., {G{\'e}ron}, T., {Gibson}, B.~J., {Goddy}, J., {Godoy-Rivera}, D., {Grabowski}, K., {Green}, P.~J., {Greener}, M., {Grier}, C.~J., {Griffith}, E., {Guo}, H., {Guy}, J., {Hadjara}, M., {Harding}, P., {Hasselquist}, S., {Hayes},
  C.~R., {Hearty}, F., {Hern{\'a}ndez}, J., {Hill}, L., {Hogg}, D.~W., {Holtzman}, J.~A., {Horta}, D., {Hsieh}, B.-C., {Hsu}, C.-H., {Hsu}, Y.-H., {Huber}, D., {Huertas-Company}, M., {Hutchinson}, B., {Hwang}, H.~S., {Ibarra-Medel}, H.~J., {Chitham}, J.~I., {Ilha}, G.~S., {Imig}, J., {Jaekle}, W., {Jayasinghe}, T., {Ji}, X., {Johnson}, J.~A., {Jones}, A., {J{\"o}nsson}, H., {Katkov}, I., {Khalatyan}, Arman, D., {Kinemuchi}, K., {Kisku}, S., {Knapen}, J.~H., {Kneib}, J.-P., {Kollmeier}, J.~A., {Kong}, M., {Kounkel}, M., {Kreckel}, K., {Krishnarao}, D., {Lacerna}, I., {Lane}, R.~R., {Langgin}, R., {Lavender}, R., {Law}, D.~R., {Lazarz}, D., {Leung}, H.~W., {Leung}, H.-H., {Lewis}, H.~M., {Li}, C., {Li}, R., {Lian}, J., {Liang}, F.-H., {Lin}, L., {Lin}, Y.-T., {Lin}, S., {Lintott}, C., {Long}, D., {Longa-Pe{\~n}a}, P., {L{\'o}pez-Cob{\'a}}, C., {Lu}, S., {Lundgren}, B.~F., {Luo}, Y., {Mackereth}, J.~T., {de la Macorra}, A., {Mahadevan}, S., {Majewski}, S.~R., {Manchado}, A., {Mandeville}, T., {Maraston}, C.,
  {Margalef-Bentabol}, B., {Masseron}, T., {Masters}, K.~L., {Mathur}, S., {McDermid}, R.~M., {Mckay}, M., {Merloni}, A., {Merrifield}, M., {Meszaros}, S., {Miglio}, A., {Di Mille}, F., {Minniti}, D., {Minsley}, R., {Monachesi}, A., {Moon}, J., {Mosser}, B., {Mulchaey}, J., {Muna}, D., {Mu{\~n}oz}, R.~R., {Myers}, A.~D., {Myers}, N., {Nadathur}, S., {Nair}, P., {Nandra}, K., {Neumann}, J., {Newman}, J.~A., {Nidever}, D.~L., {Nikakhtar}, F., {Nitschelm}, C., {O'Connell}, J.~E., {Garma-Oehmichen}, L., {Luan Souza de Oliveira}, G., {Olney}, R., {Oravetz}, D., {Ortigoza-Urdaneta}, M., {Osorio}, Y., {Otter}, J., {Pace}, Z.~J., {Padilla}, N., {Pan}, K., {Pan}, H.-A., {Parikh}, T., {Parker}, J., {Peirani}, S., {Pe{\~n}a Ram{\'\i}rez}, K., {Penny}, S., {Percival}, W.~J., {Perez-Fournon}, I., {Pinsonneault}, M., {Poidevin}, F., {Poovelil}, V.~J., {Price-Whelan}, A.~M., {B{\'a}rbara de Andrade Queiroz}, A., {Raddick}, M.~J., {Ray}, A., {Rembold}, S.~B., {Riddle}, N., {Riffel}, R.~A., {Riffel}, R., {Rix}, H.-W.,
  {Robin}, A.~C., {Rodr{\'\i}guez-Puebla}, A., {Roman-Lopes}, A., {Rom{\'a}n-Z{\'u}{\~n}iga}, C., {Rose}, B., {Ross}, A.~J., {Rossi}, G., {Rubin}, K. H.~R., {Salvato}, M., {S{\'a}nchez}, S.~F., {S{\'a}nchez-Gallego}, J.~R., {Sanderson}, R., {Santana Rojas}, F.~A., {Sarceno}, E., {Sarmiento}, R., {Sayres}, C., {Sazonova}, E., {Schaefer}, A.~L., {Schiavon}, R., {Schlegel}, D.~J., {Schneider}, D.~P., {Schultheis}, M., {Schwope}, A., {Serenelli}, A., {Serna}, J., {Shao}, Z., {Shapiro}, G., {Sharma}, A., {Shen}, Y., {Shetrone}, M., {Shu}, Y., {Simon}, J.~D., {Skrutskie}, M.~F., {Smethurst}, R., {Smith}, V., {Sobeck}, J., {Spoo}, T., {Sprague}, D., {Stark}, D.~V., {Stassun}, K.~G., {Steinmetz}, M., {Stello}, D., {Stone-Martinez}, A., {Storchi-Bergmann}, T., {Stringfellow}, G.~S., {Stutz}, A., {Su}, Y.-C., {Taghizadeh-Popp}, M., {Talbot}, M.~S., {Tayar}, J., {Telles}, E., {Teske}, J., {Thakar}, A., {Theissen}, C., {Tkachenko}, A., {Thomas}, D., {Tojeiro}, R., {Hernandez Toledo}, H., {Troup}, N.~W., {Trump}, J.~R.,
  {Trussler}, J., {Turner}, J., {Tuttle}, S., {Unda-Sanzana}, E., {V{\'a}zquez-Mata}, J.~A., {Valentini}, M., {Valenzuela}, O., {Vargas-Gonz{\'a}lez}, J., {Vargas-Maga{\~n}a}, M., {Alfaro}, P.~V., {Villanova}, S., {Vincenzo}, F., {Wake}, D., {Warfield}, J.~T., {Washington}, J.~D., {Weaver}, B.~A., {Weijmans}, A.-M., {Weinberg}, D.~H., {Weiss}, A., {Westfall}, K.~B., {Wild}, V., {Wilde}, M.~C., {Wilson}, J.~C., {Wilson}, R.~F., {Wilson}, M., {Wolf}, J., {Wood-Vasey}, W.~M., {Yan}, R., {Zamora}, O., {Zasowski}, G., {Zhang}, K., {Zhao}, C., {Zheng}, Z., {Zheng}, Z., \& {Zhu}, K., 2022.
\newblock {The Seventeenth Data Release of the Sloan Digital Sky Surveys: Complete Release of MaNGA, MaStar, and APOGEE-2 Data}, {\it \apjs\/}, {\bf 259}(2), 35.

\bibitem[{Allard} et~al.(2011){Allard}, {Homeier}, \& {Freytag}]{Allard11}
{Allard}, F., {Homeier}, D., \& {Freytag}, B., 2011.
\newblock {Model Atmospheres From Very Low Mass Stars to Brown Dwarfs}, in {\em 16th Cambridge Workshop on Cool Stars, Stellar Systems, and the Sun\/}, vol. 448 of {\bf Astronomical Society of the Pacific Conference Series}, p.~91.

\bibitem[{Allard} et~al.(2012){Allard}, {Homeier}, \& {Freytag}]{Allard12}
{Allard}, F., {Homeier}, D., \& {Freytag}, B., 2012.
\newblock {Stellar to Substellar Model Atmospheres}, in {\em From Interacting Binaries to Exoplanets: Essential Modeling Tools\/}, vol. 282, pp. 235--242.

\bibitem[Ameijeiras-Alonso et~al.(2021)Ameijeiras-Alonso, Crujeiras, \& Rodriguez-Casal]{Ameijeiras2021}
Ameijeiras-Alonso, J., Crujeiras, R.~M., \& Rodriguez-Casal, A., 2021.
\newblock multimode: An r package for mode assessment, {\it Journal of Statistical Software\/}, {\bf 97}(9), 1–32.

\bibitem[{Andrae} et~al.(2010){Andrae}, {Schulze-Hartung}, \& {Melchior}]{Andrae2010}
{Andrae}, R., {Schulze-Hartung}, T., \& {Melchior}, P., 2010.
\newblock {Dos and don'ts of reduced chi-squared}, {\it arXiv e-prints\/}, p. arXiv:1012.3754.

\bibitem[{Astropy Collaboration} et~al.(2013){Astropy Collaboration}, {Robitaille}, {Tollerud}, {Greenfield}, {Droettboom}, {Bray}, {Aldcroft}, {Davis}, {Ginsburg}, {Price-Whelan}, {Kerzendorf}, {Conley}, {Crighton}, {Barbary}, {Muna}, {Ferguson}, {Grollier}, {Parikh}, {Nair}, {Unther}, {Deil}, {Woillez}, {Conseil}, {Kramer}, {Turner}, {Singer}, {Fox}, {Weaver}, {Zabalza}, {Edwards}, {Azalee Bostroem}, {Burke}, {Casey}, {Crawford}, {Dencheva}, {Ely}, {Jenness}, {Labrie}, {Lim}, {Pierfederici}, {Pontzen}, {Ptak}, {Refsdal}, {Servillat}, \& {Streicher}]{astropy:2013}
{Astropy Collaboration}, {Robitaille}, T.~P., {Tollerud}, E.~J., {Greenfield}, P., {Droettboom}, M., {Bray}, E., {Aldcroft}, T., {Davis}, M., {Ginsburg}, A., {Price-Whelan}, A.~M., {Kerzendorf}, W.~E., {Conley}, A., {Crighton}, N., {Barbary}, K., {Muna}, D., {Ferguson}, H., {Grollier}, F., {Parikh}, M.~M., {Nair}, P.~H., {Unther}, H.~M., {Deil}, C., {Woillez}, J., {Conseil}, S., {Kramer}, R., {Turner}, J.~E.~H., {Singer}, L., {Fox}, R., {Weaver}, B.~A., {Zabalza}, V., {Edwards}, Z.~I., {Azalee Bostroem}, K., {Burke}, D.~J., {Casey}, A.~R., {Crawford}, S.~M., {Dencheva}, N., {Ely}, J., {Jenness}, T., {Labrie}, K., {Lim}, P.~L., {Pierfederici}, F., {Pontzen}, A., {Ptak}, A., {Refsdal}, B., {Servillat}, M., \& {Streicher}, O., 2013.
\newblock {Astropy: A community Python package for astronomy}, {\it \aap\/}, {\bf 558}, A33.

\bibitem[{Astropy Collaboration} et~al.(2018){Astropy Collaboration}, {Price-Whelan}, {Sip{\H{o}}cz}, {G{\"u}nther}, {Lim}, {Crawford}, {Conseil}, {Shupe}, {Craig}, {Dencheva}, {Ginsburg}, {Vand erPlas}, {Bradley}, {P{\'e}rez-Su{\'a}rez}, {de Val-Borro}, {Aldcroft}, {Cruz}, {Robitaille}, {Tollerud}, {Ardelean}, {Babej}, {Bach}, {Bachetti}, {Bakanov}, {Bamford}, {Barentsen}, {Barmby}, {Baumbach}, {Berry}, {Biscani}, {Boquien}, {Bostroem}, {Bouma}, {Brammer}, {Bray}, {Breytenbach}, {Buddelmeijer}, {Burke}, {Calderone}, {Cano Rodr{\'\i}guez}, {Cara}, {Cardoso}, {Cheedella}, {Copin}, {Corrales}, {Crichton}, {D'Avella}, {Deil}, {Depagne}, {Dietrich}, {Donath}, {Droettboom}, {Earl}, {Erben}, {Fabbro}, {Ferreira}, {Finethy}, {Fox}, {Garrison}, {Gibbons}, {Goldstein}, {Gommers}, {Greco}, {Greenfield}, {Groener}, {Grollier}, {Hagen}, {Hirst}, {Homeier}, {Horton}, {Hosseinzadeh}, {Hu}, {Hunkeler}, {Ivezi{\'c}}, {Jain}, {Jenness}, {Kanarek}, {Kendrew}, {Kern}, {Kerzendorf}, {Khvalko}, {King}, {Kirkby}, {Kulkarni},
  {Kumar}, {Lee}, {Lenz}, {Littlefair}, {Ma}, {Macleod}, {Mastropietro}, {McCully}, {Montagnac}, {Morris}, {Mueller}, {Mumford}, {Muna}, {Murphy}, {Nelson}, {Nguyen}, {Ninan}, {N{\"o}the}, {Ogaz}, {Oh}, {Parejko}, {Parley}, {Pascual}, {Patil}, {Patil}, {Plunkett}, {Prochaska}, {Rastogi}, {Reddy Janga}, {Sabater}, {Sakurikar}, {Seifert}, {Sherbert}, {Sherwood-Taylor}, {Shih}, {Sick}, {Silbiger}, {Singanamalla}, {Singer}, {Sladen}, {Sooley}, {Sornarajah}, {Streicher}, {Teuben}, {Thomas}, {Tremblay}, {Turner}, {Terr{\'o}n}, {van Kerkwijk}, {de la Vega}, {Watkins}, {Weaver}, {Whitmore}, {Woillez}, {Zabalza}, \& {Astropy Contributors}]{astropy:2018}
{Astropy Collaboration}, {Price-Whelan}, A.~M., {Sip{\H{o}}cz}, B.~M., {G{\"u}nther}, H.~M., {Lim}, P.~L., {Crawford}, S.~M., {Conseil}, S., {Shupe}, D.~L., {Craig}, M.~W., {Dencheva}, N., {Ginsburg}, A., {Vand erPlas}, J.~T., {Bradley}, L.~D., {P{\'e}rez-Su{\'a}rez}, D., {de Val-Borro}, M., {Aldcroft}, T.~L., {Cruz}, K.~L., {Robitaille}, T.~P., {Tollerud}, E.~J., {Ardelean}, C., {Babej}, T., {Bach}, Y.~P., {Bachetti}, M., {Bakanov}, A.~V., {Bamford}, S.~P., {Barentsen}, G., {Barmby}, P., {Baumbach}, A., {Berry}, K.~L., {Biscani}, F., {Boquien}, M., {Bostroem}, K.~A., {Bouma}, L.~G., {Brammer}, G.~B., {Bray}, E.~M., {Breytenbach}, H., {Buddelmeijer}, H., {Burke}, D.~J., {Calderone}, G., {Cano Rodr{\'\i}guez}, J.~L., {Cara}, M., {Cardoso}, J.~V.~M., {Cheedella}, S., {Copin}, Y., {Corrales}, L., {Crichton}, D., {D'Avella}, D., {Deil}, C., {Depagne}, {\'E}., {Dietrich}, J.~P., {Donath}, A., {Droettboom}, M., {Earl}, N., {Erben}, T., {Fabbro}, S., {Ferreira}, L.~A., {Finethy}, T., {Fox}, R.~T., {Garrison}, L.~H.,
  {Gibbons}, S.~L.~J., {Goldstein}, D.~A., {Gommers}, R., {Greco}, J.~P., {Greenfield}, P., {Groener}, A.~M., {Grollier}, F., {Hagen}, A., {Hirst}, P., {Homeier}, D., {Horton}, A.~J., {Hosseinzadeh}, G., {Hu}, L., {Hunkeler}, J.~S., {Ivezi{\'c}}, {\v{Z}}., {Jain}, A., {Jenness}, T., {Kanarek}, G., {Kendrew}, S., {Kern}, N.~S., {Kerzendorf}, W.~E., {Khvalko}, A., {King}, J., {Kirkby}, D., {Kulkarni}, A.~M., {Kumar}, A., {Lee}, A., {Lenz}, D., {Littlefair}, S.~P., {Ma}, Z., {Macleod}, D.~M., {Mastropietro}, M., {McCully}, C., {Montagnac}, S., {Morris}, B.~M., {Mueller}, M., {Mumford}, S.~J., {Muna}, D., {Murphy}, N.~A., {Nelson}, S., {Nguyen}, G.~H., {Ninan}, J.~P., {N{\"o}the}, M., {Ogaz}, S., {Oh}, S., {Parejko}, J.~K., {Parley}, N., {Pascual}, S., {Patil}, R., {Patil}, A.~A., {Plunkett}, A.~L., {Prochaska}, J.~X., {Rastogi}, T., {Reddy Janga}, V., {Sabater}, J., {Sakurikar}, P., {Seifert}, M., {Sherbert}, L.~E., {Sherwood-Taylor}, H., {Shih}, A.~Y., {Sick}, J., {Silbiger}, M.~T., {Singanamalla}, S.,
  {Singer}, L.~P., {Sladen}, P.~H., {Sooley}, K.~A., {Sornarajah}, S., {Streicher}, O., {Teuben}, P., {Thomas}, S.~W., {Tremblay}, G.~R., {Turner}, J.~E.~H., {Terr{\'o}n}, V., {van Kerkwijk}, M.~H., {de la Vega}, A., {Watkins}, L.~L., {Weaver}, B.~A., {Whitmore}, J.~B., {Woillez}, J., {Zabalza}, V., \& {Astropy Contributors}, 2018.
\newblock {The Astropy Project: Building an Open-science Project and Status of the v2.0 Core Package}, {\it \aj\/}, {\bf 156}(3), 123.

\bibitem[{Baratella} et~al.(2020){Baratella}, {D'Orazi}, {Carraro}, {Desidera}, {Randich}, {Magrini}, {Adibekyan}, {Smiljanic}, {Spina}, {Tsantaki}, {Tautvai{\v{s}}ien{\.{e}}}, {Sousa}, {Jofr{\'e}}, {Jim{\'e}nez-Esteban}, {Delgado-Mena}, {Martell}, {Van der Swaelmen}, {Roccatagliata}, {Gilmore}, {Alfaro}, {Bayo}, {Bensby}, {Bragaglia}, {Franciosini}, {Gonneau}, {Heiter}, {Hourihane}, {Jeffries}, {Koposov}, {Morbidelli}, {Prisinzano}, {Sacco}, {Sbordone}, {Worley}, {Zaggia}, \& {Lewis}]{Baratella20}
{Baratella}, M., {D'Orazi}, V., {Carraro}, G., {Desidera}, S., {Randich}, S., {Magrini}, L., {Adibekyan}, V., {Smiljanic}, R., {Spina}, L., {Tsantaki}, M., {Tautvai{\v{s}}ien{\.{e}}}, G., {Sousa}, S.~G., {Jofr{\'e}}, P., {Jim{\'e}nez-Esteban}, F.~M., {Delgado-Mena}, E., {Martell}, S., {Van der Swaelmen}, M., {Roccatagliata}, V., {Gilmore}, G., {Alfaro}, E.~J., {Bayo}, A., {Bensby}, T., {Bragaglia}, A., {Franciosini}, E., {Gonneau}, A., {Heiter}, U., {Hourihane}, A., {Jeffries}, R.~D., {Koposov}, S.~E., {Morbidelli}, L., {Prisinzano}, L., {Sacco}, G., {Sbordone}, L., {Worley}, C., {Zaggia}, S., \& {Lewis}, J., 2020.
\newblock {The Gaia-ESO Survey: a new approach to chemically characterising young open clusters. I. Stellar parameters, and iron-peak, {\ensuremath{\alpha}}-, and proton-capture elements}, {\it \aap\/}, {\bf 634}, A34.

\bibitem[{Beaton} et~al.(2021){Beaton}, {Oelkers}, {Hayes}, {Covey}, {Chojnowski}, {De Lee}, {Sobeck}, {Majewski}, {Cohen}, {Fern{\'a}ndez-Trincado}, {Longa-Pe{\~n}a}, {O'Connell}, {Santana}, {Stringfellow}, {Zasowski}, {Aerts}, {Anguiano}, {Bender}, {Ca{\~n}as}, {Cunha}, {Donor}, {Fleming}, {Frinchaboy}, {Feuillet}, {Harding}, {Hasselquist}, {Holtzman}, {Johnson}, {Kollmeier}, {Kounkel}, {Mahadevan}, {Price-Whelan}, {Rojas-Arriagada}, {Rom{\'a}n-Z{\'u}{\~n}iga}, {Schlafly}, {Schultheis}, {Shetrone}, {Simon}, {Stassun}, {Stutz}, {Tayar}, {Teske}, {Tkachenko}, {Troup}, {Albareti}, {Bizyaev}, {Bovy}, {Burgasser}, {Comparat}, {Downes}, {Geisler}, {Inno}, {Manchado}, {Ness}, {Pinsonneault}, {Prada}, {Roman-Lopes}, {Simonian}, {Smith}, {Yan}, \& {Zamora}]{Beaton21}
{Beaton}, R.~L., {Oelkers}, R.~J., {Hayes}, C.~R., {Covey}, K.~R., {Chojnowski}, S.~D., {De Lee}, N., {Sobeck}, J.~S., {Majewski}, S.~R., {Cohen}, R.~E., {Fern{\'a}ndez-Trincado}, J., {Longa-Pe{\~n}a}, P., {O'Connell}, J.~E., {Santana}, F.~A., {Stringfellow}, G.~S., {Zasowski}, G., {Aerts}, C., {Anguiano}, B., {Bender}, C., {Ca{\~n}as}, C.~I., {Cunha}, K., {Donor}, J., {Fleming}, S.~W., {Frinchaboy}, P.~M., {Feuillet}, D., {Harding}, P., {Hasselquist}, S., {Holtzman}, J.~A., {Johnson}, J.~A., {Kollmeier}, J.~A., {Kounkel}, M., {Mahadevan}, S., {Price-Whelan}, A.~M., {Rojas-Arriagada}, A., {Rom{\'a}n-Z{\'u}{\~n}iga}, C., {Schlafly}, E.~F., {Schultheis}, M., {Shetrone}, M., {Simon}, J.~D., {Stassun}, K.~G., {Stutz}, A.~M., {Tayar}, J., {Teske}, J., {Tkachenko}, A., {Troup}, N., {Albareti}, F.~D., {Bizyaev}, D., {Bovy}, J., {Burgasser}, A.~J., {Comparat}, J., {Downes}, J.~J., {Geisler}, D., {Inno}, L., {Manchado}, A., {Ness}, M.~K., {Pinsonneault}, M.~H., {Prada}, F., {Roman-Lopes}, A., {Simonian}, G. V.~A.,
  {Smith}, V.~V., {Yan}, R., \& {Zamora}, O., 2021.
\newblock {Final Targeting Strategy for the Sloan Digital Sky Survey IV Apache Point Observatory Galactic Evolution Experiment 2 North Survey}, {\it \aj\/}, {\bf 162}(6), 302.

\bibitem[{Birky} et~al.(2020){Birky}, {Hogg}, {Mann}, \& {Burgasser}]{birky20}
{Birky}, J., {Hogg}, D.~W., {Mann}, A.~W., \& {Burgasser}, A., 2020.
\newblock {Temperatures and Metallicities of M Dwarfs in the APOGEE Survey}, {\it \apj\/}, {\bf 892}(1), 31.

\bibitem[{Bohlin} et~al.(2017){Bohlin}, {M{\'e}sz{\'a}ros}, {Fleming}, {Gordon}, {Koekemoer}, \& {Kov{\'a}cs}]{Bohlin17}
{Bohlin}, R.~C., {M{\'e}sz{\'a}ros}, S., {Fleming}, S.~W., {Gordon}, K.~D., {Koekemoer}, A.~M., \& {Kov{\'a}cs}, J., 2017.
\newblock {A New Stellar Atmosphere Grid and Comparisons with HST/STIS CALSPEC Flux Distributions}, {\it \aj\/}, {\bf 153}(5), 234.

\bibitem[{Breger}(1984)]{Breger84}
{Breger}, M., 1984.
\newblock {Spectral classification and photometry of selected Pleiades stars.}, {\it \aaps\/}, {\bf 57}, 217--218.

\bibitem[{Cannon} \& {Pickering}(1993)]{Cannon93}
{Cannon}, A.~J. \& {Pickering}, E.~C., 1993.
\newblock {VizieR Online Data Catalog: Henry Draper Catalogue and Extension (Cannon+ 1918-1924; ADC 1989)}, {\it VizieR Online Data Catalog\/}, p. III/135A.

\bibitem[{Cant{\'o}} et~al.(2009){Cant{\'o}}, {Curiel}, \& {Mart{\'\i}nez-G{\'o}mez}]{Canto09}
{Cant{\'o}}, J., {Curiel}, S., \& {Mart{\'\i}nez-G{\'o}mez}, E., 2009.
\newblock {A simple algorithm for optimization and model fitting: AGA (asexual genetic algorithm)}, {\it \aap\/}, {\bf 501}(3), 1259--1268.

\bibitem[{Carbajo-Hijarrubia} et~al.(2024){Carbajo-Hijarrubia}, {Casamiquela}, {Carrera}, {Balaguer-N{\'u}{\~n}ez}, {Jordi}, {Anders}, {Gallart}, {Pancino}, {Drazdauskas}, {Stonkut{\.{e}}}, {Tautvai{\v{s}}ien{\.{e}}}, {Carrasco}, {Masana}, {Cantat-Gaudin}, \& {Blanco-Cuaresma}]{Carbajo24}
{Carbajo-Hijarrubia}, J., {Casamiquela}, L., {Carrera}, R., {Balaguer-N{\'u}{\~n}ez}, L., {Jordi}, C., {Anders}, F., {Gallart}, C., {Pancino}, E., {Drazdauskas}, A., {Stonkut{\.{e}}}, E., {Tautvai{\v{s}}ien{\.{e}}}, G., {Carrasco}, J.~M., {Masana}, E., {Cantat-Gaudin}, T., \& {Blanco-Cuaresma}, S., 2024.
\newblock {OCCASO. V. Chemical-abundance trends with Galactocentric distance and age}, {\it \aap\/}, {\bf 687}, A239.

\bibitem[{Carrera} et~al.(2019){Carrera}, {Bragaglia}, {Cantat-Gaudin}, {Vallenari}, {Balaguer-N{\'u}{\~n}ez}, {Bossini}, {Casamiquela}, {Jordi}, {Sordo}, \& {Soubiran}]{Carrera19}
{Carrera}, R., {Bragaglia}, A., {Cantat-Gaudin}, T., {Vallenari}, A., {Balaguer-N{\'u}{\~n}ez}, L., {Bossini}, D., {Casamiquela}, L., {Jordi}, C., {Sordo}, R., \& {Soubiran}, C., 2019.
\newblock {Open clusters in APOGEE and GALAH. Combining Gaia and ground-based spectroscopic surveys}, {\it \aap\/}, {\bf 623}, A80.

\bibitem[{Castelli} \& {Kurucz}(2003)]{Castelli03}
{Castelli}, F. \& {Kurucz}, R.~L., 2003.
\newblock {New Grids of ATLAS9 Model Atmospheres}, in {\em Modelling of Stellar Atmospheres\/}, vol. 210, p. A20.

\bibitem[{Coelho} et~al.(2005){Coelho}, {Barbuy}, {Mel{\'e}ndez}, {Schiavon}, \& {Castilho}]{Coelho05}
{Coelho}, P., {Barbuy}, B., {Mel{\'e}ndez}, J., {Schiavon}, R.~P., \& {Castilho}, B.~V., 2005.
\newblock {A library of high resolution synthetic stellar spectra from 300 nm to 1.8 {\ensuremath{\mu}}m with solar and {\ensuremath{\alpha}}-enhanced composition}, {\it \aap\/}, {\bf 443}(2), 735--746.

\bibitem[{Cottaar} et~al.(2014){Cottaar}, {Covey}, {Meyer}, {Nidever}, {Stassun}, {Foster}, {Tan}, {Chojnowski}, {da Rio}, {Flaherty}, {Frinchaboy}, {Skrutskie}, {Majewski}, {Wilson}, \& {Zasowski}]{Cottaar14}
{Cottaar}, M., {Covey}, K.~R., {Meyer}, M.~R., {Nidever}, D.~L., {Stassun}, K.~G., {Foster}, J.~B., {Tan}, J.~C., {Chojnowski}, S.~D., {da Rio}, N., {Flaherty}, K.~M., {Frinchaboy}, P.~M., {Skrutskie}, M., {Majewski}, S.~R., {Wilson}, J.~C., \& {Zasowski}, G., 2014.
\newblock {IN-SYNC I: Homogeneous Stellar Parameters from High-resolution APOGEE Spectra for Thousands of Pre-main Sequence Stars}, {\it \apj\/}, {\bf 794}(2), 125.

\bibitem[{Covey} et~al.(2010){Covey}, {Lada}, {Rom{\'a}n-Z{\'u}{\~n}iga}, {Muench}, {Forbrich}, \& {Ascenso}]{Covey10}
{Covey}, K.~R., {Lada}, C.~J., {Rom{\'a}n-Z{\'u}{\~n}iga}, C., {Muench}, A.~A., {Forbrich}, J., \& {Ascenso}, J., 2010.
\newblock {The Age, Stellar Content, and Star Formation Timescale of the B59 Dense Core}, {\it \apj\/}, {\bf 722}(2), 971--988.

\bibitem[{Cui} et~al.(2012){Cui}, {Zhao}, {Chu}, {Li}, {Li}, {Zhang}, {Su}, {Yao}, {Wang}, {Xing}, {Li}, {Zhu}, {Wang}, {Gu}, {Luo}, {Xu}, {Zhang}, {Liu}, {Zhang}, {Yang}, {Cao}, {Chen}, {Chen}, {Chen}, {Chen}, {Chu}, {Feng}, {Gong}, {Hou}, {Hu}, {Hu}, {Hu}, {Jia}, {Jiang}, {Jiang}, {Jiang}, {Jin}, {Li}, {Li}, {Li}, {Liu}, {Liu}, {Lu}, {Mao}, {Men}, {Qi}, {Qi}, {Shi}, {Tang}, {Tao}, {Wang}, {Wang}, {Wang}, {Wang}, {Wang}, {Wang}, {Wang}, {Wang}, {Wang}, {Wang}, {Wang}, {Wang}, {Xu}, {Xu}, {Yang}, {Yu}, {Yuan}, {Yuan}, {Zhai}, {Zhang}, {Zhang}, {Zhang}, {Zhao}, {Zhou}, {Zhou}, {Zhu}, \& {Zou}]{Lamost12}
{Cui}, X.-Q., {Zhao}, Y.-H., {Chu}, Y.-Q., {Li}, G.-P., {Li}, Q., {Zhang}, L.-P., {Su}, H.-J., {Yao}, Z.-Q., {Wang}, Y.-N., {Xing}, X.-Z., {Li}, X.-N., {Zhu}, Y.-T., {Wang}, G., {Gu}, B.-Z., {Luo}, A.~L., {Xu}, X.-Q., {Zhang}, Z.-C., {Liu}, G.-R., {Zhang}, H.-T., {Yang}, D.-H., {Cao}, S.-Y., {Chen}, H.-Y., {Chen}, J.-J., {Chen}, K.-X., {Chen}, Y., {Chu}, J.-R., {Feng}, L., {Gong}, X.-F., {Hou}, Y.-H., {Hu}, H.-Z., {Hu}, N.-S., {Hu}, Z.-W., {Jia}, L., {Jiang}, F.-H., {Jiang}, X., {Jiang}, Z.-B., {Jin}, G., {Li}, A.-H., {Li}, Y., {Li}, Y.-P., {Liu}, G.-Q., {Liu}, Z.-G., {Lu}, W.-Z., {Mao}, Y.-D., {Men}, L., {Qi}, Y.-J., {Qi}, Z.-X., {Shi}, H.-M., {Tang}, Z.-H., {Tao}, Q.-S., {Wang}, D.-Q., {Wang}, D., {Wang}, G.-M., {Wang}, H., {Wang}, J.-N., {Wang}, J., {Wang}, J.-L., {Wang}, J.-P., {Wang}, L., {Wang}, S.-Q., {Wang}, Y., {Wang}, Y.-F., {Xu}, L.-Z., {Xu}, Y., {Yang}, S.-H., {Yu}, Y., {Yuan}, H., {Yuan}, X.-Y., {Zhai}, C., {Zhang}, J., {Zhang}, Y.-X., {Zhang}, Y., {Zhao}, M., {Zhou}, F., {Zhou}, G.-H., {Zhu},
  J., \& {Zou}, S.-C., 2012.
\newblock {The Large Sky Area Multi-Object Fiber Spectroscopic Telescope (LAMOST)}, {\it Research in Astronomy and Astrophysics\/}, {\bf 12}(9), 1197--1242.

\bibitem[{Czesla} et~al.(2019){Czesla}, {Schr{\"o}ter}, {Schneider}, {Huber}, {Pfeifer}, {Andreasen}, \& {Zechmeister}]{pya}
{Czesla}, S., {Schr{\"o}ter}, S., {Schneider}, C.~P., {Huber}, K.~F., {Pfeifer}, F., {Andreasen}, D.~T., \& {Zechmeister}, M., 2019.
\newblock {PyA: Python astronomy-related packages}.

\bibitem[{Da Rio} et~al.(2016){Da Rio}, {Tan}, {Covey}, {Cottaar}, {Foster}, {Cullen}, {Tobin}, {Kim}, {Meyer}, {Nidever}, {Stassun}, {Chojnowski}, {Flaherty}, {Majewski}, {Skrutskie}, {Zasowski}, \& {Pan}]{dario16}
{Da Rio}, N., {Tan}, J.~C., {Covey}, K.~R., {Cottaar}, M., {Foster}, J.~B., {Cullen}, N.~C., {Tobin}, J.~J., {Kim}, J.~S., {Meyer}, M.~R., {Nidever}, D.~L., {Stassun}, K.~G., {Chojnowski}, S.~D., {Flaherty}, K.~M., {Majewski}, S., {Skrutskie}, M.~F., {Zasowski}, G., \& {Pan}, K., 2016.
\newblock {IN-SYNC. IV. The Young Stellar Population in the Orion A Molecular Cloud}, {\it \apj\/}, {\bf 818}(1), 59.

\bibitem[{De Silva} et~al.(2015){De Silva}, {Freeman}, {Bland-Hawthorn}, {Martell}, {de Boer}, {Asplund}, {Keller}, {Sharma}, {Zucker}, {Zwitter}, {Anguiano}, {Bacigalupo}, {Bayliss}, {Beavis}, {Bergemann}, {Campbell}, {Cannon}, {Carollo}, {Casagrande}, {Casey}, {Da Costa}, {D'Orazi}, {Dotter}, {Duong}, {Heger}, {Ireland}, {Kafle}, {Kos}, {Lattanzio}, {Lewis}, {Lin}, {Lind}, {Munari}, {Nataf}, {O'Toole}, {Parker}, {Reid}, {Schlesinger}, {Sheinis}, {Simpson}, {Stello}, {Ting}, {Traven}, {Watson}, {Wittenmyer}, {Yong}, \& {{\v{Z}}erjal}]{galah}
{De Silva}, G.~M., {Freeman}, K.~C., {Bland-Hawthorn}, J., {Martell}, S., {de Boer}, E.~W., {Asplund}, M., {Keller}, S., {Sharma}, S., {Zucker}, D.~B., {Zwitter}, T., {Anguiano}, B., {Bacigalupo}, C., {Bayliss}, D., {Beavis}, M.~A., {Bergemann}, M., {Campbell}, S., {Cannon}, R., {Carollo}, D., {Casagrande}, L., {Casey}, A.~R., {Da Costa}, G., {D'Orazi}, V., {Dotter}, A., {Duong}, L., {Heger}, A., {Ireland}, M.~J., {Kafle}, P.~R., {Kos}, J., {Lattanzio}, J., {Lewis}, G.~F., {Lin}, J., {Lind}, K., {Munari}, U., {Nataf}, D.~M., {O'Toole}, S., {Parker}, Q., {Reid}, W., {Schlesinger}, K.~J., {Sheinis}, A., {Simpson}, J.~D., {Stello}, D., {Ting}, Y.~S., {Traven}, G., {Watson}, F., {Wittenmyer}, R., {Yong}, D., \& {{\v{Z}}erjal}, M., 2015.
\newblock {The GALAH survey: scientific motivation}, {\it \mnras\/}, {\bf 449}(3), 2604--2617.

\bibitem[{Fehrenbach}(1966)]{Fehrenbach66}
{Fehrenbach}, C., 1966.
\newblock {La mesure des vitesses radiales au prisme objectif XIX,}, {\it Publications of the Observatoire Haute-Provence\/}, {\bf 8}(25), 155--214.

\bibitem[Foreman-Mackey(2016)]{corner}
Foreman-Mackey, D., 2016.
\newblock corner.py: Scatterplot matrices in python, {\it The Journal of Open Source Software\/}, {\bf 1}(2), 24.

\bibitem[Fraser(1957)]{Fraser57}
Fraser, A., 1957.
\newblock Simulation of genetic systems by automatic digital computers i. introduction, {\it Australian Journal of Biological Sciences\/}, {\bf 10}(4), 484--491, http://www.publish.csiro.au/bi/pdf/bi9570484.

\bibitem[{Gaia Collaboration} et~al.(2023){Gaia Collaboration}, {Recio-Blanco}, {Kordopatis}, {de Laverny}, {Palicio}, {Spagna}, {Spina}, {Katz}, {Re Fiorentin}, {Poggio}, {McMillan}, {Vallenari}, {Lattanzi}, {Seabroke}, {Casamiquela}, {Bragaglia}, {Antoja}, {Bailer-Jones}, {Schultheis}, {Andrae}, {Fouesneau}, {Cropper}, {Cantat-Gaudin}, {Bijaoui}, {Heiter}, {Brown}, {Prusti}, {de Bruijne}, {Arenou}, {Babusiaux}, {Biermann}, {Creevey}, {Ducourant}, {Evans}, {Eyer}, {Guerra}, {Hutton}, {Jordi}, {Klioner}, {Lammers}, {Lindegren}, {Luri}, {Mignard}, {Panem}, {Pourbaix}, {Randich}, {Sartoretti}, {Soubiran}, {Tanga}, {Walton}, {Bastian}, {Drimmel}, {Jansen}, {van Leeuwen}, {Bakker}, {Cacciari}, {Casta{\~n}eda}, {De Angeli}, {Fabricius}, {Fr{\'e}mat}, {Galluccio}, {Guerrier}, {Masana}, {Messineo}, {Mowlavi}, {Nicolas}, {Nienartowicz}, {Pailler}, {Panuzzo}, {Riclet}, {Roux}, {Sordo}, {Th{\'e}venin}, {Gracia-Abril}, {Portell}, {Teyssier}, {Altmann}, {Audard}, {Bellas-Velidis}, {Benson}, {Berthier}, {Blomme},
  {Burgess}, {Busonero}, {Busso}, {C{\'a}novas}, {Carry}, {Cellino}, {Cheek}, {Clementini}, {Damerdji}, {Davidson}, {de Teodoro}, {Nu{\~n}ez Campos}, {Delchambre}, {Dell'Oro}, {Esquej}, {Fern{\'a}ndez-Hern{\'a}ndez}, {Fraile}, {Garabato}, {Garc{\'\i}a-Lario}, {Gosset}, {Haigron}, {Halbwachs}, {Hambly}, {Harrison}, {Hern{\'a}ndez}, {Hestroffer}, {Hodgkin}, {Holl}, {Jan{\ss}en}, {Jevardat de Fombelle}, {Jordan}, {Krone-Martins}, {Lanzafame}, {L{\"o}ffler}, {Marchal}, {Marrese}, {Moitinho}, {Muinonen}, {Osborne}, {Pancino}, {Pauwels}, {Reyl{\'e}}, {Riello}, {Rimoldini}, {Roegiers}, {Rybizki}, {Sarro}, {Siopis}, {Smith}, {Sozzetti}, {Utrilla}, {van Leeuwen}, {Abbas}, {{\'A}brah{\'a}m}, {Abreu Aramburu}, {Aerts}, {Aguado}, {Ajaj}, {Aldea-Montero}, {Altavilla}, {{\'A}lvarez}, {Alves}, {Anders}, {Anderson}, {Anglada Varela}, {Baines}, {Baker}, {Balaguer-N{\'u}{\~n}ez}, {Balbinot}, {Balog}, {Barache}, {Barbato}, {Barros}, {Barstow}, {Bartolom{\'e}}, {Bassilana}, {Bauchet}, {Becciani}, {Bellazzini}, {Berihuete},
  {Bernet}, {Bertone}, {Bianchi}, {Binnenfeld}, {Blanco-Cuaresma}, {Boch}, {Bombrun}, {Bossini}, {Bouquillon}, {Bramante}, {Breedt}, {Bressan}, {Brouillet}, {Brugaletta}, {Bucciarelli}, {Burlacu}, {Butkevich}, {Buzzi}, {Caffau}, {Cancelliere}, {Carballo}, {Carlucci}, {Carnerero}, {Carrasco}, {Castellani}, {Castro-Ginard}, {Chaoul}, {Charlot}, {Chemin}, {Chiaramida}, {Chiavassa}, {Chornay}, {Comoretto}, {Contursi}, {Cooper}, {Cornez}, {Cowell}, {Crifo}, {Crosta}, {Crowley}, {Dafonte}, {Dapergolas}, {David}, {De Luise}, {De March}, {De Ridder}, {de Souza}, {de Torres}, {del Peloso}, {del Pozo}, {Delbo}, {Delgado}, {Delisle}, {Demouchy}, {Dharmawardena}, {Di Matteo}, {Diakite}, {Diener}, {Distefano}, {Dolding}, {Edvardsson}, {Enke}, {Fabre}, {Fabrizio}, {Faigler}, {Fedorets}, {Fernique}, {Figueras}, {Fournier}, {Fouron}, {Fragkoudi}, {Gai}, {Garcia-Gutierrez}, {Garcia-Reinaldos}, {Garc{\'\i}a-Torres}, {Garofalo}, {Gavel}, {Gavras}, {Gerlach}, {Geyer}, {Giacobbe}, {Gilmore}, {Girona}, {Giuffrida}, {Gomel},
  {Gomez}, {Gonz{\'a}lez-N{\'u}{\~n}ez}, {Gonz{\'a}lez-Santamar{\'\i}a}, {Gonz{\'a}lez-Vidal}, {Granvik}, {Guillout}, {Guiraud}, {Guti{\'e}rrez-S{\'a}nchez}, {Guy}, {Hatzidimitriou}, {Hauser}, {Haywood}, {Helmer}, {Helmi}, {Sarmiento}, {Hidalgo}, {H{\l}adczuk}, {Hobbs}, {Holland}, {Huckle}, {Jardine}, {Jasniewicz}, {Jean-Antoine Piccolo}, {Jim{\'e}nez-Arranz}, {Juaristi Campillo}, {Julbe}, {Karbevska}, {Kervella}, {Khanna}, {Korn}, {K{\'o}sp{\'a}l}, {Kostrzewa-Rutkowska}, {Kruszy{\'n}ska}, {Kun}, {Laizeau}, {Lambert}, {Lanza}, {Lasne}, {Le Campion}, {Lebreton}, {Lebzelter}, {Leccia}, {Leclerc}, {Lecoeur-Taibi}, {Liao}, {Licata}, {Lindstr{\o}m}, {Lister}, {Livanou}, {Lobel}, {Lorca}, {Loup}, {Madrero Pardo}, {Magdaleno Romeo}, {Managau}, {Mann}, {Manteiga}, {Marchant}, {Marconi}, {Marcos}, {Marcos Santos}, {Mar{\'\i}n Pina}, {Marinoni}, {Marocco}, {Marshall}, {Martin Polo}, {Mart{\'\i}n-Fleitas}, {Marton}, {Mary}, {Masip}, {Massari}, {Mastrobuono-Battisti}, {Mazeh}, {Messina}, {Michalik}, {Millar}, {Mints},
  {Molina}, {Molinaro}, {Moln{\'a}r}, {Monari}, {Mongui{\'o}}, {Montegriffo}, {Montero}, {Mor}, {Mora}, {Morbidelli}, {Morel}, {Morris}, {Muraveva}, {Murphy}, {Musella}, {Nagy}, {Noval}, {Oca{\~n}a}, {Ogden}, {Ordenovic}, {Osinde}, {Pagani}, {Pagano}, {Palaversa}, {Pallas-Quintela}, {Panahi}, {Payne-Wardenaar}, {Pe{\~n}alosa Esteller}, {Penttil{\"a}}, {Pichon}, {Piersimoni}, {Pineau}, {Plachy}, {Plum}, {Pr{\v{s}}a}, {Pulone}, {Racero}, {Ragaini}, {Rainer}, {Raiteri}, {Ramos}, {Ramos-Lerate}, {Regibo}, {Richards}, {Rios Diaz}, {Ripepi}, {Riva}, {Rix}, {Rixon}, {Robichon}, {Robin}, {Robin}, {Roelens}, {Rogues}, {Rohrbasser}, {Romero-G{\'o}mez}, {Rowell}, {Royer}, {Ruz Mieres}, {Rybicki}, {Sadowski}, {S{\'a}ez N{\'u}{\~n}ez}, {Sagrist{\`a} Sell{\'e}s}, {Sahlmann}, {Salguero}, {Samaras}, {Sanchez Gimenez}, {Sanna}, {Santove{\~n}a}, {Sarasso}, {Sciacca}, {Segol}, {Segovia}, {S{\'e}gransan}, {Semeux}, {Shahaf}, {Siddiqui}, {Siebert}, {Siltala}, {Silvelo}, {Slezak}, {Slezak}, {Smart}, {Snaith}, {Solano}, {Solitro},
  {Souami}, {Souchay}, {Spoto}, {Steele}, {Steidelm{\"u}ller}, {Stephenson}, {S{\"u}veges}, {Surdej}, {Szabados}, {Szegedi-Elek}, {Taris}, {Taylor}, {Teixeira}, {Tolomei}, {Tonello}, {Torra}, {Torra}, {Torralba Elipe}, {Trabucchi}, {Tsounis}, {Turon}, {Ulla}, {Unger}, {Vaillant}, {van Dillen}, {van Reeven}, {Vanel}, {Vecchiato}, {Viala}, {Vicente}, {Voutsinas}, {Weiler}, {Wevers}, {Wyrzykowski}, {Yoldas}, {Yvard}, {Zhao}, {Zorec}, {Zucker}, \& {Zwitter}]{Recio23}
{Gaia Collaboration}, {Recio-Blanco}, A., {Kordopatis}, G., {de Laverny}, P., {Palicio}, P.~A., {Spagna}, A., {Spina}, L., {Katz}, D., {Re Fiorentin}, P., {Poggio}, E., {McMillan}, P.~J., {Vallenari}, A., {Lattanzi}, M.~G., {Seabroke}, G.~M., {Casamiquela}, L., {Bragaglia}, A., {Antoja}, T., {Bailer-Jones}, C.~A.~L., {Schultheis}, M., {Andrae}, R., {Fouesneau}, M., {Cropper}, M., {Cantat-Gaudin}, T., {Bijaoui}, A., {Heiter}, U., {Brown}, A.~G.~A., {Prusti}, T., {de Bruijne}, J.~H.~J., {Arenou}, F., {Babusiaux}, C., {Biermann}, M., {Creevey}, O.~L., {Ducourant}, C., {Evans}, D.~W., {Eyer}, L., {Guerra}, R., {Hutton}, A., {Jordi}, C., {Klioner}, S.~A., {Lammers}, U.~L., {Lindegren}, L., {Luri}, X., {Mignard}, F., {Panem}, C., {Pourbaix}, D., {Randich}, S., {Sartoretti}, P., {Soubiran}, C., {Tanga}, P., {Walton}, N.~A., {Bastian}, U., {Drimmel}, R., {Jansen}, F., {van Leeuwen}, F., {Bakker}, J., {Cacciari}, C., {Casta{\~n}eda}, J., {De Angeli}, F., {Fabricius}, C., {Fr{\'e}mat}, Y., {Galluccio}, L., {Guerrier},
  A., {Masana}, E., {Messineo}, R., {Mowlavi}, N., {Nicolas}, C., {Nienartowicz}, K., {Pailler}, F., {Panuzzo}, P., {Riclet}, F., {Roux}, W., {Sordo}, R., {Th{\'e}venin}, F., {Gracia-Abril}, G., {Portell}, J., {Teyssier}, D., {Altmann}, M., {Audard}, M., {Bellas-Velidis}, I., {Benson}, K., {Berthier}, J., {Blomme}, R., {Burgess}, P.~W., {Busonero}, D., {Busso}, G., {C{\'a}novas}, H., {Carry}, B., {Cellino}, A., {Cheek}, N., {Clementini}, G., {Damerdji}, Y., {Davidson}, M., {de Teodoro}, P., {Nu{\~n}ez Campos}, M., {Delchambre}, L., {Dell'Oro}, A., {Esquej}, P., {Fern{\'a}ndez-Hern{\'a}ndez}, J., {Fraile}, E., {Garabato}, D., {Garc{\'\i}a-Lario}, P., {Gosset}, E., {Haigron}, R., {Halbwachs}, J.~L., {Hambly}, N.~C., {Harrison}, D.~L., {Hern{\'a}ndez}, J., {Hestroffer}, D., {Hodgkin}, S.~T., {Holl}, B., {Jan{\ss}en}, K., {Jevardat de Fombelle}, G., {Jordan}, S., {Krone-Martins}, A., {Lanzafame}, A.~C., {L{\"o}ffler}, W., {Marchal}, O., {Marrese}, P.~M., {Moitinho}, A., {Muinonen}, K., {Osborne}, P., {Pancino},
  E., {Pauwels}, T., {Reyl{\'e}}, C., {Riello}, M., {Rimoldini}, L., {Roegiers}, T., {Rybizki}, J., {Sarro}, L.~M., {Siopis}, C., {Smith}, M., {Sozzetti}, A., {Utrilla}, E., {van Leeuwen}, M., {Abbas}, U., {{\'A}brah{\'a}m}, P., {Abreu Aramburu}, A., {Aerts}, C., {Aguado}, J.~J., {Ajaj}, M., {Aldea-Montero}, F., {Altavilla}, G., {{\'A}lvarez}, M.~A., {Alves}, J., {Anders}, F., {Anderson}, R.~I., {Anglada Varela}, E., {Baines}, D., {Baker}, S.~G., {Balaguer-N{\'u}{\~n}ez}, L., {Balbinot}, E., {Balog}, Z., {Barache}, C., {Barbato}, D., {Barros}, M., {Barstow}, M.~A., {Bartolom{\'e}}, S., {Bassilana}, J.~L., {Bauchet}, N., {Becciani}, U., {Bellazzini}, M., {Berihuete}, A., {Bernet}, M., {Bertone}, S., {Bianchi}, L., {Binnenfeld}, A., {Blanco-Cuaresma}, S., {Boch}, T., {Bombrun}, A., {Bossini}, D., {Bouquillon}, S., {Bramante}, L., {Breedt}, E., {Bressan}, A., {Brouillet}, N., {Brugaletta}, E., {Bucciarelli}, B., {Burlacu}, A., {Butkevich}, A.~G., {Buzzi}, R., {Caffau}, E., {Cancelliere}, R., {Carballo}, R.,
  {Carlucci}, T., {Carnerero}, M.~I., {Carrasco}, J.~M., {Castellani}, M., {Castro-Ginard}, A., {Chaoul}, L., {Charlot}, P., {Chemin}, L., {Chiaramida}, V., {Chiavassa}, A., {Chornay}, N., {Comoretto}, G., {Contursi}, G., {Cooper}, W.~J., {Cornez}, T., {Cowell}, S., {Crifo}, F., {Crosta}, M., {Crowley}, C., {Dafonte}, C., {Dapergolas}, A., {David}, P., {De Luise}, F., {De March}, R., {De Ridder}, J., {de Souza}, R., {de Torres}, A., {del Peloso}, E.~F., {del Pozo}, E., {Delbo}, M., {Delgado}, A., {Delisle}, J.~B., {Demouchy}, C., {Dharmawardena}, T.~E., {Di Matteo}, P., {Diakite}, S., {Diener}, C., {Distefano}, E., {Dolding}, C., {Edvardsson}, B., {Enke}, H., {Fabre}, C., {Fabrizio}, M., {Faigler}, S., {Fedorets}, G., {Fernique}, P., {Figueras}, F., {Fournier}, Y., {Fouron}, C., {Fragkoudi}, F., {Gai}, M., {Garcia-Gutierrez}, A., {Garcia-Reinaldos}, M., {Garc{\'\i}a-Torres}, M., {Garofalo}, A., {Gavel}, A., {Gavras}, P., {Gerlach}, E., {Geyer}, R., {Giacobbe}, P., {Gilmore}, G., {Girona}, S., {Giuffrida}, G.,
  {Gomel}, R., {Gomez}, A., {Gonz{\'a}lez-N{\'u}{\~n}ez}, J., {Gonz{\'a}lez-Santamar{\'\i}a}, I., {Gonz{\'a}lez-Vidal}, J.~J., {Granvik}, M., {Guillout}, P., {Guiraud}, J., {Guti{\'e}rrez-S{\'a}nchez}, R., {Guy}, L.~P., {Hatzidimitriou}, D., {Hauser}, M., {Haywood}, M., {Helmer}, A., {Helmi}, A., {Sarmiento}, M.~H., {Hidalgo}, S.~L., {H{\l}adczuk}, N., {Hobbs}, D., {Holland}, G., {Huckle}, H.~E., {Jardine}, K., {Jasniewicz}, G., {Jean-Antoine Piccolo}, A., {Jim{\'e}nez-Arranz}, {\'O}., {Juaristi Campillo}, J., {Julbe}, F., {Karbevska}, L., {Kervella}, P., {Khanna}, S., {Korn}, A.~J., {K{\'o}sp{\'a}l}, {\'A}., {Kostrzewa-Rutkowska}, Z., {Kruszy{\'n}ska}, K., {Kun}, M., {Laizeau}, P., {Lambert}, S., {Lanza}, A.~F., {Lasne}, Y., {Le Campion}, J.~F., {Lebreton}, Y., {Lebzelter}, T., {Leccia}, S., {Leclerc}, N., {Lecoeur-Taibi}, I., {Liao}, S., {Licata}, E.~L., {Lindstr{\o}m}, H.~E.~P., {Lister}, T.~A., {Livanou}, E., {Lobel}, A., {Lorca}, A., {Loup}, C., {Madrero Pardo}, P., {Magdaleno Romeo}, A., {Managau}, S.,
  {Mann}, R.~G., {Manteiga}, M., {Marchant}, J.~M., {Marconi}, M., {Marcos}, J., {Marcos Santos}, M.~M.~S., {Mar{\'\i}n Pina}, D., {Marinoni}, S., {Marocco}, F., {Marshall}, D.~J., {Martin Polo}, L., {Mart{\'\i}n-Fleitas}, J.~M., {Marton}, G., {Mary}, N., {Masip}, A., {Massari}, D., {Mastrobuono-Battisti}, A., {Mazeh}, T., {Messina}, S., {Michalik}, D., {Millar}, N.~R., {Mints}, A., {Molina}, D., {Molinaro}, R., {Moln{\'a}r}, L., {Monari}, G., {Mongui{\'o}}, M., {Montegriffo}, P., {Montero}, A., {Mor}, R., {Mora}, A., {Morbidelli}, R., {Morel}, T., {Morris}, D., {Muraveva}, T., {Murphy}, C.~P., {Musella}, I., {Nagy}, Z., {Noval}, L., {Oca{\~n}a}, F., {Ogden}, A., {Ordenovic}, C., {Osinde}, J.~O., {Pagani}, C., {Pagano}, I., {Palaversa}, L., {Pallas-Quintela}, L., {Panahi}, A., {Payne-Wardenaar}, S., {Pe{\~n}alosa Esteller}, X., {Penttil{\"a}}, A., {Pichon}, B., {Piersimoni}, A.~M., {Pineau}, F.~X., {Plachy}, E., {Plum}, G., {Pr{\v{s}}a}, A., {Pulone}, L., {Racero}, E., {Ragaini}, S., {Rainer}, M., {Raiteri},
  C.~M., {Ramos}, P., {Ramos-Lerate}, M., {Regibo}, S., {Richards}, P.~J., {Rios Diaz}, C., {Ripepi}, V., {Riva}, A., {Rix}, H.~W., {Rixon}, G., {Robichon}, N., {Robin}, A.~C., {Robin}, C., {Roelens}, M., {Rogues}, H.~R.~O., {Rohrbasser}, L., {Romero-G{\'o}mez}, M., {Rowell}, N., {Royer}, F., {Ruz Mieres}, D., {Rybicki}, K.~A., {Sadowski}, G., {S{\'a}ez N{\'u}{\~n}ez}, A., {Sagrist{\`a} Sell{\'e}s}, A., {Sahlmann}, J., {Salguero}, E., {Samaras}, N., {Sanchez Gimenez}, V., {Sanna}, N., {Santove{\~n}a}, R., {Sarasso}, M., {Sciacca}, E., {Segol}, M., {Segovia}, J.~C., {S{\'e}gransan}, D., {Semeux}, D., {Shahaf}, S., {Siddiqui}, H.~I., {Siebert}, A., {Siltala}, L., {Silvelo}, A., {Slezak}, E., {Slezak}, I., {Smart}, R.~L., {Snaith}, O.~N., {Solano}, E., {Solitro}, F., {Souami}, D., {Souchay}, J., {Spoto}, F., {Steele}, I.~A., {Steidelm{\"u}ller}, H., {Stephenson}, C.~A., {S{\"u}veges}, M., {Surdej}, J., {Szabados}, L., {Szegedi-Elek}, E., {Taris}, F., {Taylor}, M.~B., {Teixeira}, R., {Tolomei}, L., {Tonello}, N.,
  {Torra}, F., {Torra}, J., {Torralba Elipe}, G., {Trabucchi}, M., {Tsounis}, A.~T., {Turon}, C., {Ulla}, A., {Unger}, N., {Vaillant}, M.~V., {van Dillen}, E., {van Reeven}, W., {Vanel}, O., {Vecchiato}, A., {Viala}, Y., {Vicente}, D., {Voutsinas}, S., {Weiler}, M., {Wevers}, T., {Wyrzykowski}, {\L}., {Yoldas}, A., {Yvard}, P., {Zhao}, H., {Zorec}, J., {Zucker}, S., \& {Zwitter}, T., 2023.
\newblock {Gaia Data Release 3. Chemical cartography of the Milky Way}, {\it \aap\/}, {\bf 674}, A38.

\bibitem[{Garc{\'\i}a P{\'e}rez} et~al.(2016){Garc{\'\i}a P{\'e}rez}, {Allende Prieto}, {Holtzman}, {Shetrone}, {M{\'e}sz{\'a}ros}, {Bizyaev}, {Carrera}, {Cunha}, {Garc{\'\i}a-Hern{\'a}ndez}, {Johnson}, {Majewski}, {Nidever}, {Schiavon}, {Shane}, {Smith}, {Sobeck}, {Troup}, {Zamora}, {Weinberg}, {Bovy}, {Eisenstein}, {Feuillet}, {Frinchaboy}, {Hayden}, {Hearty}, {Nguyen}, {O'Connell}, {Pinsonneault}, {Wilson}, \& {Zasowski}]{aspcap}
{Garc{\'\i}a P{\'e}rez}, A.~E., {Allende Prieto}, C., {Holtzman}, J.~A., {Shetrone}, M., {M{\'e}sz{\'a}ros}, S., {Bizyaev}, D., {Carrera}, R., {Cunha}, K., {Garc{\'\i}a-Hern{\'a}ndez}, D.~A., {Johnson}, J.~A., {Majewski}, S.~R., {Nidever}, D.~L., {Schiavon}, R.~P., {Shane}, N., {Smith}, V.~V., {Sobeck}, J., {Troup}, N., {Zamora}, O., {Weinberg}, D.~H., {Bovy}, J., {Eisenstein}, D.~J., {Feuillet}, D., {Frinchaboy}, P.~M., {Hayden}, M.~R., {Hearty}, F.~R., {Nguyen}, D.~C., {O'Connell}, R.~W., {Pinsonneault}, M.~H., {Wilson}, J.~C., \& {Zasowski}, G., 2016.
\newblock {ASPCAP: The APOGEE Stellar Parameter and Chemical Abundances Pipeline}, {\it \aj\/}, {\bf 151}(6), 144.

\bibitem[Gilhool et~al.(2017)Gilhool, Blake, Terrien, Bender, Mahadevan, \& Deshpande]{Gilhool2018}
Gilhool, S.~H., Blake, C.~H., Terrien, R.~C., Bender, C., Mahadevan, S., \& Deshpande, R., 2017.
\newblock The rotation of m dwarfs observed by the apache point galactic evolution experiment, {\it The Astronomical Journal\/}, {\bf 155}(1), 38.

\bibitem[{Gustafsson} et~al.(2008){Gustafsson}, {Edvardsson}, {Eriksson}, {J{\o}rgensen}, {Nordlund}, \& {Plez}]{MARCS08}
{Gustafsson}, B., {Edvardsson}, B., {Eriksson}, K., {J{\o}rgensen}, U.~G., {Nordlund}, {\r{A}}., \& {Plez}, B., 2008.
\newblock {A grid of MARCS model atmospheres for late-type stars. I. Methods and general properties}, {\it \aap\/}, {\bf 486}(3), 951--970.

\bibitem[{Hardorp} et~al.(1959){Hardorp}, {Rohlfs}, {Slettebak}, \& {Stock}]{Hardorp59}
{Hardorp}, J., {Rohlfs}, K., {Slettebak}, A., \& {Stock}, J., 1959.
\newblock {Luminous stars in the Northern Milky Way. Part I.}, {\it Hamburger Sternw. Warner \& Swasey Obs.\/}, {\bf C01}, 0.

\bibitem[{Haro}(1964)]{Haro64}
{Haro}, G., 1964.
\newblock {Flash stars in stellar aggregates}, in {\em The Galaxy and the Magellanic Clouds\/}, vol.~20, p.~30.

\bibitem[Harris et~al.(2020)Harris, Millman, van~der Walt, Gommers, Virtanen, Cournapeau, Wieser, Taylor, Berg, Smith, Kern, Picus, Hoyer, van Kerkwijk, Brett, Haldane, Fernández~del Río, Wiebe, Peterson, Gérard-Marchant, Sheppard, Reddy, Weckesser, Abbasi, Gohlke, \& Oliphant]{2020NumPy-Array}
Harris, C.~R., Millman, K.~J., van~der Walt, S.~J., Gommers, R., Virtanen, P., Cournapeau, D., Wieser, E., Taylor, J., Berg, S., Smith, N.~J., Kern, R., Picus, M., Hoyer, S., van Kerkwijk, M.~H., Brett, M., Haldane, A., Fernández~del Río, J., Wiebe, M., Peterson, P., Gérard-Marchant, P., Sheppard, K., Reddy, T., Weckesser, W., Abbasi, H., Gohlke, C., \& Oliphant, T.~E., 2020.
\newblock Array programming with {NumPy}, {\it Nature\/}, {\bf 585}, 357–362.

\bibitem[{He} et~al.(2019){He}, {Wang}, {Liu}, {Soria}, {Bai}, {Yang}, {Bai}, \& {Guo}]{He19}
{He}, L., {Wang}, S., {Liu}, J., {Soria}, R., {Bai}, Z., {Yang}, H., {Bai}, Y., \& {Guo}, J., 2019.
\newblock {A Combined Chandra and LAMOST Study of Stellar Activity}, {\it \apj\/}, {\bf 871}(2), 193.

\bibitem[{Heiter} et~al.(2015){Heiter}, {Jofr{\'e}}, {Gustafsson}, {Korn}, {Soubiran}, \& {Th{\'e}venin}]{Heiter15}
{Heiter}, U., {Jofr{\'e}}, P., {Gustafsson}, B., {Korn}, A.~J., {Soubiran}, C., \& {Th{\'e}venin}, F., 2015.
\newblock {Gaia FGK benchmark stars: Effective temperatures and surface gravities}, {\it \aap\/}, {\bf 582}, A49.

\bibitem[{Hern{\'a}ndez} et~al.(2004){Hern{\'a}ndez}, {Calvet}, {Brice{\~n}o}, {Hartmann}, \& {Berlind}]{Hernandez04}
{Hern{\'a}ndez}, J., {Calvet}, N., {Brice{\~n}o}, C., {Hartmann}, L., \& {Berlind}, P., 2004.
\newblock {Spectral Analysis and Classification of Herbig Ae/Be Stars}, {\it \aj\/}, {\bf 127}(3), 1682--1701.

\bibitem[{Hillenbrand}(1995)]{Hillenbrand95}
{Hillenbrand}, L.~A., 1995.
\newblock {\it {Herbig Ae/Be Stars: An Investigation of Molecular Environments and Associated Stellar Populations}\/}, Ph.D. thesis, University of California System; University of Texas, Austin, Department of Astronomy; -; -.

\bibitem[{Holtzman} et~al.(2015){Holtzman}, {Shetrone}, {Johnson}, {Allende Prieto}, {Anders}, {Andrews}, {Beers}, {Bizyaev}, {Blanton}, {Bovy}, {Carrera}, {Chojnowski}, {Cunha}, {Eisenstein}, {Feuillet}, {Frinchaboy}, {Galbraith-Frew}, {Garc{\'\i}a P{\'e}rez}, {Garc{\'\i}a-Hern{\'a}ndez}, {Hasselquist}, {Hayden}, {Hearty}, {Ivans}, {Majewski}, {Martell}, {Meszaros}, {Muna}, {Nidever}, {Nguyen}, {O'Connell}, {Pan}, {Pinsonneault}, {Robin}, {Schiavon}, {Shane}, {Sobeck}, {Smith}, {Troup}, {Weinberg}, {Wilson}, {Wood-Vasey}, {Zamora}, \& {Zasowski}]{Holtzman15}
{Holtzman}, J.~A., {Shetrone}, M., {Johnson}, J.~A., {Allende Prieto}, C., {Anders}, F., {Andrews}, B., {Beers}, T.~C., {Bizyaev}, D., {Blanton}, M.~R., {Bovy}, J., {Carrera}, R., {Chojnowski}, S.~D., {Cunha}, K., {Eisenstein}, D.~J., {Feuillet}, D., {Frinchaboy}, P.~M., {Galbraith-Frew}, J., {Garc{\'\i}a P{\'e}rez}, A.~E., {Garc{\'\i}a-Hern{\'a}ndez}, D.~A., {Hasselquist}, S., {Hayden}, M.~R., {Hearty}, F.~R., {Ivans}, I., {Majewski}, S.~R., {Martell}, S., {Meszaros}, S., {Muna}, D., {Nidever}, D., {Nguyen}, D.~C., {O'Connell}, R.~W., {Pan}, K., {Pinsonneault}, M., {Robin}, A.~C., {Schiavon}, R.~P., {Shane}, N., {Sobeck}, J., {Smith}, V.~V., {Troup}, N., {Weinberg}, D.~H., {Wilson}, J.~C., {Wood-Vasey}, W.~M., {Zamora}, O., \& {Zasowski}, G., 2015.
\newblock {Abundances, Stellar Parameters, and Spectra from the SDSS-III/APOGEE Survey}, {\it \aj\/}, {\bf 150}(5), 148.

\bibitem[{Hunter}(2007)]{Matplotlib}
{Hunter}, J.~D., 2007.
\newblock Matplotlib: A 2d graphics environment, {\it Computing in Science Engineering\/}, {\bf 9}(3), 90--95.

\bibitem[{Husser} et~al.(2013){Husser}, {Wende-von Berg}, {Dreizler}, {Homeier}, {Reiners}, {Barman}, \& {Hauschildt}]{Husser13}
{Husser}, T.~O., {Wende-von Berg}, S., {Dreizler}, S., {Homeier}, D., {Reiners}, A., {Barman}, T., \& {Hauschildt}, P.~H., 2013.
\newblock {A new extensive library of PHOENIX stellar atmospheres and synthetic spectra}, {\it \aap\/}, {\bf 553}, A6.

\bibitem[{Ishida}(1970)]{Ishida70}
{Ishida}, K., 1970.
\newblock {Spectrographic Observations of IC 1805}, {\it \pasj\/}, {\bf 22}, 277.

\bibitem[{J{\"o}nsson} et~al.(2020){J{\"o}nsson}, {Holtzman}, {Allende Prieto}, {Cunha}, {Garc{\'\i}a-Hern{\'a}ndez}, {Hasselquist}, {Masseron}, {Osorio}, {Shetrone}, {Smith}, {Stringfellow}, {Bizyaev}, {Edvardsson}, {Majewski}, {M{\'e}sz{\'a}ros}, {Souto}, {Zamora}, {Beaton}, {Bovy}, {Donor}, {Pinsonneault}, {Poovelil}, \& {Sobeck}]{Jonsson20}
{J{\"o}nsson}, H., {Holtzman}, J.~A., {Allende Prieto}, C., {Cunha}, K., {Garc{\'\i}a-Hern{\'a}ndez}, D.~A., {Hasselquist}, S., {Masseron}, T., {Osorio}, Y., {Shetrone}, M., {Smith}, V., {Stringfellow}, G.~S., {Bizyaev}, D., {Edvardsson}, B., {Majewski}, S.~R., {M{\'e}sz{\'a}ros}, S., {Souto}, D., {Zamora}, O., {Beaton}, R.~L., {Bovy}, J., {Donor}, J., {Pinsonneault}, M.~H., {Poovelil}, V.~J., \& {Sobeck}, J., 2020.
\newblock {APOGEE Data and Spectral Analysis from SDSS Data Release 16: Seven Years of Observations Including First Results from APOGEE-South}, {\it \aj\/}, {\bf 160}(3), 120.

\bibitem[{Kiminki} et~al.(2015){Kiminki}, {Kim}, {Bagley}, {Sherry}, \& {Rieke}]{Kiminki15}
{Kiminki}, M.~M., {Kim}, J.~S., {Bagley}, M.~B., {Sherry}, W.~H., \& {Rieke}, G.~H., 2015.
\newblock {The O- and B-Type Stellar Population in W3: Beyond the High-Density Layer}, {\it \apj\/}, {\bf 813}(1), 42.

\bibitem[{Koenig} \& {Allen}(2011)]{Koenig11}
{Koenig}, X.~P. \& {Allen}, L.~E., 2011.
\newblock {Disk Evolution in W5: Intermediate-mass Stars at 2-5 Myr}, {\it \apj\/}, {\bf 726}(1), 18.

\bibitem[{Kollmeier} et~al.(2017){Kollmeier}, {Zasowski}, {Rix}, {Johns}, {Anderson}, {Drory}, {Johnson}, {Pogge}, {Bird}, {Blanc}, {Brownstein}, {Crane}, {De Lee}, {Klaene}, {Kreckel}, {MacDonald}, {Merloni}, {Ness}, {O'Brien}, {Sanchez-Gallego}, {Sayres}, {Shen}, {Thakar}, {Tkachenko}, {Aerts}, {Blanton}, {Eisenstein}, {Holtzman}, {Maoz}, {Nandra}, {Rockosi}, {Weinberg}, {Bovy}, {Casey}, {Chaname}, {Clerc}, {Conroy}, {Eracleous}, {G{\"a}nsicke}, {Hekker}, {Horne}, {Kauffmann}, {McQuinn}, {Pellegrini}, {Schinnerer}, {Schlafly}, {Schwope}, {Seibert}, {Teske}, \& {van Saders}]{Kollmeier17}
{Kollmeier}, J.~A., {Zasowski}, G., {Rix}, H.-W., {Johns}, M., {Anderson}, S.~F., {Drory}, N., {Johnson}, J.~A., {Pogge}, R.~W., {Bird}, J.~C., {Blanc}, G.~A., {Brownstein}, J.~R., {Crane}, J.~D., {De Lee}, N.~M., {Klaene}, M.~A., {Kreckel}, K., {MacDonald}, N., {Merloni}, A., {Ness}, M.~K., {O'Brien}, T., {Sanchez-Gallego}, J.~R., {Sayres}, C.~C., {Shen}, Y., {Thakar}, A.~R., {Tkachenko}, A., {Aerts}, C., {Blanton}, M.~R., {Eisenstein}, D.~J., {Holtzman}, J.~A., {Maoz}, D., {Nandra}, K., {Rockosi}, C., {Weinberg}, D.~H., {Bovy}, J., {Casey}, A.~R., {Chaname}, J., {Clerc}, N., {Conroy}, C., {Eracleous}, M., {G{\"a}nsicke}, B.~T., {Hekker}, S., {Horne}, K., {Kauffmann}, J., {McQuinn}, K. B.~W., {Pellegrini}, E.~W., {Schinnerer}, E., {Schlafly}, E.~F., {Schwope}, A.~D., {Seibert}, M., {Teske}, J.~K., \& {van Saders}, J.~L., 2017.
\newblock {SDSS-V: Pioneering Panoptic Spectroscopy}, {\it arXiv e-prints\/}, p. arXiv:1711.03234.

\bibitem[Kosiorowski \& Zawadzki(2022)]{DepthProc}
Kosiorowski, D. \& Zawadzki, Z., 2022.
\newblock {\it DepthProc An R Package for Robust Exploration of Multidimensional Economic Phenomena\/}.

\bibitem[{Kounkel} et~al.(2019){Kounkel}, {Covey}, {Moe}, {Kratter}, {Su{\'a}rez}, {Stassun}, {Rom{\'a}n-Z{\'u}{\~n}iga}, {Hernandez}, {Kim}, {Pe{\~n}a Ram{\'\i}rez}, {Roman-Lopes}, {Stringfellow}, {Jaehnig}, {Borissova}, {Tofflemire}, {Krolikowski}, {Rizzuto}, {Kraus}, {Badenes}, {Longa-Pe{\~n}a}, {G{\'o}mez Maqueo Chew}, {Barba}, {Nidever}, {Brown}, {De Lee}, {Pan}, {Bizyaev}, {Oravetz}, \& {Oravetz}]{Kounkel19}
{Kounkel}, M., {Covey}, K., {Moe}, M., {Kratter}, K.~M., {Su{\'a}rez}, G., {Stassun}, K.~G., {Rom{\'a}n-Z{\'u}{\~n}iga}, C., {Hernandez}, J., {Kim}, J.~S., {Pe{\~n}a Ram{\'\i}rez}, K., {Roman-Lopes}, A., {Stringfellow}, G.~S., {Jaehnig}, K.~O., {Borissova}, J., {Tofflemire}, B., {Krolikowski}, D., {Rizzuto}, A., {Kraus}, A., {Badenes}, C., {Longa-Pe{\~n}a}, P., {G{\'o}mez Maqueo Chew}, Y., {Barba}, R., {Nidever}, D.~L., {Brown}, C., {De Lee}, N., {Pan}, K., {Bizyaev}, D., {Oravetz}, D., \& {Oravetz}, A., 2019.
\newblock {Close Companions around Young Stars}, {\it \aj\/}, {\bf 157}(5), 196.

\bibitem[{Li} et~al.(2022){Li}, {Zhao}, {Chen}, {Liang}, \& {Zhao}]{LAMOSTDR8}
{Li}, Z., {Zhao}, G., {Chen}, Y., {Liang}, X., \& {Zhao}, J., 2022.
\newblock {The stellar parameters and elemental abundances from low-resolution spectra - I. 1.2 million giants from LAMOST DR8}, {\it \mnras\/}, {\bf 517}(4), 4875--4891.

\bibitem[{L{\'o}pez-Valdivia} et~al.(2024){L{\'o}pez-Valdivia}, {Adame}, {Zagala Lagunas}, {Rom{\'a}n-Z{\'u}{\~n}iga}, {Hern{\'a}ndez}, {S{\'a}nchez}, {Fern{\'a}ndez-Trincado}, {Carigi}, {Kounkel}, {Lane}, {Stassun}, \& {Villanova}]{Ricardo24}
{L{\'o}pez-Valdivia}, R., {Adame}, L., {Zagala Lagunas}, E., {Rom{\'a}n-Z{\'u}{\~n}iga}, C.~G., {Hern{\'a}ndez}, J., {S{\'a}nchez}, E., {Fern{\'a}ndez-Trincado}, J.~G., {Carigi}, L., {Kounkel}, M., {Lane}, R.~R., {Stassun}, K.~G., \& {Villanova}, S., 2024.
\newblock {Atmospheric parameters and chemical abundances within 100 pc: a sample of G, K, and M main-sequence stars}, {\it \mnras\/}, {\bf 533}(1), 395--412.

\bibitem[{Magic} et~al.(2015){Magic}, {Chiavassa}, {Collet}, \& {Asplund}]{Magic15}
{Magic}, Z., {Chiavassa}, A., {Collet}, R., \& {Asplund}, M., 2015.
\newblock {The Stagger-grid: A grid of 3D stellar atmosphere models. IV. Limb darkening coefficients}, {\it \aap\/}, {\bf 573}, A90.

\bibitem[{Ma{\'\i}z Apell{\'a}niz} et~al.(2019){Ma{\'\i}z Apell{\'a}niz}, {Trigueros P{\'a}ez}, {Negueruela}, {Barb{\'a}}, {Sim{\'o}n-D{\'\i}az}, {Lorenzo}, {Sota}, {Gamen}, {Fari{\~n}a}, {Salas}, {Caballero}, {Morrell}, {Pellerin}, {Alfaro}, {Herrero}, {Arias}, \& {Marco}]{Maiz19}
{Ma{\'\i}z Apell{\'a}niz}, J., {Trigueros P{\'a}ez}, E., {Negueruela}, I., {Barb{\'a}}, R.~H., {Sim{\'o}n-D{\'\i}az}, S., {Lorenzo}, J., {Sota}, A., {Gamen}, R.~C., {Fari{\~n}a}, C., {Salas}, J., {Caballero}, J.~A., {Morrell}, N.~I., {Pellerin}, A., {Alfaro}, E.~J., {Herrero}, A., {Arias}, J.~I., \& {Marco}, A., 2019.
\newblock {MONOS: Multiplicity Of Northern O-type Spectroscopic systems. I. Project description and spectral classifications and visual multiplicity of previously known objects}, {\it \aap\/}, {\bf 626}, A20.

\bibitem[{Majewski} et~al.(2017){Majewski}, {Schiavon}, {Frinchaboy}, {Allende Prieto}, {Barkhouser}, {Bizyaev}, {Blank}, {Brunner}, {Burton}, {Carrera}, {Chojnowski}, {Cunha}, {Epstein}, {Fitzgerald}, {Garc{\'\i}a P{\'e}rez}, {Hearty}, {Henderson}, {Holtzman}, {Johnson}, {Lam}, {Lawler}, {Maseman}, {M{\'e}sz{\'a}ros}, {Nelson}, {Nguyen}, {Nidever}, {Pinsonneault}, {Shetrone}, {Smee}, {Smith}, {Stolberg}, {Skrutskie}, {Walker}, {Wilson}, {Zasowski}, {Anders}, {Basu}, {Beland}, {Blanton}, {Bovy}, {Brownstein}, {Carlberg}, {Chaplin}, {Chiappini}, {Eisenstein}, {Elsworth}, {Feuillet}, {Fleming}, {Galbraith-Frew}, {Garc{\'\i}a}, {Garc{\'\i}a-Hern{\'a}ndez}, {Gillespie}, {Girardi}, {Gunn}, {Hasselquist}, {Hayden}, {Hekker}, {Ivans}, {Kinemuchi}, {Klaene}, {Mahadevan}, {Mathur}, {Mosser}, {Muna}, {Munn}, {Nichol}, {O'Connell}, {Parejko}, {Robin}, {Rocha-Pinto}, {Schultheis}, {Serenelli}, {Shane}, {Silva Aguirre}, {Sobeck}, {Thompson}, {Troup}, {Weinberg}, \& {Zamora}]{apo2}
{Majewski}, S.~R., {Schiavon}, R.~P., {Frinchaboy}, P.~M., {Allende Prieto}, C., {Barkhouser}, R., {Bizyaev}, D., {Blank}, B., {Brunner}, S., {Burton}, A., {Carrera}, R., {Chojnowski}, S.~D., {Cunha}, K., {Epstein}, C., {Fitzgerald}, G., {Garc{\'\i}a P{\'e}rez}, A.~E., {Hearty}, F.~R., {Henderson}, C., {Holtzman}, J.~A., {Johnson}, J.~A., {Lam}, C.~R., {Lawler}, J.~E., {Maseman}, P., {M{\'e}sz{\'a}ros}, S., {Nelson}, M., {Nguyen}, D.~C., {Nidever}, D.~L., {Pinsonneault}, M., {Shetrone}, M., {Smee}, S., {Smith}, V.~V., {Stolberg}, T., {Skrutskie}, M.~F., {Walker}, E., {Wilson}, J.~C., {Zasowski}, G., {Anders}, F., {Basu}, S., {Beland}, S., {Blanton}, M.~R., {Bovy}, J., {Brownstein}, J.~R., {Carlberg}, J., {Chaplin}, W., {Chiappini}, C., {Eisenstein}, D.~J., {Elsworth}, Y., {Feuillet}, D., {Fleming}, S.~W., {Galbraith-Frew}, J., {Garc{\'\i}a}, R.~A., {Garc{\'\i}a-Hern{\'a}ndez}, D.~A., {Gillespie}, B.~A., {Girardi}, L., {Gunn}, J.~E., {Hasselquist}, S., {Hayden}, M.~R., {Hekker}, S., {Ivans}, I., {Kinemuchi},
  K., {Klaene}, M., {Mahadevan}, S., {Mathur}, S., {Mosser}, B., {Muna}, D., {Munn}, J.~A., {Nichol}, R.~C., {O'Connell}, R.~W., {Parejko}, J.~K., {Robin}, A.~C., {Rocha-Pinto}, H., {Schultheis}, M., {Serenelli}, A.~M., {Shane}, N., {Silva Aguirre}, V., {Sobeck}, J.~S., {Thompson}, B., {Troup}, N.~W., {Weinberg}, D.~H., \& {Zamora}, O., 2017.
\newblock {The Apache Point Observatory Galactic Evolution Experiment (APOGEE)}, {\it \aj\/}, {\bf 154}(3), 94.

\bibitem[{McCuskey}(1974)]{McCuskey74}
{McCuskey}, S.~W., 1974.
\newblock {The space density of A stars in a region in Cassiopeia.}, {\it \aj\/}, {\bf 79}, 107--115.

\bibitem[{M}c{K}inney(2010)]{Pandas10}
{M}c{K}inney, W., 2010.
\newblock {D}ata {S}tructures for {S}tatistical {C}omputing in {P}ython, in {\em {P}roceedings of the 9th {P}ython in {S}cience {C}onference\/}, pp. 56 -- 61.

\bibitem[{Mendoza V.}(1956)]{Mendoza56}
{Mendoza V.}, E.~E., 1956.
\newblock {A Spectroscopic Study of the Pleiades.}, {\it \apj\/}, {\bf 123}, 54.

\bibitem[Mori et~al.(2023)Mori, moshi, Kim, Staffuzza, Fontal, Matsen, Stachelek, \& liuzj039]{Patchwork}
Mori, H., moshi, Kim, Staffuzza, R.~V., Fontal, A., Matsen, E., Stachelek, J., \& liuzj039, 2023.
\newblock ponnhide/patchworklib: v0.6.3.

\bibitem[{Mosler}(2013)]{Mosler2013}
{Mosler}, K., 2013.
\newblock {\it {Depth Statistics}\/}, pp. 17--34, {Springer Berlin Heidelberg}, {Berlin, Heidelberg}.

\bibitem[{Nesterov} et~al.(1995){Nesterov}, {Kuzmin}, {Ashimbaeva}, {Volchkov}, {R{\"o}ser}, \& {Bastian}]{Nesterov95}
{Nesterov}, V.~V., {Kuzmin}, A.~V., {Ashimbaeva}, N.~T., {Volchkov}, A.~A., {R{\"o}ser}, S., \& {Bastian}, U., 1995.
\newblock {The Henry Draper Extension Charts: A catalogue of accurate positions, proper motions, magnitudes and spectral types of 86933 stars}, {\it \aaps\/}, {\bf 110}, 367.

\bibitem[{Netopil} et~al.(2016){Netopil}, {Paunzen}, {Heiter}, \& {Soubiran}]{Netopil16}
{Netopil}, M., {Paunzen}, E., {Heiter}, U., \& {Soubiran}, C., 2016.
\newblock On the metallicity of open clusters - iii. homogenised sample, {\it A\&A\/}, {\bf 585}, A150.

\bibitem[Newton et~al.(2015)Newton, Charbonneau, Irwin, \& Mann]{Newton15}
Newton, E.~R., Charbonneau, D., Irwin, J., \& Mann, A.~W., 2015.
\newblock An empirical calibration to estimate cool dwarf fundamental parameters from h-band spectra, {\it The Astrophysical Journal\/}, {\bf 800}(2), 85, http://arxiv.org/pdf/1412.2758.

\bibitem[Oja(1983)]{Oja83}
Oja, H., 1983.
\newblock Descriptive statistics for multivariate distributions, {\it Statistics \& Probability Letters\/}, {\bf 1}(6), 327--332.

\bibitem[Oller-Moreno(2020)]{parmap}
Oller-Moreno, S., 2020.
\newblock parmap: Easy to use map and starmap python equivalents.

\bibitem[{Olney} et~al.(2020){Olney}, {Kounkel}, {Schillinger}, {Scoggins}, {Yin}, {Howard}, {Covey}, {Hutchinson}, \& {Stassun}]{anet}
{Olney}, R., {Kounkel}, M., {Schillinger}, C., {Scoggins}, M.~T., {Yin}, Y., {Howard}, E., {Covey}, K.~R., {Hutchinson}, B., \& {Stassun}, K.~G., 2020.
\newblock {APOGEE Net: Improving the Derived Spectral Parameters for Young Stars through Deep Learning}, {\it \aj\/}, {\bf 159}(4), 182.

\bibitem[{Origlia} et~al.(1993){Origlia}, {Moorwood}, \& {Oliva}]{Origlia93}
{Origlia}, L., {Moorwood}, A.~F.~M., \& {Oliva}, E., 1993.
\newblock {The 1.5-1.7 mu.m spectrum of cool stars : line identifications, indices for spectral classification and the stellar content of the Seyfert galaxy NGC 1068.}, {\it \aap\/}, {\bf 280}, 536--550.

\bibitem[{Pecaut} \& {Mamajek}(2013)]{PecautMamajek}
{Pecaut}, M.~J. \& {Mamajek}, E.~E., 2013.
\newblock {Intrinsic Colors, Temperatures, and Bolometric Corrections of Pre-main-sequence Stars}, {\it \apjs\/}, {\bf 208}(1), 9.

\bibitem[Pedregosa et~al.(2011)Pedregosa, Varoquaux, Gramfort, Michel, Thirion, Grisel, Blondel, Prettenhofer, Weiss, Dubourg, Vanderplas, Passos, Cournapeau, Brucher, Perrot, \& Duchesnay]{scikit-learn}
Pedregosa, F., Varoquaux, G., Gramfort, A., Michel, V., Thirion, B., Grisel, O., Blondel, M., Prettenhofer, P., Weiss, R., Dubourg, V., Vanderplas, J., Passos, A., Cournapeau, D., Brucher, M., Perrot, M., \& Duchesnay, E., 2011.
\newblock Scikit-learn: Machine learning in {P}ython, {\it Journal of Machine Learning Research\/}, {\bf 12}, 2825--2830.

\bibitem[{Porto de Mello} et~al.(2014){Porto de Mello}, {da Silva}, {da Silva}, \& {de Nader}]{Porto14}
{Porto de Mello}, G.~F., {da Silva}, R., {da Silva}, L., \& {de Nader}, R.~V., 2014.
\newblock {A photometric and spectroscopic survey of solar twin stars within 50 parsecs of the Sun. I. Atmospheric parameters and color similarity to the Sun}, {\it \aap\/}, {\bf 563}, A52.

\bibitem[{Prosser} et~al.(1991){Prosser}, {Stauffer}, \& {Kraft}]{Prosser91}
{Prosser}, C.~F., {Stauffer}, J., \& {Kraft}, R.~P., 1991.
\newblock {The Search for Faint Members of the Pleiades. II. Colors, Spectral Types, and H-alpha Emission Line Strengths for M Dwarfs in the Pleiades, Hyades, and Gliese Field}, {\it \aj\/}, {\bf 101}, 1361.

\bibitem[{Pr{\v{s}}a} et~al.(2016){Pr{\v{s}}a}, {Harmanec}, {Torres}, {Mamajek}, {Asplund}, {Capitaine}, {Christensen-Dalsgaard}, {Depagne}, {Haberreiter}, {Hekker}, {Hilton}, {Kopp}, {Kostov}, {Kurtz}, {Laskar}, {Mason}, {Milone}, {Montgomery}, {Richards}, {Schmutz}, {Schou}, \& {Stewart}]{Prsa16}
{Pr{\v{s}}a}, A., {Harmanec}, P., {Torres}, G., {Mamajek}, E., {Asplund}, M., {Capitaine}, N., {Christensen-Dalsgaard}, J., {Depagne}, {\'E}., {Haberreiter}, M., {Hekker}, S., {Hilton}, J., {Kopp}, G., {Kostov}, V., {Kurtz}, D.~W., {Laskar}, J., {Mason}, B.~D., {Milone}, E.~F., {Montgomery}, M., {Richards}, M., {Schmutz}, W., {Schou}, J., \& {Stewart}, S.~G., 2016.
\newblock {Nominal Values for Selected Solar and Planetary Quantities: IAU 2015 Resolution B3}, {\it \aj\/}, {\bf 152}(2), 41.

\bibitem[{R Core Team}(2023)]{rsoft}
{R Core Team}, 2023.
\newblock {\it R: A Language and Environment for Statistical Computing\/}, R Foundation for Statistical Computing, Vienna, Austria.

\bibitem[{Raddi} et~al.(2013){Raddi}, {Drew}, {Fabregat}, {Steeghs}, {Wright}, {Sale}, {Farnhill}, {Barlow}, {Greimel}, {Sabin}, {Corradi}, \& {Drake}]{Raddi13}
{Raddi}, R., {Drew}, J.~E., {Fabregat}, J., {Steeghs}, D., {Wright}, N.~J., {Sale}, S.~E., {Farnhill}, H.~J., {Barlow}, M.~J., {Greimel}, R., {Sabin}, L., {Corradi}, R.~M.~L., \& {Drake}, J.~J., 2013.
\newblock {First results of an H{\ensuremath{\alpha}} based search of classical Be stars in the Perseus Arm and beyond}, {\it \mnras\/}, {\bf 430}(3), 2169--2187.

\bibitem[{Ram{\'\i}rez-Preciado} et~al.(2020){Ram{\'\i}rez-Preciado}, {Roman-Lopes}, {Rom{\'a}n-Z{\'u}{\~n}iga}, {Hern{\'a}ndez}, {Garc{\'\i}a-Hern{\'a}ndez}, {Stassun}, {Stringfellow}, \& {Kim}]{RamirezPreciado20}
{Ram{\'\i}rez-Preciado}, V.~G., {Roman-Lopes}, A., {Rom{\'a}n-Z{\'u}{\~n}iga}, C.~G., {Hern{\'a}ndez}, J., {Garc{\'\i}a-Hern{\'a}ndez}, D.~A., {Stassun}, K., {Stringfellow}, G.~S., \& {Kim}, J.~S., 2020.
\newblock {Spectral Classification of B Stars: The Empirical Sequence Using SDSS-IV/APOGEE Near-IR Data}, {\it \apj\/}, {\bf 894}(1), 5.

\bibitem[{Roman-Lopes} et~al.(2018){Roman-Lopes}, {Rom{\'a}n-Z{\'u}{\~n}iga}, {Tapia}, {Chojnowski}, {G{\'o}mez Maqueo Chew}, {Garc{\'\i}a-Hern{\'a}ndez}, {Borissova}, {Minniti}, {Covey}, {Longa-Pe{\~n}a}, {Fernandez-Trincado}, {Zamora}, \& {Nitschelm}]{RomanLopes18}
{Roman-Lopes}, A., {Rom{\'a}n-Z{\'u}{\~n}iga}, C., {Tapia}, M., {Chojnowski}, D., {G{\'o}mez Maqueo Chew}, Y., {Garc{\'\i}a-Hern{\'a}ndez}, D.~A., {Borissova}, J., {Minniti}, D., {Covey}, K.~R., {Longa-Pe{\~n}a}, P., {Fernandez-Trincado}, J.~G., {Zamora}, O., \& {Nitschelm}, C., 2018.
\newblock {Massive Stars in the SDSS-IV/APOGEE SURVEY. I. OB Stars}, {\it \apj\/}, {\bf 855}(1), 68.

\bibitem[{Roman-Lopes} et~al.(2019){Roman-Lopes}, {Rom{\'a}n-Z{\'u}{\~n}iga}, {Tapia}, {Hern{\'a}ndez}, {Ram{\'\i}rez-Preciado}, {Stringfellow}, {Ybarra}, {Kim}, {Minniti}, {Covey}, {Kounkel}, {Su{\'a}rez}, {Borissova}, {Garc{\'\i}a-Hern{\'a}ndez}, {Zamora}, \& {Trujillo}]{RomanLopes19}
{Roman-Lopes}, A., {Rom{\'a}n-Z{\'u}{\~n}iga}, C.~G., {Tapia}, M., {Hern{\'a}ndez}, J., {Ram{\'\i}rez-Preciado}, V., {Stringfellow}, G.~S., {Ybarra}, J.~E., {Kim}, J.~S., {Minniti}, D., {Covey}, K.~R., {Kounkel}, M., {Su{\'a}rez}, G., {Borissova}, J., {Garc{\'\i}a-Hern{\'a}ndez}, D.~A., {Zamora}, O., \& {Trujillo}, J.~D., 2019.
\newblock {Massive Stars in the SDSS-IV/APOGEE-2 Survey. II. OB-stars in the W345 Complexes}, {\it \apj\/}, {\bf 873}(1), 66.

\bibitem[{Rom{\'a}n-Z{\'u}{\~n}iga} et~al.(2023){Rom{\'a}n-Z{\'u}{\~n}iga}, {Kounkel}, {Hern{\'a}ndez}, {Pe{\~n}a Ram{\'\i}rez}, {L{\'o}pez-Valdivia}, {Covey}, {Stutz}, {Roman-Lopes}, {Campbell}, {Khilfeh}, {Tapia}, {Stringfellow}, {Downes}, {Stassun}, {Minniti}, {Bayo}, {Kim}, {Su{\'a}rez}, {Ybarra}, {Fern{\'a}ndez-Trincado}, {Longa-Pe{\~n}a}, {Ram{\'\i}rez-Preciado}, {Serna}, {Lane}, {Garc{\'\i}a-Hern{\'a}ndez}, {Beaton}, {Bizyaev}, \& {Pan}]{roman23}
{Rom{\'a}n-Z{\'u}{\~n}iga}, C.~G., {Kounkel}, M., {Hern{\'a}ndez}, J., {Pe{\~n}a Ram{\'\i}rez}, K., {L{\'o}pez-Valdivia}, R., {Covey}, K.~R., {Stutz}, A.~M., {Roman-Lopes}, A., {Campbell}, H., {Khilfeh}, E., {Tapia}, M., {Stringfellow}, G.~S., {Downes}, J.~J., {Stassun}, K.~G., {Minniti}, D., {Bayo}, A., {Kim}, J.~S., {Su{\'a}rez}, G., {Ybarra}, J.~E., {Fern{\'a}ndez-Trincado}, J.~G., {Longa-Pe{\~n}a}, P., {Ram{\'\i}rez-Preciado}, V., {Serna}, J., {Lane}, R.~R., {Garc{\'\i}a-Hern{\'a}ndez}, D.~A., {Beaton}, R.~L., {Bizyaev}, D., \& {Pan}, K., 2023.
\newblock {Stellar Properties for a Comprehensive Collection of Star-forming Regions in the SDSS APOGEE-2 Survey}, {\it \aj\/}, {\bf 165}(2), 51.

\bibitem[{Santos} et~al.(2008){Santos}, {Melo}, {James}, {Gameiro}, {Bouvier}, \& {Gomes}]{Santos08}
{Santos}, N.~C., {Melo}, C., {James}, D.~J., {Gameiro}, J.~F., {Bouvier}, J., \& {Gomes}, J.~I., 2008.
\newblock {Chemical abundances in six nearby star-forming regions. Implications for galactic evolution and planet searches around very young stars}, {\it \aap\/}, {\bf 480}(3), 889--897.

\bibitem[{Sarmento} et~al.(2020){Sarmento}, {Delgado Mena}, {Rojas-Ayala}, \& {Blanco-Cuaresma}]{Sarmento2020}
{Sarmento}, P., {Delgado Mena}, E., {Rojas-Ayala}, B., \& {Blanco-Cuaresma}, S., 2020.
\newblock {Derivation of parameters for 3748 FGK stars using H-band spectra from APOGEE Data Release 14}, {\it \aap\/}, {\bf 636}, A85.

\bibitem[Schwarz(1978)]{Schwarz78}
Schwarz, G., 1978.
\newblock {Estimating the Dimension of a Model}, {\it The Annals of Statistics\/}, {\bf 6}(2), 461--464.

\bibitem[Scrucca et~al.(2023)Scrucca, Fraley, Murphy, \& Raftery]{mclust}
Scrucca, L., Fraley, C., Murphy, T.~B., \& Raftery, A.~E., 2023.
\newblock {\it Model-Based Clustering, Classification, and Density Estimation Using {mclust} in {R}\/}, Chapman and Hall/CRC.

\bibitem[Segaert et~al.(2023)Segaert, Hubert, Rousseeuw, \& Raymaekers]{mrfDepth}
Segaert, P., Hubert, M., Rousseeuw, P., \& Raymaekers, J., 2023.
\newblock {\it mrfDepth: Depth Measures in Multivariate, Regression and Functional Settings\/}, R package version 1.0.15.

\bibitem[Silverman(1981)]{Silverman81}
Silverman, B.~W., 1981.
\newblock Using kernel density estimates to investigate multimodality, {\it Journal of the Royal Statistical Society: Series B (Methodological)\/}, {\bf 43}(1), 97--99.

\bibitem[{Silverman}(1986)]{Silverman86}
{Silverman}, B.~W., 1986.
\newblock {\it {Density estimation for statistics and data analysis}\/}.

\bibitem[Small(1990)]{Small90}
Small, C.~G., 1990.
\newblock A survey of multidimensional medians, {\it International Statistical Review / Revue Internationale de Statistique\/}, {\bf 58}(3), 263--277.

\bibitem[{Soubiran} et~al.(2016){Soubiran}, {Le Campion}, {Brouillet}, \& {Chemin}]{Soubiran16}
{Soubiran}, C., {Le Campion}, J.-F., {Brouillet}, N., \& {Chemin}, L., 2016.
\newblock {The PASTEL catalogue: 2016 version}, {\it \aap\/}, {\bf 591}, A118.

\bibitem[{Spina} et~al.(2014){Spina}, {Randich}, {Palla}, {Biazzo}, {Sacco}, {Alfaro}, {Franciosini}, {Magrini}, {Morbidelli}, {Frasca}, {Adibekyan}, {Delgado-Mena}, {Sousa}, {Gonz{\'a}lez Hern{\'a}ndez}, {Montes}, {Tabernero}, {Tautvai{\v{s}}ien{\.{e}}}, {Bonito}, {Lanzafame}, {Gilmore}, {Jeffries}, {Vallenari}, {Bensby}, {Bragaglia}, {Flaccomio}, {Korn}, {Pancino}, {Recio-Blanco}, {Smiljanic}, {Bergemann}, {Costado}, {Damiani}, {Hill}, {Hourihane}, {Jofr{\'e}}, {de Laverny}, {Lardo}, {Masseron}, {Prisinzano}, \& {Worley}]{Spina14}
{Spina}, L., {Randich}, S., {Palla}, F., {Biazzo}, K., {Sacco}, G.~G., {Alfaro}, E.~J., {Franciosini}, E., {Magrini}, L., {Morbidelli}, L., {Frasca}, A., {Adibekyan}, V., {Delgado-Mena}, E., {Sousa}, S.~G., {Gonz{\'a}lez Hern{\'a}ndez}, J.~I., {Montes}, D., {Tabernero}, H., {Tautvai{\v{s}}ien{\.{e}}}, G., {Bonito}, R., {Lanzafame}, A.~C., {Gilmore}, G., {Jeffries}, R.~D., {Vallenari}, A., {Bensby}, T., {Bragaglia}, A., {Flaccomio}, E., {Korn}, A.~J., {Pancino}, E., {Recio-Blanco}, A., {Smiljanic}, R., {Bergemann}, M., {Costado}, M.~T., {Damiani}, F., {Hill}, V., {Hourihane}, A., {Jofr{\'e}}, P., {de Laverny}, P., {Lardo}, C., {Masseron}, T., {Prisinzano}, L., \& {Worley}, C.~C., 2014.
\newblock {The Gaia-ESO Survey: Metallicity of the Chamaeleon I star-forming region}, {\it \aap\/}, {\bf 568}, A2.

\bibitem[{Spina} et~al.(2017){Spina}, {Randich}, {Magrini}, {Jeffries}, {Friel}, {Sacco}, {Pancino}, {Bonito}, {Bravi}, {Franciosini}, {Klutsch}, {Montes}, {Gilmore}, {Vallenari}, {Bensby}, {Bragaglia}, {Flaccomio}, {Koposov}, {Korn}, {Lanzafame}, {Smiljanic}, {Bayo}, {Carraro}, {Casey}, {Costado}, {Damiani}, {Donati}, {Frasca}, {Hourihane}, {Jofr{\'e}}, {Lewis}, {Lind}, {Monaco}, {Morbidelli}, {Prisinzano}, {Sousa}, {Worley}, \& {Zaggia}]{Spina17}
{Spina}, L., {Randich}, S., {Magrini}, L., {Jeffries}, R.~D., {Friel}, E.~D., {Sacco}, G.~G., {Pancino}, E., {Bonito}, R., {Bravi}, L., {Franciosini}, E., {Klutsch}, A., {Montes}, D., {Gilmore}, G., {Vallenari}, A., {Bensby}, T., {Bragaglia}, A., {Flaccomio}, E., {Koposov}, S.~E., {Korn}, A.~J., {Lanzafame}, A.~C., {Smiljanic}, R., {Bayo}, A., {Carraro}, G., {Casey}, A.~R., {Costado}, M.~T., {Damiani}, F., {Donati}, P., {Frasca}, A., {Hourihane}, A., {Jofr{\'e}}, P., {Lewis}, J., {Lind}, K., {Monaco}, L., {Morbidelli}, L., {Prisinzano}, L., {Sousa}, S.~G., {Worley}, C.~C., \& {Zaggia}, S., 2017.
\newblock {The Gaia-ESO Survey: the present-day radial metallicity distribution of the Galactic disc probed by pre-main-sequence clusters}, {\it \aap\/}, {\bf 601}, A70.

\bibitem[{Sprague} et~al.(2022){Sprague}, {Culhane}, {Kounkel}, {Olney}, {Covey}, {Hutchinson}, {Lingg}, {Stassun}, {Rom{\'a}n-Z{\'u}{\~n}iga}, {Roman-Lopes}, {Nidever}, {Beaton}, {Borissova}, {Stutz}, {Stringfellow}, {Ram{\'\i}rez}, {Ram{\'\i}rez-Preciado}, {Hern{\'a}ndez}, {Kim}, \& {Lane}]{anet2}
{Sprague}, D., {Culhane}, C., {Kounkel}, M., {Olney}, R., {Covey}, K.~R., {Hutchinson}, B., {Lingg}, R., {Stassun}, K.~G., {Rom{\'a}n-Z{\'u}{\~n}iga}, C.~G., {Roman-Lopes}, A., {Nidever}, D., {Beaton}, R.~L., {Borissova}, J., {Stutz}, A., {Stringfellow}, G.~S., {Ram{\'\i}rez}, K.~P., {Ram{\'\i}rez-Preciado}, V., {Hern{\'a}ndez}, J., {Kim}, J.~S., \& {Lane}, R.~R., 2022.
\newblock {APOGEE Net: An Expanded Spectral Model of Both Low-mass and High-mass Stars}, {\it \aj\/}, {\bf 163}(4), 152.

\bibitem[Straumit et~al.(2022)Straumit, Tkachenko, Gebruers, Audenaert, Xiang, Zari, Aerts, Johnson, Kollmeier, Rix, Beaton, Saders, Teske, Roman-Lopes, Ting, \& Román-Zúñiga]{zpayne}
Straumit, I., Tkachenko, A., Gebruers, S., Audenaert, J., Xiang, M., Zari, E., Aerts, C., Johnson, J.~A., Kollmeier, J.~A., Rix, H.-W., Beaton, R.~L., Saders, J. L.~V., Teske, J., Roman-Lopes, A., Ting, Y.-S., \& Román-Zúñiga, C.~G., 2022.
\newblock Zeta-payne: A fully automated spectrum analysis algorithm for the milky way mapper program of the sdss-v survey, {\it The Astronomical Journal\/}, {\bf 163}(5), 236.

\bibitem[{Takeda} et~al.(2002){Takeda}, {Ohkubo}, \& {Sadakane}]{Takeda02}
{Takeda}, Y., {Ohkubo}, M., \& {Sadakane}, K., 2002.
\newblock {Spectroscopic Determination of Atmospheric Parameters of Solar-Type Stars: Description of the Method and Application to the Sun}, {\it \pasj\/}, {\bf 54}, 451--462.

\bibitem[{Taylor}(2005)]{topcat}
{Taylor}, M.~B., 2005.
\newblock {TOPCAT \& STIL: Starlink Table/VOTable Processing Software}, in {\em Astronomical Data Analysis Software and Systems XIV\/}, vol. 347 of {\bf Astronomical Society of the Pacific Conference Series}, p.~29.

\bibitem[Van~Rossum \& Drake(2009)]{python}
Van~Rossum, G. \& Drake, F.~L., 2009.
\newblock {\it Python 3 Reference Manual\/}, CreateSpace, Scotts Valley, CA.

\bibitem[Virtanen et~al.(2020)Virtanen, Gommers, Oliphant, Haberland, Reddy, Cournapeau, Burovski, Peterson, Weckesser, Bright, {van der Walt}, Brett, Wilson, Millman, Mayorov, Nelson, Jones, Kern, Larson, Carey, Polat, Feng, Moore, {VanderPlas}, Laxalde, Perktold, Cimrman, Henriksen, Quintero, Harris, Archibald, Ribeiro, Pedregosa, {van Mulbregt}, \& {SciPy 1.0 Contributors}]{2020SciPy-NMeth}
Virtanen, P., Gommers, R., Oliphant, T.~E., Haberland, M., Reddy, T., Cournapeau, D., Burovski, E., Peterson, P., Weckesser, W., Bright, J., {van der Walt}, S.~J., Brett, M., Wilson, J., Millman, K.~J., Mayorov, N., Nelson, A. R.~J., Jones, E., Kern, R., Larson, E., Carey, C.~J., Polat, {\.I}., Feng, Y., Moore, E.~W., {VanderPlas}, J., Laxalde, D., Perktold, J., Cimrman, R., Henriksen, I., Quintero, E.~A., Harris, C.~R., Archibald, A.~M., Ribeiro, A.~H., Pedregosa, F., {van Mulbregt}, P., \& {SciPy 1.0 Contributors}, 2020.
\newblock {{SciPy} 1.0: Fundamental Algorithms for Scientific Computing in Python}, {\it Nature Methods\/}, {\bf 17}, 261--272.

\bibitem[{Voroshilov} et~al.(1985){Voroshilov}, {Guseva}, {Kalandadze}, {Kolesnik}, {Kuznetsov}, {Metreveli}, \& {Shapovalov}]{Voroshilov85}
{Voroshilov}, V.~I., {Guseva}, N.~G., {Kalandadze}, N.~B., {Kolesnik}, L.~N., {Kuznetsov}, V.~I., {Metreveli}, M.~D., \& {Shapovalov}, A.~N., 1985.
\newblock {\it {Catalogue of BV magnitudes and spectral classes for 6000 stars. Ukrainian Acad. Nauk, Kiev, 1-140.}\/}, "".

\bibitem[Waskom et~al.(2020)Waskom, Gelbart, Botvinnik, Ostblom, Hobson, Lukauskas, Gemperline, Augspurger, Halchenko, Warmenhoven, Cole, de~Ruiter, Vanderplas, Hoyer, Pye, Miles, Swain, Meyer, Martin, Bachant, Quintero, Kunter, Villalba, Brian, Fitzgerald, Evans, Williams, O'Kane, Yarkoni, \& Brunner]{seaborn}
Waskom, M., Gelbart, M., Botvinnik, O., Ostblom, J., Hobson, P., Lukauskas, S., Gemperline, D.~C., Augspurger, T., Halchenko, Y., Warmenhoven, J., Cole, J.~B., de~Ruiter, J., Vanderplas, J., Hoyer, S., Pye, C., Miles, A., Swain, C., Meyer, K., Martin, M., Bachant, P., Quintero, E., Kunter, G., Villalba, S., Brian, Fitzgerald, C., Evans, C., Williams, M.~L., O'Kane, D., Yarkoni, T., \& Brunner, T., 2020.
\newblock mwaskom/seaborn: v0.11.1 (december 2020).

\bibitem[{Wenger} et~al.(2000){Wenger}, {Ochsenbein}, {Egret}, {Dubois}, {Bonnarel}, {Borde}, {Genova}, {Jasniewicz}, {Lalo{\"e}}, {Lesteven}, \& {Monier}]{Simbad}
{Wenger}, M., {Ochsenbein}, F., {Egret}, D., {Dubois}, P., {Bonnarel}, F., {Borde}, S., {Genova}, F., {Jasniewicz}, G., {Lalo{\"e}}, S., {Lesteven}, S., \& {Monier}, R., 2000.
\newblock {The SIMBAD astronomical database. The CDS reference database for astronomical objects}, {\it \aaps\/}, {\bf 143}, 9--22.

\bibitem[{Wilson} et~al.(2010){Wilson}, {Hearty}, {Skrutskie}, {Majewski}, {Schiavon}, {Eisenstein}, {Gunn}, {Blank}, {Henderson}, {Smee}, {Barkhouser}, {Harding}, {Fitzgerald}, {Stolberg}, {Arns}, {Nelson}, {Brunner}, {Burton}, {Walker}, {Lam}, {Maseman}, {Barr}, {Leger}, {Carey}, {MacDonald}, {Horne}, {Young}, {Rieke}, {Rieke}, {O'Brien}, {Hope}, {Krakula}, {Crane}, {Zhao}, {Carr}, {Harrison}, {Stoll}, {Vernieri}, {Holtzman}, {Shetrone}, {Allende-Prieto}, {Johnson}, {Frinchaboy}, {Zasowski}, {Bizyaev}, {Gillespie}, \& {Weinberg}]{Wilson10}
{Wilson}, J.~C., {Hearty}, F., {Skrutskie}, M.~F., {Majewski}, S., {Schiavon}, R., {Eisenstein}, D., {Gunn}, J., {Blank}, B., {Henderson}, C., {Smee}, S., {Barkhouser}, R., {Harding}, A., {Fitzgerald}, G., {Stolberg}, T., {Arns}, J., {Nelson}, M., {Brunner}, S., {Burton}, A., {Walker}, E., {Lam}, C., {Maseman}, P., {Barr}, J., {Leger}, F., {Carey}, L., {MacDonald}, N., {Horne}, T., {Young}, E., {Rieke}, G., {Rieke}, M., {O'Brien}, T., {Hope}, S., {Krakula}, J., {Crane}, J., {Zhao}, B., {Carr}, M., {Harrison}, C., {Stoll}, R., {Vernieri}, M.~A., {Holtzman}, J., {Shetrone}, M., {Allende-Prieto}, C., {Johnson}, J., {Frinchaboy}, P., {Zasowski}, G., {Bizyaev}, D., {Gillespie}, B., \& {Weinberg}, D., 2010.
\newblock {The Apache Point Observatory Galactic Evolution Experiment (APOGEE) high-resolution near-infrared multi-object fiber spectrograph}, in {\em Ground-based and Airborne Instrumentation for Astronomy III\/}, vol. 7735 of {\bf Society of Photo-Optical Instrumentation Engineers (SPIE) Conference Series}, p. 77351C.

\bibitem[{Wilson} et~al.(2012){Wilson}, {Hearty}, {Skrutskie}, {Majewski}, {Schiavon}, {Eisenstein}, {Gunn}, {Holtzman}, {Nidever}, {Gillespie}, {Weinberg}, {Blank}, {Henderson}, {Smee}, {Barkhouser}, {Harding}, {Hope}, {Fitzgerald}, {Stolberg}, {Arns}, {Nelson}, {Brunner}, {Burton}, {Walker}, {Lam}, {Maseman}, {Barr}, {Leger}, {Carey}, {MacDonald}, {Ebelke}, {Beland}, {Horne}, {Young}, {Rieke}, {Rieke}, {O'Brien}, {Crane}, {Carr}, {Harrison}, {Stoll}, {Vernieri}, {Shetrone}, {Allende-Prieto}, {Johnson}, {Frinchaboy}, {Zasowski}, {Garcia Perez}, {Bizyaev}, {Cunha}, {Smith}, {Meszaros}, {Zhao}, {Hayden}, {Chojnowski}, {Andrews}, {Loomis}, {Owen}, {Klaene}, {Brinkmann}, {Stauffer}, {Long}, {Jordan}, {Holder}, {Cope}, {Naugle}, {Pfaffenberger}, {Schlegel}, {Blanton}, {Muna}, {Weaver}, {Snedden}, {Pan}, {Brewington}, {Malanushenko}, {Malanushenko}, {Simmons}, {Oravetz}, {Mahadevan}, \& {Halverson}]{Wilson12}
{Wilson}, J.~C., {Hearty}, F., {Skrutskie}, M.~F., {Majewski}, S.~R., {Schiavon}, R., {Eisenstein}, D., {Gunn}, J., {Holtzman}, J., {Nidever}, D., {Gillespie}, B., {Weinberg}, D., {Blank}, B., {Henderson}, C., {Smee}, S., {Barkhouser}, R., {Harding}, A., {Hope}, S., {Fitzgerald}, G., {Stolberg}, T., {Arns}, J., {Nelson}, M., {Brunner}, S., {Burton}, A., {Walker}, E., {Lam}, C., {Maseman}, P., {Barr}, J., {Leger}, F., {Carey}, L., {MacDonald}, N., {Ebelke}, G., {Beland}, S., {Horne}, T., {Young}, E., {Rieke}, G., {Rieke}, M., {O'Brien}, T., {Crane}, J., {Carr}, M., {Harrison}, C., {Stoll}, R., {Vernieri}, M., {Shetrone}, M., {Allende-Prieto}, C., {Johnson}, J., {Frinchaboy}, P., {Zasowski}, G., {Garcia Perez}, A., {Bizyaev}, D., {Cunha}, K., {Smith}, V.~V., {Meszaros}, S., {Zhao}, B., {Hayden}, M., {Chojnowski}, S.~D., {Andrews}, B., {Loomis}, C., {Owen}, R., {Klaene}, M., {Brinkmann}, J., {Stauffer}, F., {Long}, D., {Jordan}, W., {Holder}, D., {Cope}, F., {Naugle}, T., {Pfaffenberger}, B., {Schlegel}, D.,
  {Blanton}, M., {Muna}, D., {Weaver}, B., {Snedden}, S., {Pan}, K., {Brewington}, H., {Malanushenko}, E., {Malanushenko}, V., {Simmons}, A., {Oravetz}, D., {Mahadevan}, S., \& {Halverson}, S., 2012.
\newblock {Performance of the Apache Point Observatory Galactic Evolution Experiment (APOGEE) high-resolution near-infrared multi-object fiber spectrograph}, in {\em Ground-based and Airborne Instrumentation for Astronomy IV\/}, vol. 8446 of {\bf Society of Photo-Optical Instrumentation Engineers (SPIE) Conference Series}, p. 84460H.

\bibitem[{Yao} et~al.(2018){Yao}, {Meyer}, {Covey}, {Tan}, \& {Da Rio}]{yao18}
{Yao}, Y., {Meyer}, M.~R., {Covey}, K.~R., {Tan}, J.~C., \& {Da Rio}, N., 2018.
\newblock {IN-SYNC. VIII. Primordial Disk Frequencies in NGC 1333, IC 348, and the Orion A Molecular Cloud}, {\it \apj\/}, {\bf 869}(1), 72.

\bibitem[Yu(2021)]{mixr}
Yu, Y., 2021.
\newblock {\it mixR: Finite Mixture Modeling for Raw and Binned Data\/}, R package version 0.2.0.

\bibitem[{Zasowski} et~al.(2017){Zasowski}, {Cohen}, {Chojnowski}, {Santana}, {Oelkers}, {Andrews}, {Beaton}, {Bender}, {Bird}, {Bovy}, {Carlberg}, {Covey}, {Cunha}, {Dell'Agli}, {Fleming}, {Frinchaboy}, {Garc{\'\i}a-Hern{\'a}ndez}, {Harding}, {Holtzman}, {Johnson}, {Kollmeier}, {Majewski}, {M{\'e}sz{\'a}ros}, {Munn}, {Mu{\~n}oz}, {Ness}, {Nidever}, {Poleski}, {Rom{\'a}n-Z{\'u}{\~n}iga}, {Shetrone}, {Simon}, {Smith}, {Sobeck}, {Stringfellow}, {Szigeti{\'a}ros}, {Tayar}, \& {Troup}]{Zasowski17}
{Zasowski}, G., {Cohen}, R.~E., {Chojnowski}, S.~D., {Santana}, F., {Oelkers}, R.~J., {Andrews}, B., {Beaton}, R.~L., {Bender}, C., {Bird}, J.~C., {Bovy}, J., {Carlberg}, J.~K., {Covey}, K., {Cunha}, K., {Dell'Agli}, F., {Fleming}, S.~W., {Frinchaboy}, P.~M., {Garc{\'\i}a-Hern{\'a}ndez}, D.~A., {Harding}, P., {Holtzman}, J., {Johnson}, J.~A., {Kollmeier}, J.~A., {Majewski}, S.~R., {M{\'e}sz{\'a}ros}, S., {Munn}, J., {Mu{\~n}oz}, R.~R., {Ness}, M.~K., {Nidever}, D.~L., {Poleski}, R., {Rom{\'a}n-Z{\'u}{\~n}iga}, C., {Shetrone}, M., {Simon}, J.~D., {Smith}, V.~V., {Sobeck}, J.~S., {Stringfellow}, G.~S., {Szigeti{\'a}ros}, L., {Tayar}, J., \& {Troup}, N., 2017.
\newblock {Target Selection for the SDSS-IV APOGEE-2 Survey}, {\it \aj\/}, {\bf 154}(5), 198.

\end{thebibliography}

\appendix
\section{Pseudo-codes}\label{appendix:codes}
\subsection{Spectrum normalisation}
The pseudo-codes implementing the iterative normalisation of the observed spectrum are given below (\texttt{ normalisesSpectrum} --- Algorithm~\ref{alg:normspec}, and \texttt{ SigmaClippingNorm} --- Algorithm~\ref{alg:zeroclip}).
We use the functions \texttt{ polyfit}, \texttt{ poly1d} and \texttt{ polyval} from \texttt{ NumPy}\citep{2020NumPy-Array} to fit polynomials to each chip separately, selecting the degree of the polynomial with the smallest Bayesian Information Criterion \citep[BIC,][]{Schwarz78}. 
To remove absorption lines and other outliers from the {sought} continuum, we apply a non-symmetrical $\sigma$-clipping each time we normalise the spectrum. 
For this, we use the function \texttt{ sigma\_clip} provided in \texttt{ AstroPy} \citep{astropy:2013, astropy:2018}. 
The algorithm detailing this normalisation and $\sigma$-clipping is detailed in \texttt{ SigmaClippingNorm} (Algorithm~\ref{alg:zeroclip}).
We repeat the procedures of normalisation and $\sigma$-clip until the normalised spectrum of the previous iteration is equal to the normalised spectrum obtained in a given repeat (procedure \texttt{ normalisesSpectrum}, Algorithm~\ref{alg:normspec}).

\begin{figure}
\includegraphics[alt={Graphs showing the sigma clipping procedure of an spectrum.},width=1\columnwidth]{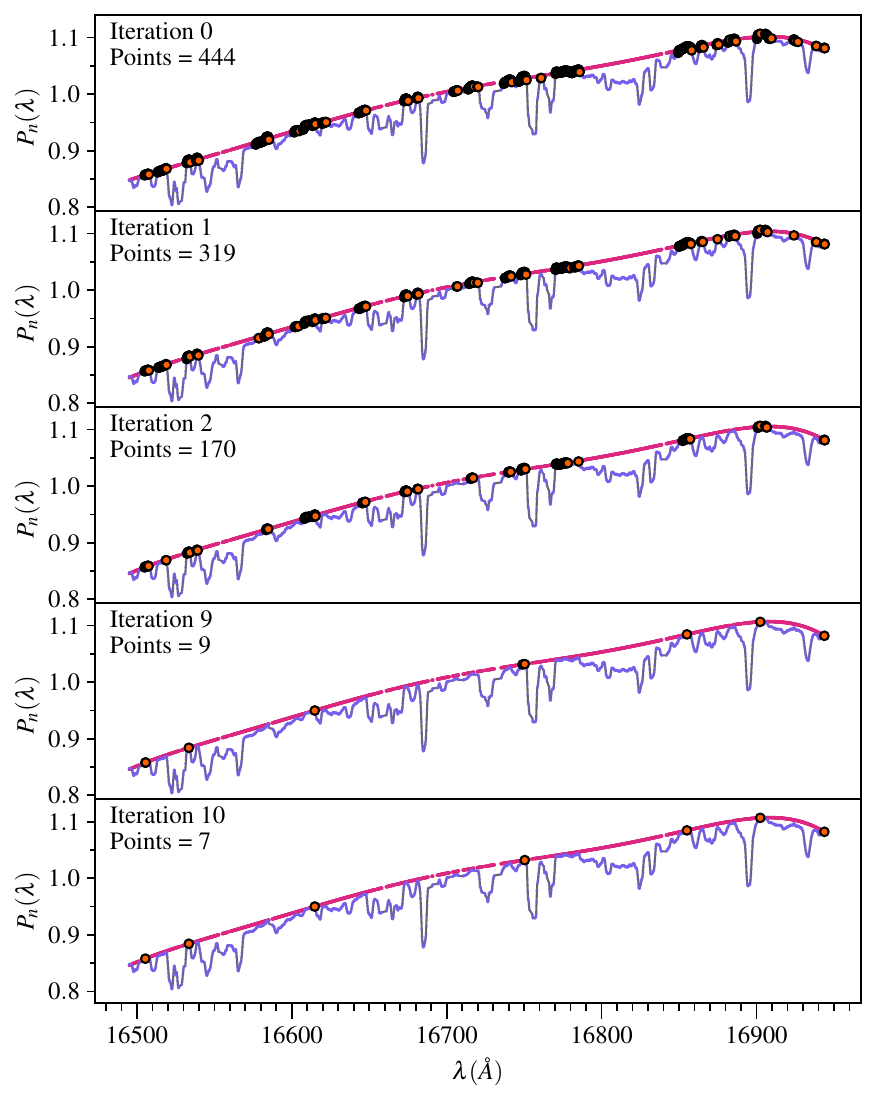}
\caption{{The polynomial $P_n(\lambda)$ constructed (\textit{magenta} line) following the $\sigma$-clipped observed spectrum (\textit{orange} circles) in each iteration $n$. The \textit{original} solar spectrum is shown in the \textit{purple} line.} }\label{fig:polinomioiterations}
\end{figure}

\begin{figure*}
\includegraphics[alt={Graphs showing the continuum normalisation following the sigma clipping procedure of an spectrum.}]{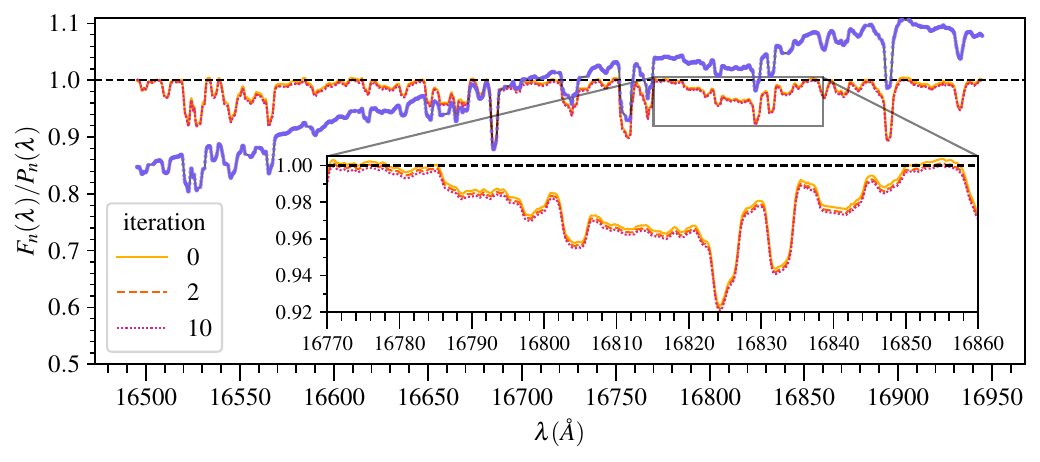}
\caption{{The continuum normalised spectrum $F_n(\lambda)/P_n(\lambda)$ in three iterations. The \textit{original} solar spectrum is shown in the \textit{purple} line. The inset shows the region around the hydrogen absorption line Br-$\lambda$ ($n=4\to n=11$).}}\label{fig:normfluxiter}
\end{figure*}

{We show in Figure~\ref{fig:polinomioiterations} the polynomial $P_n(\lambda)$ constructed from the $\sigma$-clipped \textit{red} chip of the solar spectrum, for selected  iterations. 
In each iteration, the number of wavelengths in the $\sigma$-clipped spectrum diminishes, as expected, since we aim to remove any noise or absorption feature to reveal the underlying featureless continuum. 
In each iteration, we construct 16 polynomials $P^i(\lambda)=\Sigma_i C_i\lambda^i$, with degrees $i=1$ to $16$, for the resultant $\sigma$-clipped spectrum.  
We select the polynomial with the smallest BIC value as the  best-fitting polynomial $P_n(\lambda)$, which is shown in Fig.~\ref{fig:polinomioiterations} as a magenta lines. 
Then, we continuum-normalise the observed spectra $S^{obs}(\lambda)$, dividing it by the recently found best-fitting polynomial of the iteration $P_n(\lambda)$.
Figure~\ref{fig:normfluxiter} shows this continuum normalised spectrum for three iterations. 
The inset at Fig.~\ref{fig:normfluxiter} shows the continuum normalised region around the conspicuous hydrogen absorption line Br$-\lambda$; differences between the continuum-normalised spectra are almost negligible.}

\begin{algorithm}
\caption{Normalisation of the Observed Spectrum}\label{alg:normspec}
\begin{algorithmic}[1]
\Procedure{normalisesSpectrum}{Observed spectrum $S^{\text{obs}}_\lambda$} 
\State $\hat{S}_{b}^{*}(\lambda)$, $\hat{S}_{g}^{*}(\lambda)$, $\hat{S}_{r}^{*}(\lambda)$, $n_b$, $n_g$, $n_r\gets$ \Call{SigmaClippingNorm}{$S^{\text{obs}}_{\lambda}$}\Comment{Algorithm \ref{alg:zeroclip}}
\Repeat
    \For{each chip}
        \State $P_n(\lambda)\gets$\Call{ConstructPolynomial}{$\hat{S}_{c}^{*}(\lambda)$, $n_c$} \Comment{Fits a polynomial of degree $n_c$ from the previous iteration, using \texttt{ polyfit} and \texttt{ poly1d,}}
    \EndFor
    \State $P_0(\lambda)\gets P_{b}(\lambda)+P_{g}(\lambda)+P_{r}(\lambda)$ \Comment{Piecewise polynomial of the previous iteration.}
    \State $\hat{S}_{(b,i)}^{*}(\lambda)$, $\hat{S}_{(g,i)}^{*}(\lambda)$, $\hat{S}_{(r,i)}^{*}(\lambda)$, $n_{(b,i)}$, $n_{(g,i)}$, $n_{(r,i)}\gets$ \Call{SigmaClippingNorm}{$\hat{S}^{*}_{\lambda}$}\Comment{Algorithm \ref{alg:zeroclip}}
    \For{each chip}
        \State $P_{(n,i)}(\lambda)\gets$\Call{ConstructPolynomial}{$\hat{S}_{(c,i)}^{*}(\lambda)$, $n_{(c,i)}$}
    \EndFor
    \State $P_{(0,i)}(\lambda)\gets P_{(b,i)}(\lambda)+P_{(g,i)}(\lambda)+P_{(r,i)}(\lambda)$ \Comment{Piecewise polynomial of this iteration.}
\Until{$P_{(0,i)}=P_{0}$}
\EndProcedure\\
\Return{normalised spectrum $P_{(0,i)}$}
\end{algorithmic}
\end{algorithm}

\begin{algorithm}
\caption{Normalisation and Clipping of the Observed Spectrum}\label{alg:zeroclip}
\begin{algorithmic}[1]
\Procedure{SigmaClippingNorm}{Observed spectrum $S(\lambda)$ or previously normalised spectrum $\hat{S}(\lambda)$}
\For{each chip}\Comment{Each chip is treated separately from the other two.} 
\For{$n=1,\ldots16$}
\State $P_n(\lambda)\gets$
    \Call{PolynomialFitting}{$S_c(\lambda)$, $n$} \Comment{Fit a polynomial of degree $n$ to the chip, \texttt{ polyfit} function from \texttt{ numpy}.}
\State $R_n\gets\sum\limits_{\lambda} \bigl(1-\frac{S^{\text{obs}}_{ c}(\lambda)}{P_n(\lambda)}\bigr)^2$ \Comment{Residuals sum of squares}
\State $BIC(P(n))\gets N\log(R_n/N)+(n+1)\log N$ \Comment{Bayesian Information Criterion for RSS}
\EndFor
\State $P_s(\lambda)\gets$ Select $P_n(\lambda)$ with the smallest $BIC$.
\State $\hat{S}_c(\lambda)\gets S_c(\lambda)/P_s(\lambda)$ \Comment{pixel per pixel}
\State $\hat{S}_c^*(\lambda)\gets$\Call{Sigma\_clip}{$\hat{S}_c(\lambda)$, $\sigma_l=1.2$, $\sigma_u=3$} \Comment{\texttt{ sigma\_clip} function from \texttt{ astropy.stats}.}
\EndFor
\EndProcedure  \\
\Return {$\hat{S}_{b}^{*}$, $\hat{S}_{g}^{*}$, $\hat{S}_{r}^{*}$, $n_b$, $n_g$, $n_r$} \Comment{Returns preliminary continuum-normalised spectrum  and the degree of the fitting polynomial with the lowest BIC for each chip.}
\end{algorithmic}
\end{algorithm}

\subsection{The asexual genetic algorithm implementation}\label{ap:aga}
The pseudo-codes of the implementation of AGA \citep{Canto09} are given below. 
The main routine, \texttt{AGATonalli}, generates the $N_0$ parents of the zero-$\mathrm{th}$ generation, from which $N_p$ individuals will be selected based on their fitness to become the parents of the subsequent generation.
The search hyper-volume decreases as generations are created. 
The offspring, with their respective parameters, is fathered by each parent in the subroutine \texttt{ AsexualReproduction}.
The fittest individuals in the generation are then selected and they carry  their attributes (the parameters) down to the next generation. 
The routine \texttt{ AGATonalli} ends when the convergence criteria is meet.
The fittest individual from the last generation is then the best-fitting model for the observed spectrum.

\begin{algorithm}
\caption{Implementation of AGA in \tonalli}\label{alg:agatonalli}
\begin{algorithmic}[1]
\Procedure{AGATonalli}{Observed spectrum $S^{\text{obs}}_\lambda$, Library $S_\lambda$, $N_0$, $N_0$, $p$} 
\Require $p \in (0,1)$, $N_0\ge N_p$, $N_p\ge2$.
\State $n=0$. \Comment{Zero-$\mathrm{th}$ generation}
\State $x_{(i,\text{min})},x_{(i,\text{max})}\gets$\Call{Read}{$S_\lambda$}. 
\Comment{Read limits from either the Library or user input. Subscript $i=1\ldots6$ denotes the stellar parameters.}
\For{$j=1,\ldots,N_0$} \Comment{zero-$\mathrm{th}$ Generation; subscript $j$ denotes an individual.}
    \State $x_{(i,j)}\gets$\Call{Random}{$x_{(i,\text{min})}, x_{(i,\text{max})}$} $\forall i$. 
    \Comment{Randomly generates $x_i$ from a uniform distribution within the intervals $[x_{(i,\text{min})}, x_{(i,\text{max})}]$.}  
    \State $\vec{X}_j \gets (x_{(1,j)}, \ldots, x_{(6,j)})$
    \Comment{Vector parameter $\vec{X}_j$ of the $j-\mathrm{th}$ individual.} 
    \State $S_{j}(\lambda)\gets$\Call{Interpolation}{$S^*_{\lambda}; x_{(1,j)},\ldots, x_{(4,j)}$}
    \Comment{Interpolation from a limited number of synthetic spectra, $S^*_{\lambda}$, to obtain the spectrum for the individual with stellar parameters $x_{(1,j)}, \ldots, x_{(4,j)}$.}
    \State $S_{j}(\lambda)\gets$\Call{Broadening}{$S_{j}(\lambda)$}.
    \Comment{Rotational broadening of the synthetic spectrum}
    \State $S_{j}(\lambda)\gets$\Call{DopplerShift}{$S_{j}(\lambda)$}
    \Comment{Doppler shift of synthetic spectrum}.
    \State $\chi^2_j\gets$\Call{Fitness}{$S_{j}(\lambda)$}
    \Comment{Fitness of $S_{j}(\lambda)$, equation~(\ref{eq:FOM})}
\EndFor
\State $V_0\gets(\vec{X_j}, S_{j}(\lambda), \chi_j^2)$. \Comment{Information of the $N_0$ individuals in a matrix $V_0$ of $N_0\times 3$ dimensions.}
\State $V_{(0,\text{best})}\gets$\Call{Sort}{$V_0$ by $\chi^2$} 
\Comment{Sort $V_0$ select the first $N_p$ rows. Matrix $V_{(0,\text{best})}$ of $N_p\times 3$ dimensions}
\While{convergence criteria are not meet} \Comment{Subsequent generations}
    \State $V_{\text{best}}\gets V_{n}$ \Comment{Parents are the fittest individuals of the previous generation.}
    \State $n\gets n+1$ \Comment{Number of generation}
    \State $(\Delta x_{i})_n\gets\bigl(\Delta x_{i}\bigr)_0 p^n$ \Comment{Equation~(\ref{eq:deltax})}
    \For{$k=1,\ldots,N_p$}
    \Comment{$k-\mathrm{th}$ best-fitting individual of the generation $n$: parent}
        \State $x_{(ik,\text{min})}\gets x_{(ik,\text{best})}-(\Delta x_{i})_n/2$, $x_{(ik,\text{max})}\gets x_{(ik,\text{best})}+(\Delta x_{i})_n/2$.  \Comment{Sides of the hyper-volume centred in $x_{(ik,\text{best})}$}
        \State $V_k\gets$\Call{AsexualReproduction}{$S^{\text{obs}}(\lambda)$, $S^*(\lambda)$, $N_p$, $x_{(ik,\text{min})}$, $x_{(ik,\text{max})}$}\Comment{Asexual reproduction of the $k-\mathrm{th}$ individual, see Algorithm \ref{alg:agtonalli}} 
    \EndFor
    \State $V_n\gets V_k$ $\forall k$  \Comment{Information of the $(N_p-1)\times N_p$ children of the $N_p$ parents, parents included.}
    \State $V_{(n,\text{best})}\gets$\Call{Sort}{$V_n$ by $\chi^2$} 
\EndWhile
\State Select the fittest element from $V_{\text{best}}$ of the last generation: best-fitting model for the observed spectrum.
\EndProcedure
\end{algorithmic}
\end{algorithm}

\begin{algorithm}
\caption{Asexual Reproduction in \tonalli}\label{alg:agtonalli}
\begin{algorithmic}[1]
\Procedure{AsexualReproduction}{Observed spectrum $S^{\text{obs}}(\lambda)$, Library $S^*(\lambda)$, $N_p$, $x_{(i,\text{min})}$, $x_{(i,\text{max})}$} 
        \For{$j=1,\ldots, N_p-1$} 
            \State $x_{(i,j)}\gets$\Call{Random}{$x_{(ik,\text{min})}, x_{(ik,\text{max})}$} $\forall i$
            \State $\vec{X}_j \gets (x_{(1,j)}, \ldots, x_{(6,j)})$
            \State $S_{j}(\lambda)\gets$\Call{Interpolation}{$S^*(\lambda); x_{(1,j)}, \ldots, x_{(4,j)}$}
            \State $S_{j}(\lambda)\gets$\Call{Broadening}{$S_{j}(\lambda)$, $x_{(5,j)}$}.
            \State $S_{j}(\lambda)\gets$\Call{DopplerShift}{$S_{j}(\lambda)$, $x_{(6,j)}$}
            \State $\chi^2_j\gets$\Call{Fitness}{$S^{\text{obs}}(\lambda)$,$S_{j}(\lambda)$}
        \EndFor
    \State $V\gets(\vec{X_j}, S_{j}(\lambda), \chi_{j}^2)$.
    \Return $V$ 
    \Comment{Information of the $(N_p-1)$ children of the parent, parent included.}
\EndProcedure  
\end{algorithmic}
\end{algorithm}

\clearpage

\section{Influence of the input parameters in the recovery of the solar atmospheric parameters}\label{appendix:inputparameters}

The following experiments were conducted with the \apo~ DR17 solar spectrum reflected by Vesta as our sample spectrum.

\subsection{Parameters controlling AGA: optimisation}
We perform 108 optimisations (i.e. minimizing $\chi^2$) with \tonalli, varying the input parameters $N_0$, $N_p$, and $p$ as follows: $N_0=[25,\,50,\,75,\,100,\,250,\,500,\,750,\,1000,\,2500]$; $N_p=[5,\,10,\,15,\,20]$ (which is equivalent to $N_p^2+N_p=[30,\,110,\,240,\,420]$ individuals in each asexual generation), and $p=[0.4,\,0.6,\,0.8]$. For this experiment, the number of spectra used in the interpolations was fixed to $N_{\text{interpol}}=2\times2\times2\times2=16$ for the \textit{coarse} interpolation, and to $N_{\text{interpol}}=2\times2\times3\times3=36$ for the \textit{fine} interpolation.

\begin{figure}
\includegraphics[alt={Graph demonstraing the increase of the total population generated in a tonalli run with the number of parents per generation and the convergency factor.}, width=1\columnwidth]{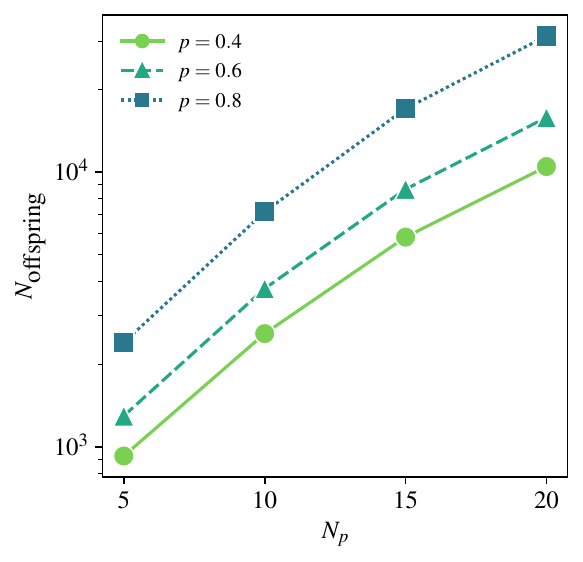}
\caption{{The dependency of the total offspring population $N_{\text{offspring}}$ as function of the number of parents per generation $N_p$ ($x$-axis) and the convergency factor $p$ (scatter symbols).}}\label{fig:noffvsnpp}
\end{figure}

We introduce the variables $N_{\text{tot}}$, defined as the zero-$\mathrm{th}$ generation parents and their total offspring, that is, all the individuals created in all generations of AGA in a single \tonalli~ run, and $N_{\text{offspring}}$, the total offspring (all the individuals created after the Monte Carlo zeroth generation), $N_{\text{tot}}\equiv N_0+N_{\text{offspring}}$.
We note that neither of the preceding variables are input parameters, but the result of the interplay between the input parameters $N_0$, $N_p$, and $p$.
We recall from Section~\ref{sec:parameters} that the convergence factor $p$ controls the hyper-volume decrement per generation: smaller values of $p$ imply fewer created generations, thus restricting the offspring $N_{\text{offspring}}$. 
For our models, given $N_p$, the models with $p=0.8$ are between $\sim2.5$ and $\sim3$ times larger than the total offspring of the models with $p=0.4$.
Figure~\ref{fig:noffvsnpp} shows the non-linear relationship between $p$, $N_p$ and $N_{\text{offspring}}$.

To qualify the accuracy of \tonalli~at the optimisation run, we compute the differences between the expected solar value of a given grid parameter and the parameter of the best-fitting model against $N_{\text{offspring}}$, $X_\odot-X_{\tonalli}$. 
The $108\times4$ differences are shown the scatter plots of Figure~\ref{fig:paramntot}: the larger $N_{\text{offspring}}$ is, the difference between the solar value and the \tonalli~value decrease, irrespective of the input parameters $N_p$, $p$, and $N_{\text{0}}$, which is the desired outcome.

We explore the combined effect of $N_{\text{offspring}}$ (which serves as a proxy of both $N_p$ and $p$) and the input Monte Carlo zero-$\mathrm{th}$ generation $N_0$, plotting in Figure~\ref{fig:paramntot} the differences $X_\odot-X_\tonalli$ now as function of the total number of individuals created in a \tonalli~run, $N_{\text{total}}$. 
The scatter in the differences decreases with sufficiently high $N_{\text{total}}$. 
For the parameters $[\text{M/H}]$, $[\alpha/\text{M}]$, and $\log(g)$, $N_{\text{total}}$ can be as low as $\sim4000$ individuals. 
However, the scatter in the effective temperature differences shows the need of a large total population ($N_{\text{total}}\gtrsim10000$) to obtain an accurate result.

We remind the reader that the above results were obtained assuming one Monte Carlo realisation and adopting the minimum $\chi^2$ as the FOM, which was explained in detail in Section~\ref{sec:algorithm}. 
In practice, we perform several Monte Carlo realisations, as explained in Sections~\ref{Section:MinimumExp}, \ref{subs:montecarlo1}, and \ref{Subsection:VestaMinRep}, to obtain a \textit{best-value} and a \textit{credible interval} (see Section~\ref{sec:modeltonallivesta}) for the solar atmospheric parameters.
In this case, the wall-clock time is a variable to take into account with limited computational resources and/or when trying to obtain the stellar parameters for a large number of spectra. 
Figure~\ref{fig:timetot} exhibits how the input parameters impact the total computing time (the wall-clock time), with 25 allocated CPUs of the multicore CPU (AMD Ryzen  3990X 64-Core Processor), while running simultaneously four models. Each model in \tonalli~ ran in parallel \citep[using the python module \tt{parmap},][]{parmap}. 
For models with $N_p=5$, the wall-clock time scales linearly with $N_{\text{total}}$, whereas for larger $N_p$,  time scales as a fractional power of the total number of individuals, $t\propto \sqrt{N_{\text{total}}}$. 
At any rate, we aim for running times sufficiently low enough to ensure total short wall-clock times when we run \tonalli~ to obtain credible intervals for the stellar atmospheric parameters, that is, running \tonalli with several Monte Carlo realisations.

The input parameters $p=0.4$, $N_p=10$ and $N_0=250$ (close to or adopted input parameters $p=0.4$, $N_p=10$ and $N_0=250$) procures accurate results within an acceptable wall-clock time. 

\begin{figure}
\includegraphics[alt={Graphs showing the bias of the results for the Solar spectrum in a single tonalli run, for several combinations of the asexual genetic algorithm hyper parameters, as function of the offspring computed in the asexual reproductions.}, width=\columnwidth]{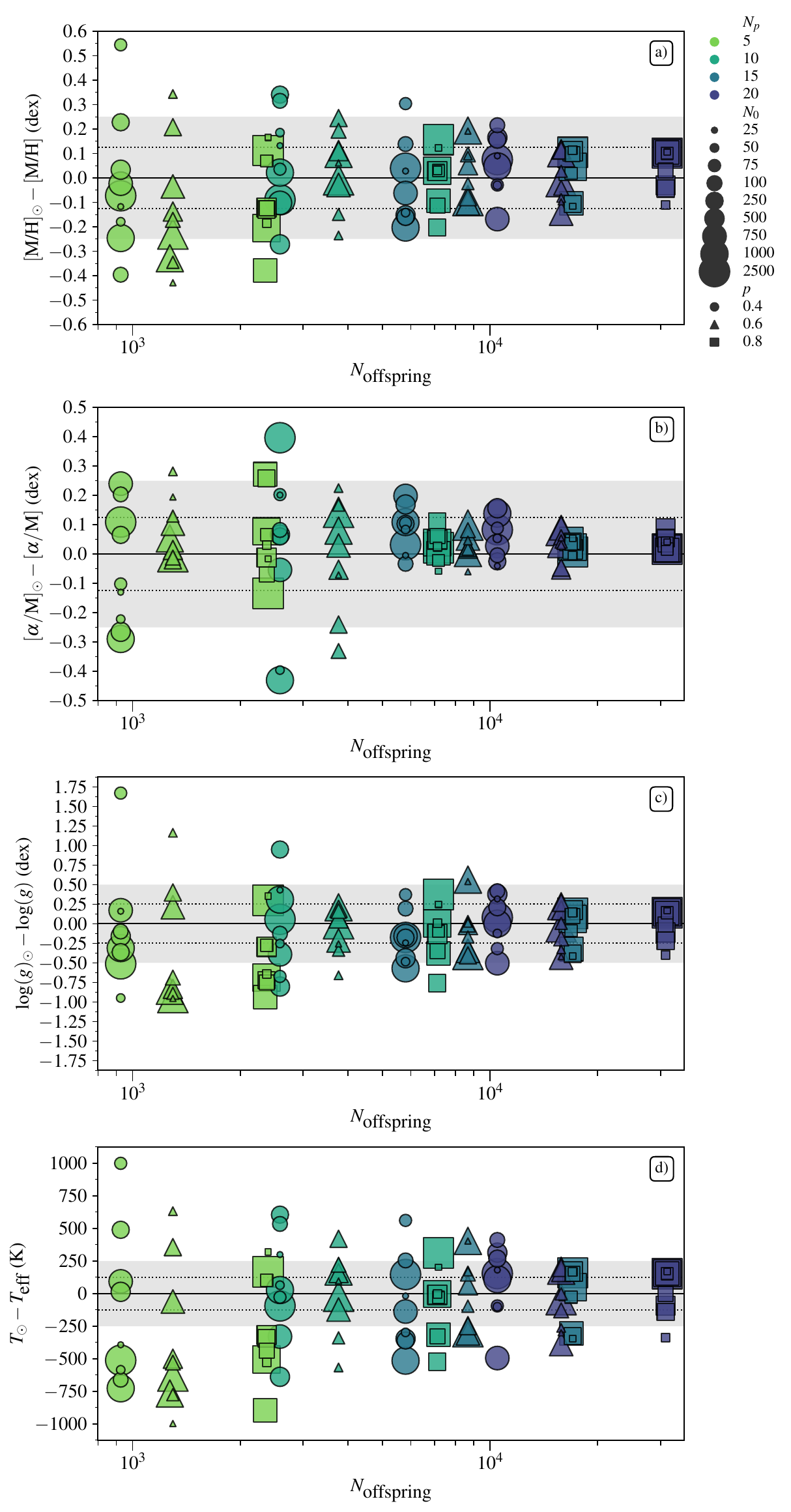}
\caption{{Difference between the expected parameter value $X_\odot$ and the parameter value obtained by \tonalli~ as function of the total offspring $N_{\text{offspring}}$. For all the subplots, the number of parents per generation $N_p$ are depicted by the color (\textit{green} hues indicate smaller $N_p$, \textit{purple} hues correspond to larger $N_p$); the size of the symbols portrays the \textit{zero-$\mathrm{th}$} generation $N_0$ (a larger symbol corresponds to a larger $N_0$, while the symbol represents the \textit{convergence factor} $p$ (\textit{circle}: $p=0.4$, \textit{triangle}: $p=0.6$, \textit{square}: $p=0.8$. The \textit{solid horizontal line} shows $X_\odot-X_\tonalli=0$; the \textit{dotted horizontal lines} display $\pm1/2\Delta_X$ (half the grid step of the synthetic library), while the \textit{grey rectangular area} comprises differences within $\pm\Delta_X$ .  \textit{Panel a)}: difference in metal abundance, assuming $[M/H]_\odot=0$. \textit{Panel b)}: difference in $\alpha$-elements abundance, assuming $[\alpha/\text{M}]_\odot=0$. \textit{Panel c)}: difference in logarithm of the surface gravity, assuming $\log(g)_\odot=4.44$~dex \textit{Panel d)}: difference in effective temperature, assuming $T_\odot=5777$~K.}}\label{fig:paramnoff}
\end{figure}

\begin{figure}
\includegraphics[alt={Graphs showing the bias of the results for the Solar spectrum for several combinations of the asexual genetic algorithm hyper parameters, as function of the total population generated in a single tonalli run.}, width=\columnwidth]{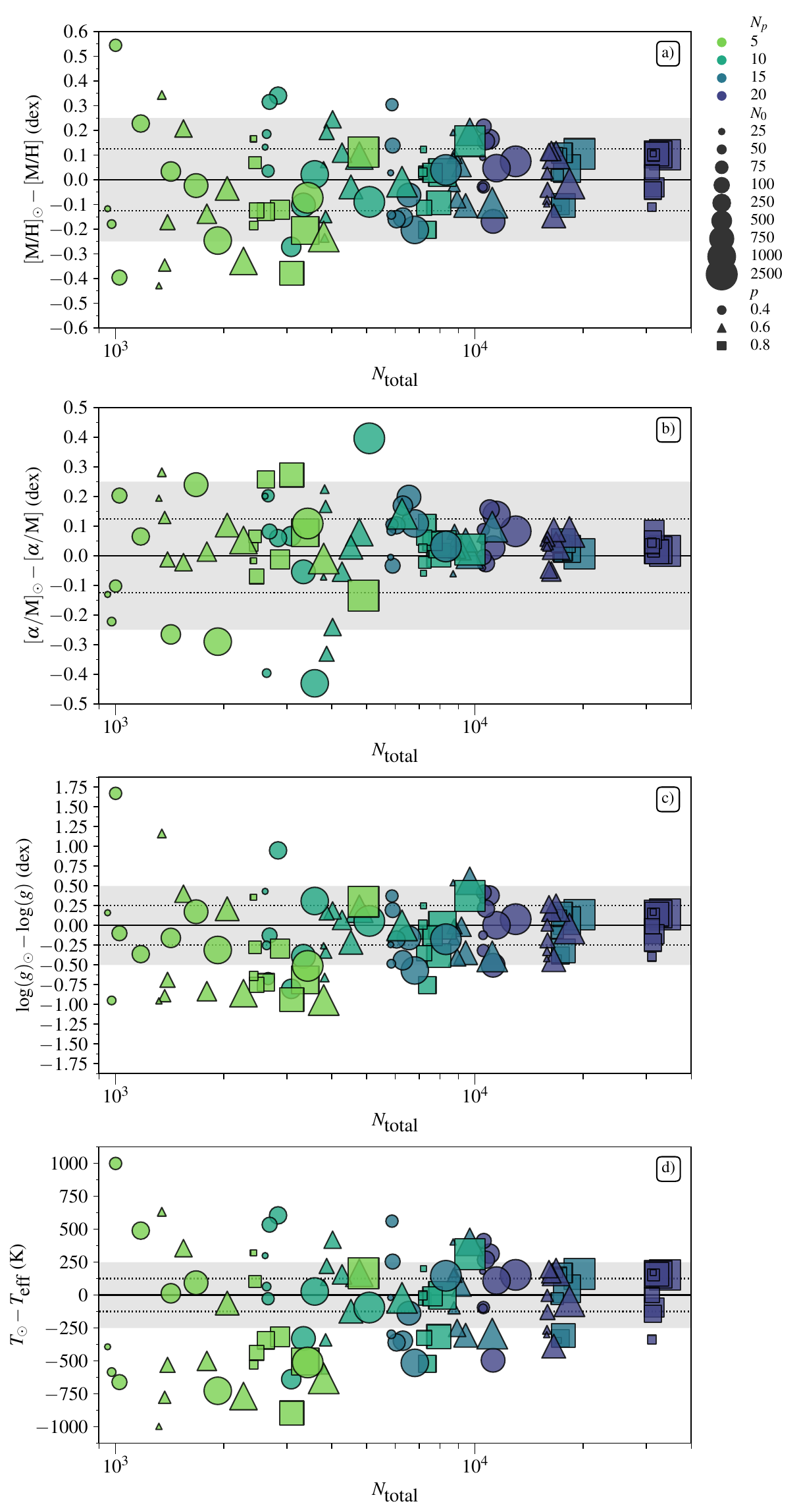}
\caption{{Difference between the expected parameter value $X_\odot$ and the parameter value obtained by \tonalli~ as function of the total individuals $N_{\text{total}}$ created in a \tonalli~run. Symbols are the same as in Fig.~\ref{fig:paramnoff}.}}\label{fig:paramntot}
\end{figure}

\begin{figure}
\includegraphics[alt={Graphs showing computing time of a single tonalli run for several combinations of the asexual genetic algorithm hyper parameters.}, width=1\columnwidth]{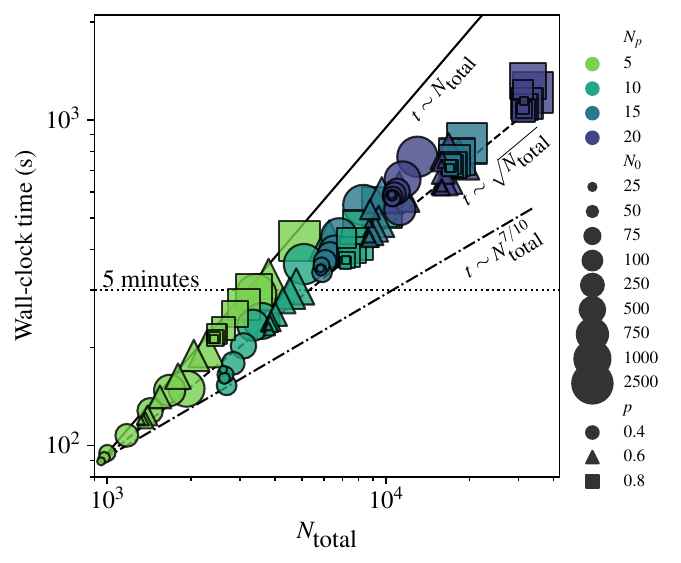}
\caption{{Wall-clock time $t$ of \tonalli~ as a function of $N_{\text{total}}$, with 25 (out of 128) allocated CPUs. The \textit{dotted} horizontal line marks five minutes of elapsed real time. We include three relationships of the wall-clock time with the total individuals $N_{\text{total}}$: $t\propto N_{\text{total}}$ (\textit{solid} line), $t\propto \sqrt{N_{\text{total}}}$ (\textit{dashed} line), and $t\propto N^{7/10}_{\text{total}}$ (\textit{dot-dashed} line).  
See \ref{fig:paramnoff} for details of the symbols.}}\label{fig:timetot}
\end{figure}

\subsection{Parameters controlling AGA: Monte Carlo realisations}\label{app:montecarlo}
We now examine the effects of the input parameters in the accuracy of \tonalli~ when dealing with multiple Monte Carlo realisations.
We repeat the above experiments, switching on the Monte Carlo procedure to obtain the so-called \textit{best-fitting} solar parameters and their associated \textit{credible interval}, as done in Section~\ref{Section:Results}, with 50 Monte Carlo realisations. 
We adopt the univariate median and the interquartile range IQR of the Monte Carlo distributions, as explained in Section~\ref{Section:univariate}.

\begin{figure}
\includegraphics[alt={Graphs showing the bias of the results for the Solar spectrum for several combinations of the asexual genetic algorithm hyper parameters, as function of the total offspring computed in the Monte Carlo realisations of tonalli.}, width=\columnwidth]{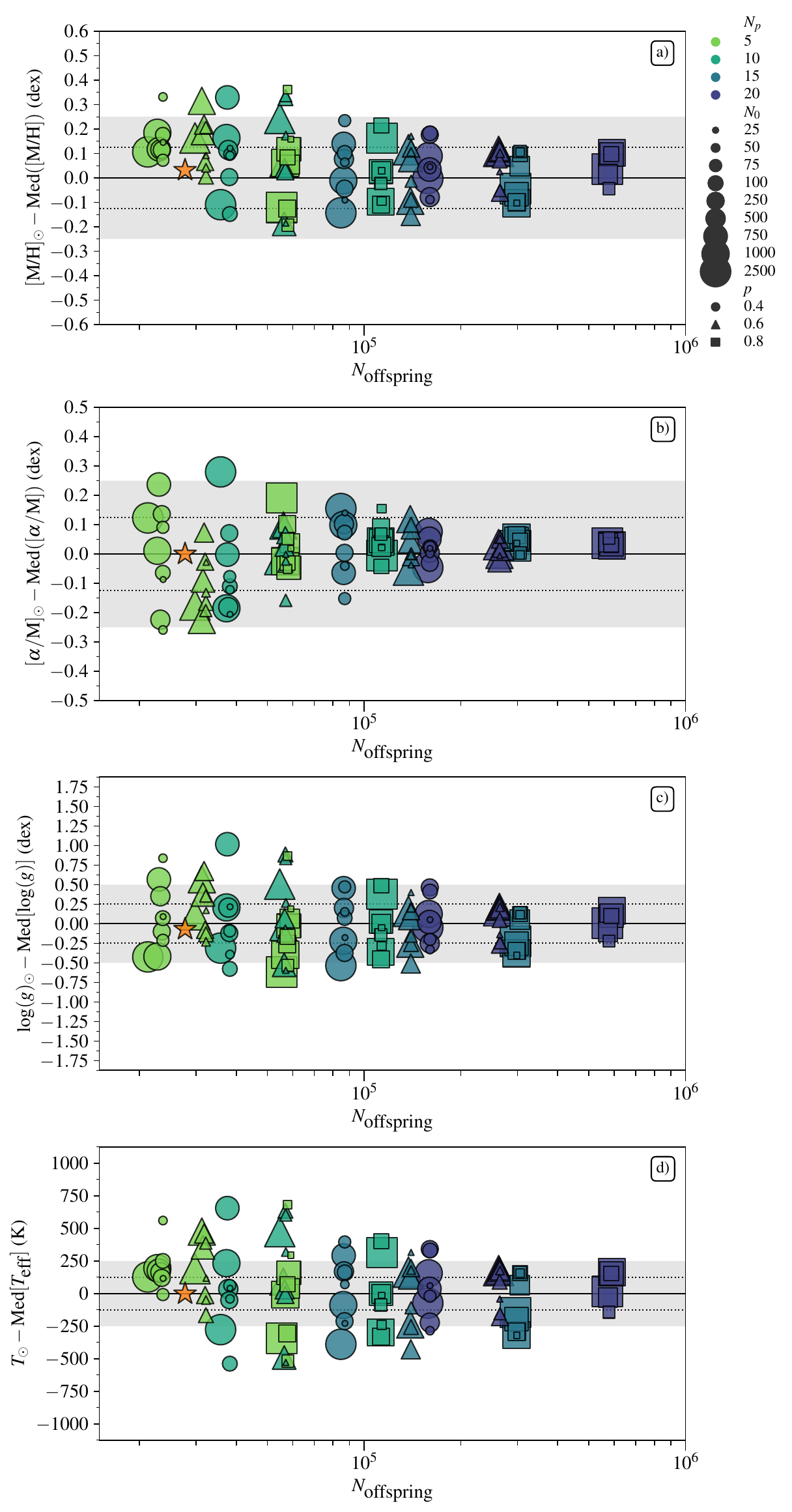}
\caption{{Difference between the expected parameter value $X_\odot$ and the \textit{median} of the 1D Monte Carlo distribution of the solar parameter obtained by \tonalli~ as function of the total offspring $N_{\text{offspring}}$. The \textit{orange star} points the median value obtained in Section~\ref{Section:univariate}. \ref{fig:paramnoff} for details of the symbols.}}\label{fig:paramnoffMC}
\end{figure}

\begin{figure}
\includegraphics[alt={Graphs showing the half of the interquartile ranges of the results for the Solar spectrum for several combinations of the asexual genetic algorithm hyper parameters, as function of the total offspring computed in the Monte Carlo realisations of tonalli.}, width=\columnwidth]{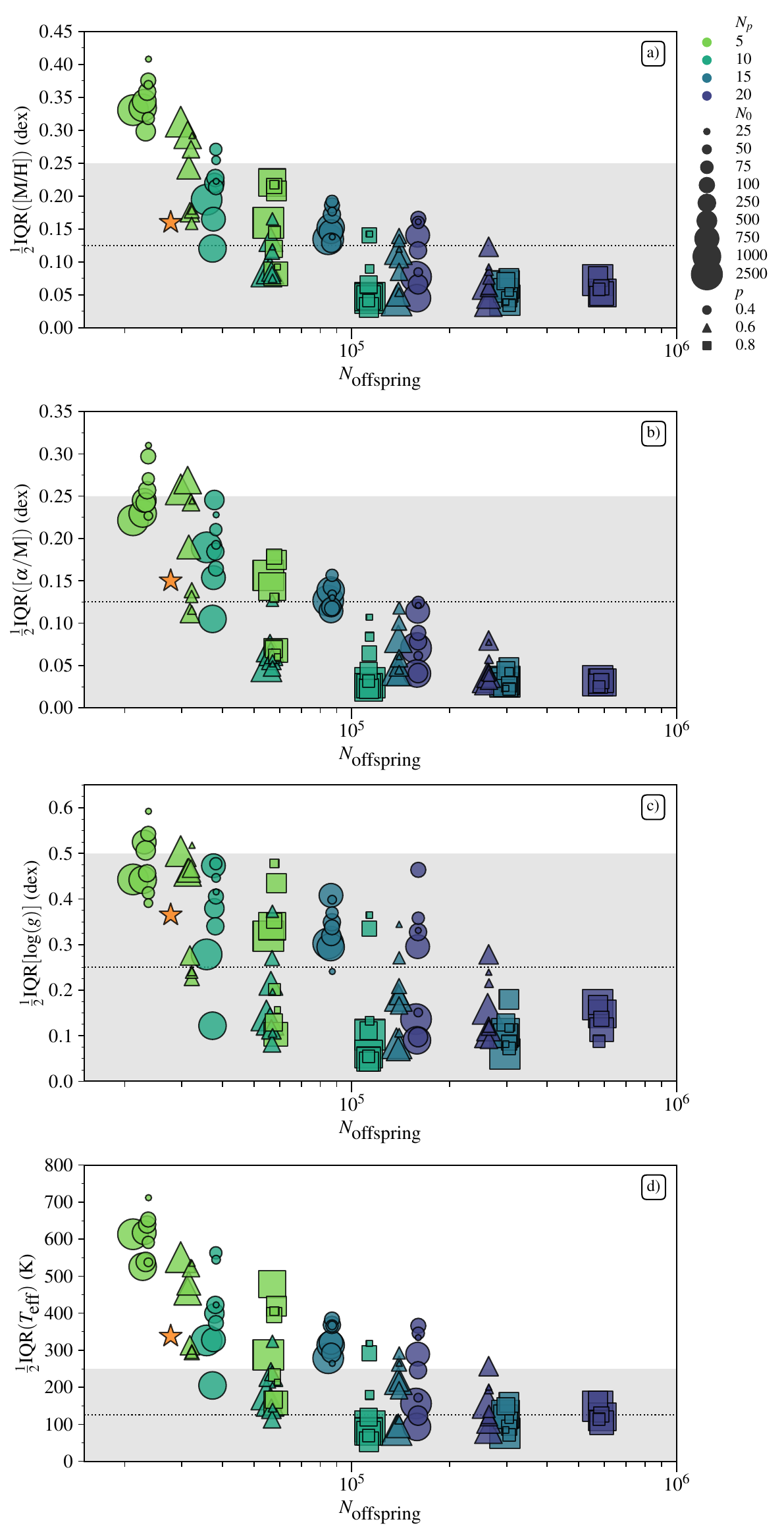}
\caption{{Half the interquartile range of the 1D Monte Carlo distribution of the solar parameter  obtained by \tonalli~ as function of the total offspring $N_{\text{offspring}}$. The \textit{orange star} points half the IQR of the 1D distribution of the parameter obtained in Section~\ref{Section:univariate}. See \ref{fig:paramnoff} for details of the symbols.}}\label{fig:IQRMC}
\end{figure}

The results are shown in Figures~\ref{fig:paramnoffMC} and ~\ref{fig:IQRMC}. 
We just show the dependence of the medians and of the IQR values on the offspring population, since $N_{\text{offspring}}\sim N_{\text{total}}$. 
We observe in Figure~\ref{fig:paramnoffMC} that the difference between the solar expected values and the median parameters are again slightly inaccurate for certain combinations of $N_p$ and $p$, that is, for small $N_{\text{offspring}}$. However, the accuracy improves respect to the optimisation best-fitting results (compare with ~\ref{fig:paramnoff}).
Regarding the credible interval, Figure~\ref{fig:IQRMC} shows a clear, not unexpected, negative correlation of the precision of our method with the total offspring $N_{\text{offspring}}$, which is actually a correlation with both $N_p$ and $p$. 
At any rate, for accurate ($X_\odot-\text{Median}[X]\sim 0$) and precise ($\frac{1}{2}\text{IQR}[X]\lesssim \frac{1}{2}\Delta_X$, where $\Delta_X$ is the \marcs~ grid step) determination of the solar atmospheric parameters, \tonalli~ requires combinations of the input parameters that result in $N_{\text{offspring}}$ in excess of $\sim10^5$ individuals.

The orange star point in the plots shown in Figures~\ref{fig:paramnoffMC} and ~\ref{fig:IQRMC} represents the univariate median and the IQR of the Monte Carlo simulation distributions as described in Section~\ref{Section:univariate}, with the following input parameters adopted values : $N_0=250$, $N_p=6$, $p=0.4$. With this choice of $N_p$ and $p$, we sacrifice precision (in the effective temperature and to some extent, in the logarithm of the surface gravity estimations) for computational time.


\bsp	
\label{lastpage}
\end{document}